\documentclass[a4paper]{ar-1col} 
\usepackage{ulem, url, amssymb, amsmath, natbib, caption, subcaption}  

\newcommand{\bfx}{{\boldsymbol{x}}}
\newcommand{\Mpc}{{\rm Mpc}}

\newcommand{\muK}{\mu {\rm \;K}}
\newcommand{\Msun}{{M_{\odot}}}

\newcommand{\HI}{H{\sc \,i}}
\newcommand{\HII}{H{\sc \,ii}}
\newcommand{\HeI}{He{\sc \,i}}
\newcommand{\HeII}{He{\sc \,ii}}
\newcommand{\HeIII}{He{\sc \,iii}}
\newcommand{\CIV}{C{\sc \,iv}}
\newcommand{\OVI}{O{\sc \,vi}}
\newcommand{\OVII}{O{\sc \,vii}}
\newcommand{\OVIII}{O{\sc \,viii}}
\newcommand{\OI}{O{\sc \,i}}

\newcommand{\SIV}{Si{\sc \, IV}}

\begin{document}


\jname{Annu. Rev. in Astronomy \& Astrophysics}
\jyear{2016}
\jvol{1}

\title{The Evolution of the Intergalactic Medium}

\markboth{Matthew McQuinn}{intergalactic medium}

\author{Matthew McQuinn
\affil{Department of Astronomy, University of Washington, Seattle, WA 98195, USA; mcquinn@uw.edu}}

\begin{abstract}
 The bulk of cosmic matter resides in a dilute reservoir that fills the space between galaxies, the intergalactic medium (IGM). The history of this reservoir is intimately tied to the cosmic histories of structure formation, star formation, and supermassive black hole accretion.  Our models for the IGM at intermediate redshifts ($2 \lesssim z \lesssim 5$) are a tremendous success, quantitatively explaining the statistics of Ly$\alpha$ absorption of intergalactic hydrogen.  However, at both lower and higher redshifts (and around galaxies) much is still unknown about the IGM.  We review the theoretical models and measurements that form the basis for the modern understanding of the IGM, and we discuss unsolved puzzles (ranging from the largely unconstrained process of reionization at high-$z$ to the missing baryon problem at low-$z$), highlighting the efforts that have the potential to solve them.
\end{abstract}


\maketitle

\tableofcontents

\section{introduction}
Most cosmic matter resides in the void between galaxies known as the intergalactic medium (IGM).  At present, roughly half of the dark matter resides in structures that can more or less be thought of as intergalactic (i.e. unvirialized objects and dark matter halos likely too small to contain a galaxy, taken here to be $<10^{9}\Msun$).  The present fraction of baryons that are intergalactic is probably much higher.  At earlier cosmic times, the intergalactic reservoir was even larger as fewer massive halos had coalesced from it.  By $z=3$\,(6), a much larger $80\%\,$(95\%) of the dark matter was extragalactic.  Eventually, before $z = 20$ or so, the first galaxies had yet to form, and so the term ``intergalactic'' is no longer applicable.  This review starts at this time, a time for which observations are scant.  It ends at the present, again an epoch with few observations, as the IGM has become so tenuous that only its densest constituents are visible.  A major focus of this review will be the intermediate epochs, when the Universe was one to a few billion years old.  During these epochs a rich set of observations provide a fabulous test of IGM models.

An understanding of the IGM is relevant for many astrophysical disciplines.  For the cosmologist, the IGM has been used to test our models of structure formation on the smallest comoving scales \citep{viel05, 2005PhRvD..71j3515S}, it induces anisotropies in the cosmic microwave background \citep{ostriker86, hu00}, and the astrophysical processes that shape the IGM can bias cosmological parameter inferences from galaxy clustering \citep{pritchard07, wyithe11}.  For the large community interested in galaxy formation, the IGM is the trough from which galaxies feed, setting the minimum mass of galaxies \citep{1986MNRAS.218P..25R, 1992MNRAS.256P..43E, 1996ApJ...465..608T}, and feeding some galaxies more than others \citep[e.g.][]{keres05}.  In addition, the IGM can be used to measure the total ionizing photon production of galaxies \citep{haardt96, miralda03, faucher08, becker13} as well as galaxies' stellar nucleosynthetic yield, as  a large fraction is ejected into the IGM \citep[e.g.][]{peeples14}.  For those who focus on our own Milky Way, aside from the obvious connection that the Milky Way too is fed by the same intergalactic spigot, we note that many of the physical processes at play in the IGM are simplified versions of those encountered in the interstellar medium (ISM).

An appeal of studying the IGM is that we have an excellent understanding of the cosmological initial conditions and of the equations that govern much of the subsequent evolution of what becomes the IGM.  While uncertain astrophysical processes certainly enter into the equations, for times well after reionization and densities less than ten times or so the cosmic mean, the consensus view is that astrophysical effects can essentially be ignored (as astrophysical radiation backgrounds are approximately uniform and galactic mechanical feedback is mostly localized in the dense regions around the sources).  This view (which has been tested extensively in the \HI\ Ly$\alpha$ forest) enables intergalactic studies to search for subtle effects, potentially improving constraints on the cosmological initial conditions, on the sources of intergalactic heating, and on how galactic winds operate.

The aim of this review is to highlight the developments that form the modern understanding of the IGM.  This means covering a lot of ground (spanning from the cosmic dark ages to the present day) at the expense of only being able to discuss the major developments on each topic.  However, this approach allows us to present a more holistic understanding of the IGM, which we think is particularly important as, e.g., the properties of the IGM at any epoch are connected to those at other epochs.   For uncovered topics, we point the reader to the more comprehensive recent review of \citet{meiksin09}.  There are also reviews from the last decade that overlap with topics covered in this review: \citet{fan-review} on reionization,  \citet{2007ARA&A..45..221B} on the missing baryon problem, \citet{2010ARA&A..48..127M} on the high-redshift $21\,$cm signal, and \citet{bromm11} on the first galaxies.  We find, largely because our focus on the history and properties of the IGM, that the overlap is not so substantial.
  
This review adopts a narrow definition of the IGM as being anything outside of the virial radius of galaxies and clusters (the medium between halos rather than the medium between galaxies).  In terms of density, this means we are considering gas that has densities less than $\sim 200$ times the mean cosmic density.   We will not cover the literature on damped Ly$\alpha$ systems \citep[][which are primarily on the outskirts of galactic disks]{wolfe05}, on the intracluster medium \citep{rosati02, kravtsov12}, and on the circumgalactic medium \citep{stocke13, werk14}.  For many observations, however, the distinction between intergalactic and galactic/circumgalactic can be difficult. In fact there is debate as to whether almost all of the ``intergalactic'' metal line absorption at $z\sim 0$ could actually arise from within virialized regions \citep{shull10, 2011ApJ...740...91P}.  We use this pretext to opt out of a detailed summary of the extensive literature on metal absorption lines in quasar spectra, focusing primarily on pixel optical depth constraints on metal absorption (which are unquestionably probing intergalactic gas).
 
The calculations presented in this review assume the concordance flat $\Lambda$CDM cosmological model with $\Omega_m \approx 0.3$, $\Omega_b \approx 0.045$, $\sigma_8 \approx 0.8$,  $n_s \approx 0.96$, and $Y_{\rm He} = 0.25$ \citep{planck}, although the precise value of these parameters depends on the study being summarized.  We also assume the standard Fourier convention used in cosmology in which the $(2\pi)$'s only appear under the $dk$'s.  We now briefly overview the history of the IGM, and we use this overview to serve as an outline for ensuing sections.

\subsection{a brief overview and review outline}

\begin{figure}[t!]
  \centering
    \begin{subfigure}[t]{0.49\textwidth}
    \centering
    \includegraphics[width=1\textwidth]{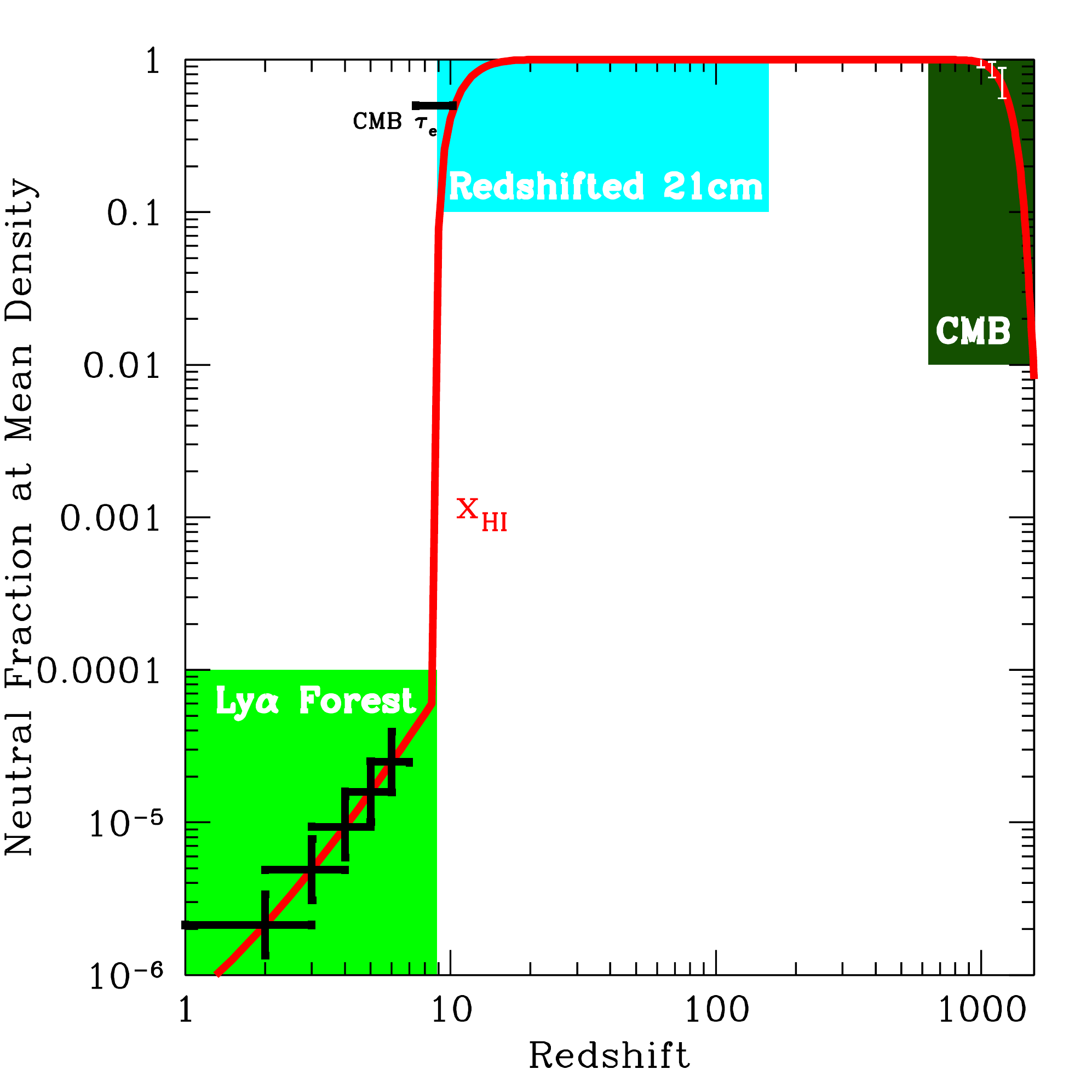}
\end{subfigure}
    \begin{subfigure}[t]{.49\textwidth}
    \centering
\includegraphics[width=1\textwidth]{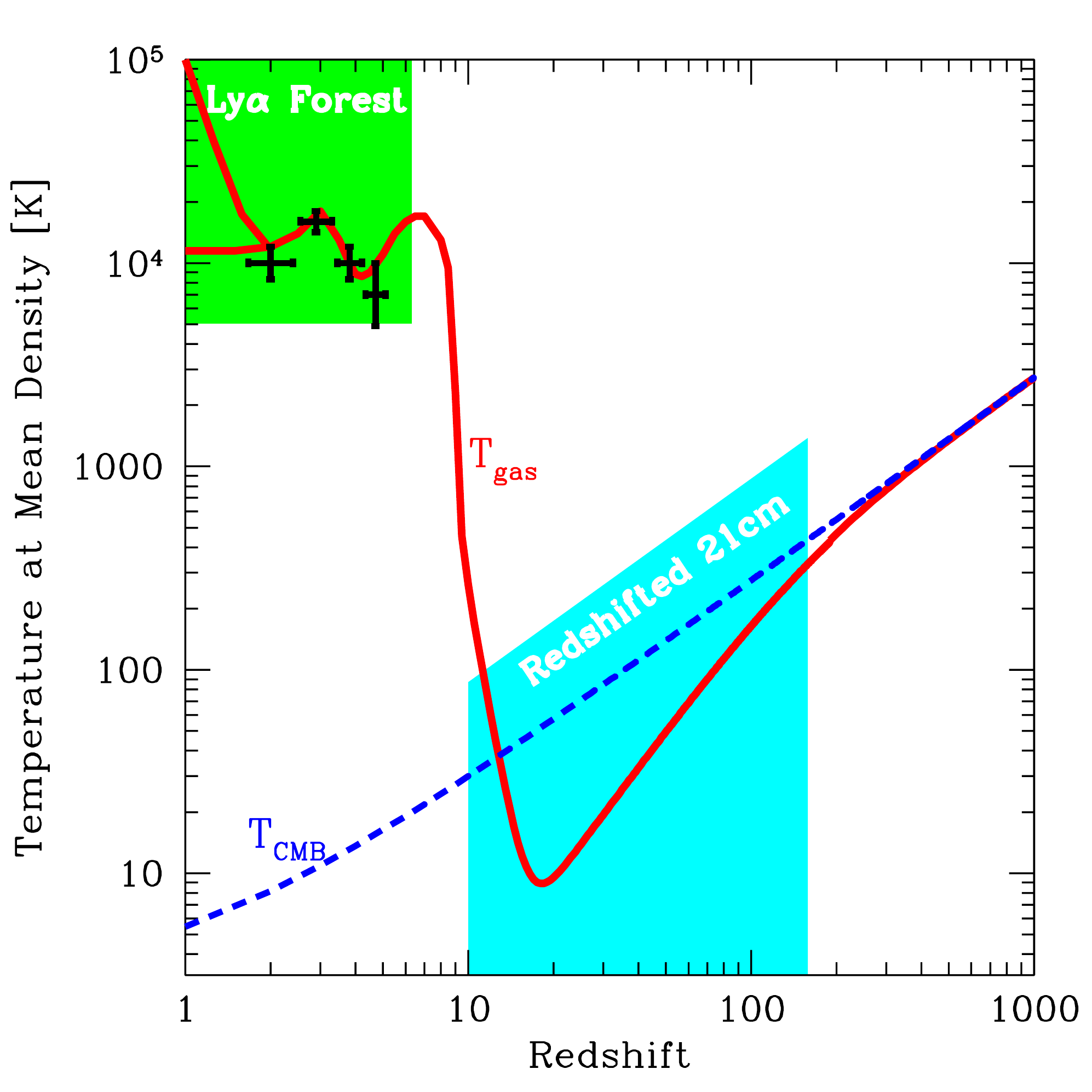} 
\end{subfigure}
\caption{Cartoon showing the ionization and thermal history of intergalactic gas. The red curves show a model of intergalactic gas.  Error bars symbolize existing constraints, and the highlighted regions illustrate the potential purview of the named cosmological probe.   In the temperature panel, the model curve bifurcates at low redshifts to indicate the IGM temperature becoming multiphase. 
}
\label{fig:thermalev}
\end{figure}

Nearly scale invariant and Gaussian potential fluctuations (evolved from early times in a $\Lambda$CDM background Friedmann-Lema\^itre-Robertson-Walker cosmology with a photon-to-baryon ratio of a billion) explain the statistical properties of the cosmic microwave background (CMB) and the clustering of galaxies \citep{peebles, dodelson}, not to mention the nucleosynthetic yields of the Big Bang \citep{BBN}.  Also from CMB observations we know that the cosmic gas ``recombined'' and became neutral around the Universe's $400,000^{\rm th}$ birthday ($z\approx 1100$).  Subsequently this gas cooled with the expansion of the Universe.  It is from these cold conditions that the IGM emerged, its structure largely the result of gravity acting on the primordial matter fluctuations.

The history of the IGM is tied to the history of star and supermassive black hole formation as these objects are thought to be the primary sources of intergalactic heating, ionization, and metal enrichment. The first star likely formed in the rarest peak in the cosmological density field at a redshift of $\sim 70$ \citep{naoz06}.  This star and its more immediate brethren are thought to have formed in $10^5-10^7\Msun$ halos -- halos whose virialized gas is hot enough to cool and condense by exciting molecular hydrogen transitions.  Gradually, more and more stars formed in the Universe.  \citet{trenti09} estimate that the Universe reached a saturation-level of POPIII star formation in minihalos of $\sim 10^{-5}-10^{-4}\Msun$~yr$^{-1}$~comoving~Mpc$^{-3}$ at $z\sim 20-30$, with saturation set by their $11-13~$eV radiative backgrounds destroying the molecular hydrogen coolant that had created them \citep{haiman00}.  Such small star formation rates are insufficient to ionize but a small fraction of the then neutral IGM, even for a top-heavy stellar population that simulations find is most likely \citep{2002Sci...295...93A, greif11}.  If some of these stars ended their lives in pair instability supernovae or in other exotic stellar deaths, then they may be detectable in the IGM via their enrichment patterns \citep[][ \S~\ref{ss:metals}]{heger02, kulkarni13}.
  
 The first galaxies formed later in halos with masses of $\gtrsim 10^8 \Msun$, halos that could cool by more robust atomic transitions and that formed in abundance around $z\sim 10$ \citep{barkana01, bromm11}. Unlike the diminutive dark matter halos that hosted the first stars, which likely formed only a handful of stars \citep{2002Sci...295...93A, greif11, bromm04}, these halos were able to harbor more sustained star formation \citep{1999ApJ...513..142M, bromm11}.  It is thought that even our soon-to-be-launched large space telescope, the James Web Space Telescope, will not be able to directly image galaxies in halos with $< 10^{9-10} \Msun$ at $z>6$ \citep{kuhlen12, behroozi15} and so the first galaxies may be most detectable through their impact on the IGM.  
 
 Once the first galaxies emerged, there was not much time before the IGM was affected by the associated radiative backgrounds. First, their $\sim 10\,$eV and soft $X$-ray backgrounds respectively pumped the hyperfine states of hydrogen and heated the gas in manners that are potentially observable with the $21\;$cm line of atomic hydrogen \citep[e.g.][]{madau97, furl-rev}.  Next, their greater than $13.6\;$eV photons photoionized nearly all the intergalactic hydrogen.  This cosmological ``reionization'' also heated the IGM to tens of thousands of degrees Kelvin, smoothing the cosmic distribution of gas and affecting the subsequent formation of galaxies.  \emph{We discuss observations that constrain reionization as well as theoretical models of this process in \S~\ref{sec:highz}.}

Once the cosmic hydrogen became reionized, a largely uniform metagalactic ionizing background quickly pervaded space and kept the intergalactic hydrogen highly ionized.  It is thought that at high redshifts this background was sourced by stars, but that by $z\sim 3$ quasars became important, if not dominant.  Over $2<z<5$, owing to several fortuitous factors, there is a wealth of absorption line data for intergalactic hydrogen, helium, and select metals, far more intergalactic data than at other cosmic epochs.  This data has been used to show that the structure of the low-density IGM is in quantitative agreement with cosmological simulations of the $\Lambda$CDM cosmology, to understand the evolution of the ionizing background, to measure the temperature of the IGM, and to constrain the intergalactic enrichment history.   \emph{We discuss these observations of the intermediate redshift IGM in \S~\ref{sec:intermediate}.}

By lower redshifts still ($z\lesssim2$), it again becomes difficult to observe the bulk of the intergalactic volume as cosmological expansion has further diluted most cosmic gas.  In addition, structure formation shocks becomes more efficient at heating intergalactic gas, putting $\sim 50\%$ in a warm-hot $10^{5-6}~$K phase by the present \citep{cen99}, a phase which is even more difficult to probe observationally than the unshocked $\sim10^{4}~$K photoionized one.  The low redshift IGM has had much more time to be affected by astrophysical processes, and these processes in turn shape gas accretion onto galaxies.  \emph{Section~\ref{sec:lowzIGM} describes our understanding of the present-day IGM.}  

The solid curves in Figure~\ref{fig:thermalev} show a theoretical model for the thermal and ionization history of the Universe.  The highlighted regions represent the space that is potentially constrained by different cosmological probes.  Error bars symbolize existing constraints.  This review will cover the IGM probes that appear in this figure as well as a few others.

\section{the IGM at intermediate redshifts, $z=2-5$}
\label{sec:intermediate}
Redshifts of $z=2-5$ set the foundation for our understanding of intergalactic matter.  This redshift range has been so important firstly because at $z\gtrsim 2$ the Ly$\alpha$ line (as well as some of the most useful metal lines, particularly \CIV~$\lambda \lambda$1548, 1551\AA\ and \OVI~$\lambda\lambda$1032, 1038\AA) has redshifted sufficiently that it can be observed with ground-based optical telescopes and, secondly, because there are plenty of bright quasars that enable high signal-to-noise ($S/N$) spectra at high-resolution.  In addition, at these times the density and ionization state of the IGM are ideal for probing gas near the cosmic mean density with the \HI\ Ly$\alpha$ forest.  The Ly$\alpha$ forest spectral region of hundreds of quasars has been observed at high resolution with $10\,$m telescopes \citep[see][]{kodiaq}, and the number has surpassed a hundred thousand at medium resolution on the $2.5\,$m Sloan telescope \citep{lee13}.  These quasar spectra have been used to conduct precision tests of the Ly$\alpha$ forest (\S~\ref{ss:Lyaforest}), to measure the \HI\ column density distribution (\S~\ref{sec:dNdz}), to constrain the thermal history of the IGM (\S~\ref{ss:thermalhistory}), to constrain the metagalactic ultraviolet background (\S~\ref{ss:uvbmodels}), to study the \HeII\ Ly$\alpha$ forest  (\S~\ref{ss:HeIIforest}), and to measure the enrichment of intergalactic gas (\S~\ref{ss:metals}). This grab bag of measurements and tests, whose summary comprises this section, form the basis for our understanding of the IGM at intermediate redshifts.

\subsection{the Ly$\alpha$ forest}
\label{ss:Lyaforest}
Soon after \citet{schmidt65} detected several quasi-stellar objects at significant redshifts, it was realized that there should be hydrogen Ly$
\alpha$ absorption in their spectra from intervening intergalactic gas \citep{gunn65,scheuer65, bahcall65}.  However it took until the mid-1990s for the study of the Ly$\alpha$ forest to reach maturity, following the  commissioning of the HIRES spectrograph on the Keck telescope \citep{1994SPIE.2198..362V}, which enabled resolved studies of the Ly$\alpha$ forest absorption \citep{hu95,kirkman97, kim97}, and also following the advent of cosmological hydrodynamics simulations in the emerging cold dark matter model \citep{cen92, katz92}.  (See \citealt{1998ARA&A..36..267R} for more on the history.)  Indeed, calculations of mock Ly$\alpha$ forest spectra using the first cold dark matter simulations showed a forest of absorption much like that in the actual observations \citep{cen94, zhang95, hernquist96, miralda96}, reenforcing this suggestion from earlier analytic models \citep{bi92}.  This success led to the modern paradigm that the ``trees'' in the forest are the highly photoionized sheets, filaments, and halos that result from cosmic structure formation in a Universe with gravitationally dominant cold dark matter and with an approximately uniform ionizing background.  This paradigm has been subjected to a battery of tests with hardly a chink in its armor.  In what follows we describe the latest such comparisons and how these relate to our present understanding of the low density IGM.

\begin{figure}
\includegraphics[width=1.0\textwidth]{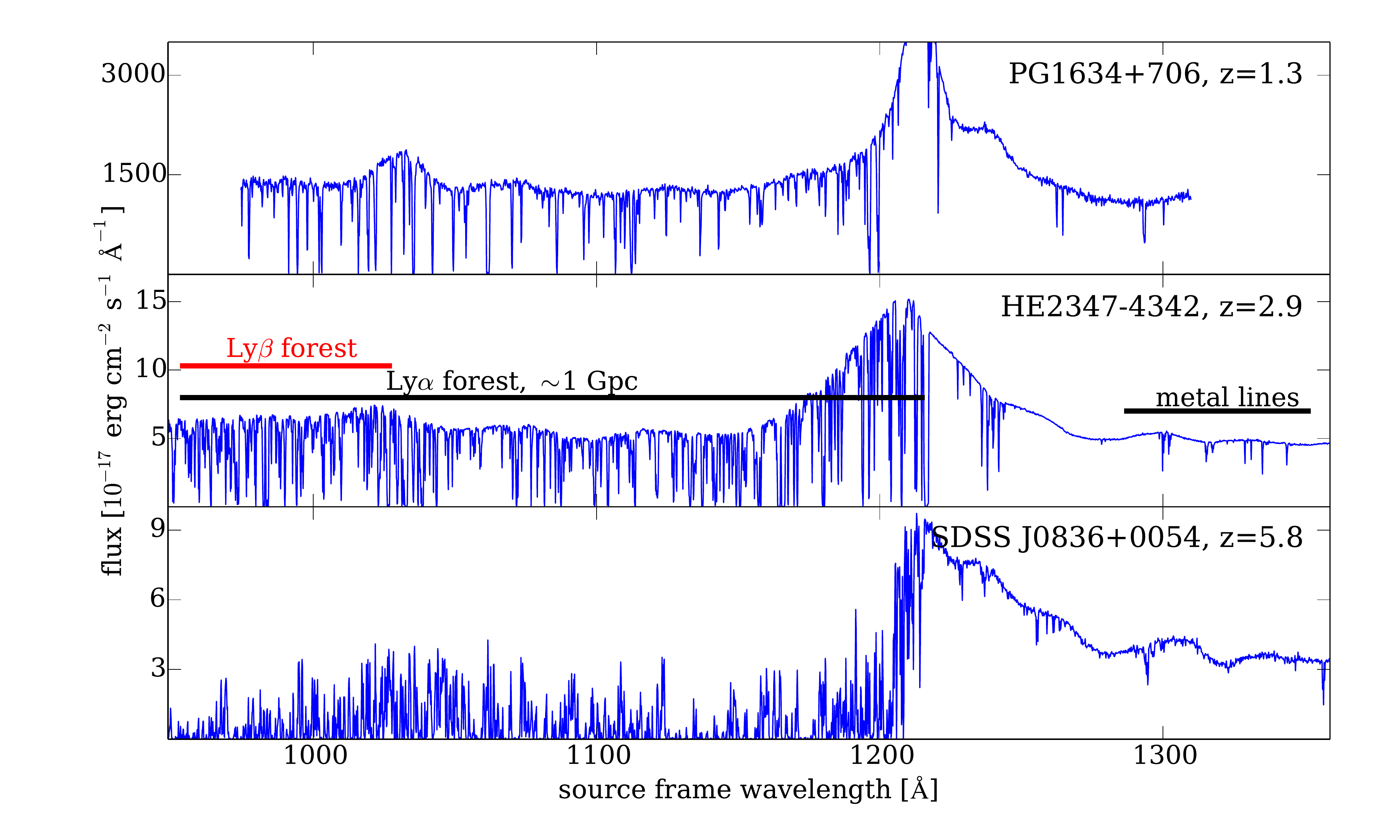}
\caption{Ly$\alpha$ forest spectral region for three quasars chosen to span a large range in redshift.  The HST/STIS spectrum of PG1634+706 was provided by X.~Prochaska, the VLT/UVES spectrum of HE2347-4342 by C. Fechner \citep{fechner07}, and the VLT/X-Shooter spectrum by G. Becker \citep{2013MNRAS.435.1198D}. \label{fig:forestex}}
\end{figure}

Figure~\ref{fig:forestex} shows spectra of three of the brightest quasars at their respective redshift, focusing on the Ly$\alpha$ forest spectral region. The hydrogen Ly$\alpha$ forest is a region in the spectrum of all high-redshift sources that appears blueward of the \HI\ Ly$\alpha$ resonance in the frame of the source, observed at $1216 \,(1+z) \,$\AA\ where $z$ is the redshift of the quasar.  It corresponds to the absorption of intergalactic neutral hydrogen ``clouds'' along the quasar's sightline, with a cloud at redshift $z_1<z$ absorbing in Ly$\alpha$ at $1216 \,(1+z_1) \,$\AA.  It is typically studied in quasar spectra, with each spectrum probing the absorption of gas that lies $\sim 1\;$Gpc in front of the quasar (as foreground \HI\ continuum absorption tends to eliminate the ultraviolet flux blueward of $912$\AA, especially for $z\gtrsim 2$ sightlines).  At each location $z$, the Ly$\alpha$ optical depth corresponding to gas at a fixed density with a smooth line-of-sight gradient, $dv/dx$, in velocity (including the Hubble contribution) is given by 
\begin{equation}
\tau_{\rm Ly\alpha}(z) = 1.3 \, \Delta_b\, \left( \frac{x_{\rm HI}}{10^{-5}} \right)  \left(\frac{1+z}{4} \right)^{3/2} \left(\frac{dv/dx}{H(z)/(1+z)} \right)^{-1},
\label{eqn:tau}
\end{equation}
where the optical depth is related to the absorption probability via $P = \exp(-\tau_{\rm Ly\alpha})$.\footnote{The approximation of setting $dv/dx = H(z)/(1+z)$ in equation~(\ref{eqn:tau}), known as the ``fluctuating Gunn-Peterson approximation'', is relatively accurate and allows one to calculate the absorption from just density skewers, ignoring peculiar velocities \citep{1997seim.proc..133W}.}  Here, $\Delta_b$ is the baryonic density in units of the cosmic mean, and $x_{\rm HI}$ is the fraction of hydrogen that is neutral. Similarly, there are Ly$\beta$, Ly$\gamma$, Ly$\delta$, etc. \HI\ absorption forests, corresponding to absorption into a progressively higher-$n$ Rydberg state.  With increasing $n$, the associated forest spans a progressively shorter path length (and falls on top of lower redshift, smaller-$n$ forests) and is less absorbed (owing to smaller oscillator strengths). 

  Equation~(\ref{eqn:tau}) shows that the Ly$\alpha$ forest is sensitive to $x_{\rm HI} \sim 10^{-5}$ at $z=3$, which translates to astonishingly low \HI\ number densities of $n_{\rm HI} \sim 10^{-10}~$cm$^{-3}$.  It turns out that over much of cosmic time such number densities occur in the low-density IGM  (as a apparent from the spectra in Fig.~\ref{fig:forestex}).  In the post-reionization IGM, $x_{\rm HI}$ is physically set by the balance between photoionization and recombination and is given by 
\begin{equation}
x_{\rm HI} = \frac{\alpha_A \,n_e}{\Gamma},
\label{eqn:xHI}
\end{equation}
except in regions that have been shock heated to $\gtrsim 10^5-10^6~$K such that collisional ionization becomes important.  Equation~(\ref{eqn:xHI}) assumes photoionization equilibrium and that $x_{\rm HI}\ll 1$, which are both likely hold as the photoionization time $\Gamma^{-1} \sim 30,000~$yr (a number that, remarkably, is valid over a large range in redshift, $1\lesssim z\lesssim 5.5$) is much shorter than the recombination time $(\alpha_A\, n_e)^{-1} \sim 10^{10} \Delta_b^{-1} [(1+z)/4]^{-3}$~yr.  The photoionization time is also the timescale to reach equilibrium.

\begin{textbox}
\subsubsection{Cosmological simulations of the IGM}
At early times the matter overdensity fluctuations in the Universe, $\delta \equiv \Delta_b-1$, were small such that $\delta \ll 1$, with properties that are understood from observations of the CMB as well as other probes of large-scale structure.  Their smallness allows one to solve for their evolution using simple perturbative equations.  The matter overdensity fluctuations grow with time and eventually become nonperturbative, requiring a simulation.  The typical collisionless matter-only (or ``$N$-body'') cosmological simulation starts with a grid of particles for the dark matter at time zero (the limit in which $\delta=0$) and, assuming periodic boundary conditions, displaces them using perturbation theory to their positions at redshifts of a hundred or so (when the fluctuations in the density at the grid resolution are ${\cal O} (0.1)$ such that perturbation theory is still applicable; \citealt{1983MNRAS.204..891K, 1983ApJ...274L...1W, 2012PDU.....1...50K}).   From there, the evolution of the dark matter particles is evolved using the full nonlinear dynamics.   In linear perturbation theory (which is the order used to set the initial displacements in the majority of simulations), the Fourier series of the initial displacements equals $-i \tilde \delta_k/k$ \citep{zeldovich70}, where the $\delta_k$ are the Nyquist-sampled Fourier series expansion of $\delta(\bfx)$ in the box.  Each of the $\tilde \delta_k$ in the simulation box is a Gaussian deviate with standard deviation in modulus of $P_L(k,z) V^{-1}$ and random phase, where $V$ is the simulation volume and $P_L(k,z)$ is the linear-theory matter power spectrum (which can be easily calculated with widely used codes such as CAMB; \url{http://camb.info}).

When the dynamics of the gas is included, in addition to following the trajectories of collisionless dark matter particles, a hydro solver is employed for the gaseous evolution.  To study the IGM, the gas also needs to be heated and cooled appropriately.  Except for the small set of simulations that attempts to model reionization with radiative transfer, the bulk of simulations employ a uniform ionizing background that determines the ionization states and temperatures of the gas (see \S~\ref{ss:uvbmodels}).  These simulations also incorporate the cooling rates for all processes relevant for primordial gas \citep{katz92}, with some also incorporating cooling due to metals \citep[e.g.][]{2009MNRAS.393...99W}.   

Accurate simulations are necessary for modern Ly$\alpha$ forest analyses. While different methods for solving the hydrodynamics and gravity yield nearly identical predictions \citep{regan07, bird13, 2015MNRAS.446.3697L}, numerical convergence in box size and in resolution are notoriously tricky  \citep{meiksin04, mcdonald05b, tytler09, bolton09, lidz10, 2015MNRAS.446.3697L}.
A typical modern simulation used to study the Ly$\alpha$ forest has of order $N_{\rm side}^3 = 1000^3$ gas resolution elements and dark matter particles, although some have reached a few times larger $N_{\rm side}$ employing Eulerian hydro grids \citep[e.g.][]{2015MNRAS.446.3697L}.  With $1000^3$ elements, a box size of $\approx 20~$comoving Mpc is required to resolve $10^6\Msun$ structures, roughly $10^{-4}$ of the Jeans' mass for mean density gas at $z=3$ -- the mass resolution studies have found is required for $1-10\%$ accuracy in many forest statistics \citep{bolton09, lidz10, 2015MNRAS.446.3697L}.  The standard deviation of $\delta$ when smoothed over a $20\;$Mpc cubic volume is $\approx 0.2$ at $z=3$ and so at this level a $20\;$Mpc box is not representative of the Universe.  Large simulations (or extrapolations based on moderate-sized ones) are often necessary for robust inferences.
\end{textbox}

\begin{figure}
\includegraphics[width=.9\textwidth]{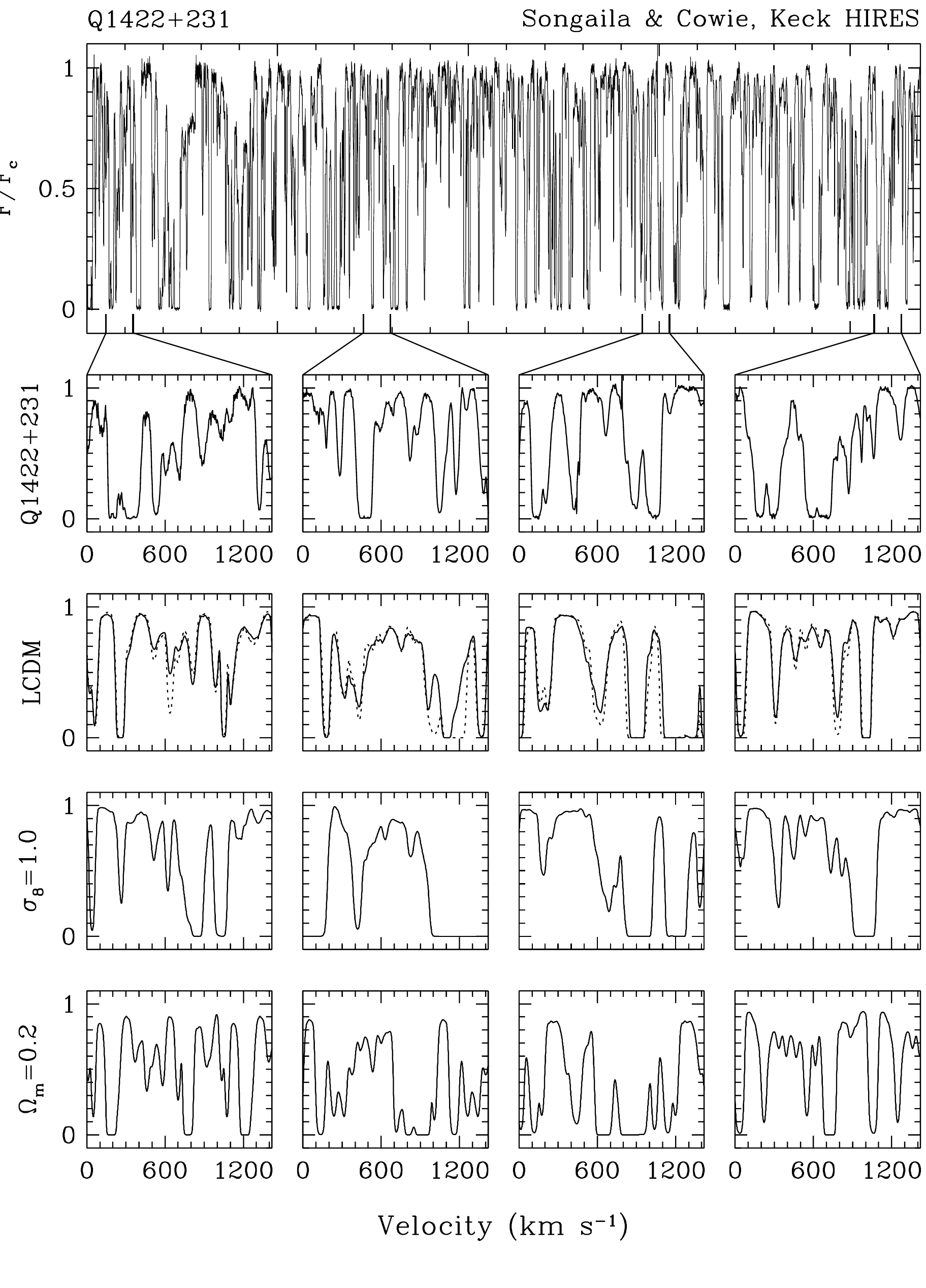}
\caption{Top two rows show the continuum-normalized Ly$\alpha$ forest spectrum taken with KECK/HIRES of quasar Q1422+231, zooming in on four select regions.  The bottom three rows show the absorption in hydrodynamic simulations using three different cosmologies, where the stated parameter is varied from the $\Lambda$CDM case taking $\sigma_8=0.8$ and $\Omega_m =0.4$.  The same random numbers are used to initialize all simulations.  The dotted curves in the $\Lambda$CDM panels are from an $N$-body simulation without gas; the absorption of this case is computed assuming that gas traces the $N$-body matter distribution.  From \citet{weinberg}.    \label{fig:Lyaex}}
\end{figure}

Inference from the Ly$\alpha$ forest spectra is complicated by there being no reliable analytic model for the mildly nonlinear densities probed by the forest.  All analyses require a comparison with large cosmological simulations. Figure~\ref{fig:Lyaex}, from \citet{weinberg}, shows just such a comparison.  The top two panels show the Ly$\alpha$ forest transmission of a real quasar observed with the HIRES instrument on the Keck~I telescope, zooming in on several select regions.  The other panels show the Ly$\alpha$ forest extracted from cosmological hydrodynamic simulations in three different cosmologies.  In addition, the dotted curves in the $\Lambda$CDM panels are computed assuming that the gas traces the dark matter, an approximation that nearly reproduces the absorption seen in the full hydrodynamic calculation (solid curve).  Thus, the absorption structures in the Ly$\alpha$ forest largely follow the underlying voids, sheets, and filaments in the dark matter \citep[e.g.][]{2001MNRAS.324..141M}.

Figure~\ref{fig:Lyaex} shows that the $\Lambda$CDM model predicts a forest that is qualitatively similar to that seen in observations.  To make this comparison more quantitative (as well as to appreciate the implications), we need to understand how the simulations model the density structures, the photoionization rates, and the gas temperatures.  In addition, analyses must account for common contaminants of the Ly$\alpha$ forest signal.  There are a standard set of prescriptions for dealing with each of these:
\begin{enumerate}
\item The statistical properties of the gas density field are assumed to be those expected from evolving (via the equations of hydrodynamics and gravity) the cosmological initial conditions subject to a uniform ionizing background that turns on at $z\sim10$.  This posits that galactic feedback processes -- which are known to blow baryons out of galaxies -- do not significantly impact the low density gas seen in the Ly$\alpha$ forest.  This supposition is supported by simulations with simple feedback prescriptions \citep{theuns02, mcdonald05, bertone06, kawata07}, although it may not hold to the forecasted precision of upcoming Ly$\alpha$ forest analyses \citep{viel13}.  

\item The \HI\ photoionization rate $\Gamma$ (and hence the ionizing background) is assumed to be spatially uniform, which is motivated by the much longer mean free path of ionizing photons relative to the mean distance between sources -- a scenario that suppresses fluctuations \citep{croft04, mcdonald05}.   In addition, the amplitude of $\Gamma$ is adjusted in post processing (an approximation justified in \S~\ref{ss:uvbmodels}) until the simulations match the observed mean amount of absorption in the forest.  A byproduct of this adjustment is a ``flux decrement'' measurement of $\Gamma$, with \citet{becker13} finding a remarkably constant $\Gamma \approx 10^{-12\pm 0.3}~$s$^{-1}$ over $2<z<5$ \citep[see also][]{1997ApJ...489....7R, 2003MNRAS.342.1205M, bolton05, faucher07}.

\item The temperature is assumed to be a power-law function of density as motivated in \citet{hui97}, an assumption that should not apply during and soon after reionization processes \citep{trac08, furlanetto09, mcquinn09}.  The power-law assumption is in many studies implicit as they use hydrodynamic simulations with a uniform ionizing background that results in a near power-law relation.  Often this power-law is parameterized in terms of the density in units of the cosmic mean, $\Delta_b$, as
\begin{equation}
T(\Delta_b) = T_0 \Delta_b^{\gamma-1}, 
\label{eqn:TDrelation}
\end{equation}
with $T_0$ and $\gamma-1$ being parameters that are varied to fit the observations.  We discuss the physics of what sets $T_0$ and $\gamma-1$ as well as the latest constraints on the temperature in \S~\ref{ss:thermalhistory}.  Many studies adjust the temperatures in post-processing rather than use the temperatures generated in the simulation, which captures temperature's effect on thermal broadening but not on the gas pressure.  (The smoothing from gas pressure is the smaller of these effects for most Ly$\alpha$ forest statistics.)  Others inject heat into the simulated gas in a way that achieves a specified $T_0$ and $\gamma$ \citep[e.g.][]{bolton08}.  
\item  It is assumed that the Ly$\alpha$ forest absorption can be extracted sufficiently well that metal line contamination, errors in the estimate of the quasar's intrinsic continuum, and damping wing absorption do not bias inference.  Each of these contaminants come with their own set of issues and techniques for addressing them: 
\begin{description}
\item[Metal lines] are responsible for $10\%$ of the absorption in the Ly$\alpha$ forest spectral region at $z=2$, with a decreasing fraction towards higher redshifts \citep[ \S~\ref{ss:metals}]{pod2, 2005MNRAS.360.1373K}.  Metal absorption can be isolated or corrected for by studying the absorption redward of the forest or, for high-quality data, within the forest itself, often using the doublet structure of prominent ions.
 \item[The quasar continuum] is fit by finding low-absorption points and interpolating between them in high quality spectra.  This method leads to quantifiable errors \citep[e.g.][]{bolton05,faucher07}, which increase with redshift and which for many analyses are small enough to safely ignore.  
 \item[Damping wing absorption] becomes significant for dense systems with $N_{\rm HI} \gtrsim 10^{19}~$cm$^{-2}$, with one system occurring every several Ly$\alpha$ forest spectra at $z\sim 3$.  Systems with significant damping wings are not captured reliably in standard cosmological simulations.  These systems often can be removed visually.
 \end{description}
 In high $S/N$, high resolution spectra, all of these issues are easier to diagnose/eliminate.  For Sloan spectra, more sophisticated techniques than those listed above are often required \citep[e.g.][]{mcdonald05b}.
\end{enumerate}

\begin{figure*}[t!]
  \centering
    \begin{subfigure}[t]{0.45\textwidth}
    \centering
\includegraphics[width=1\textwidth]{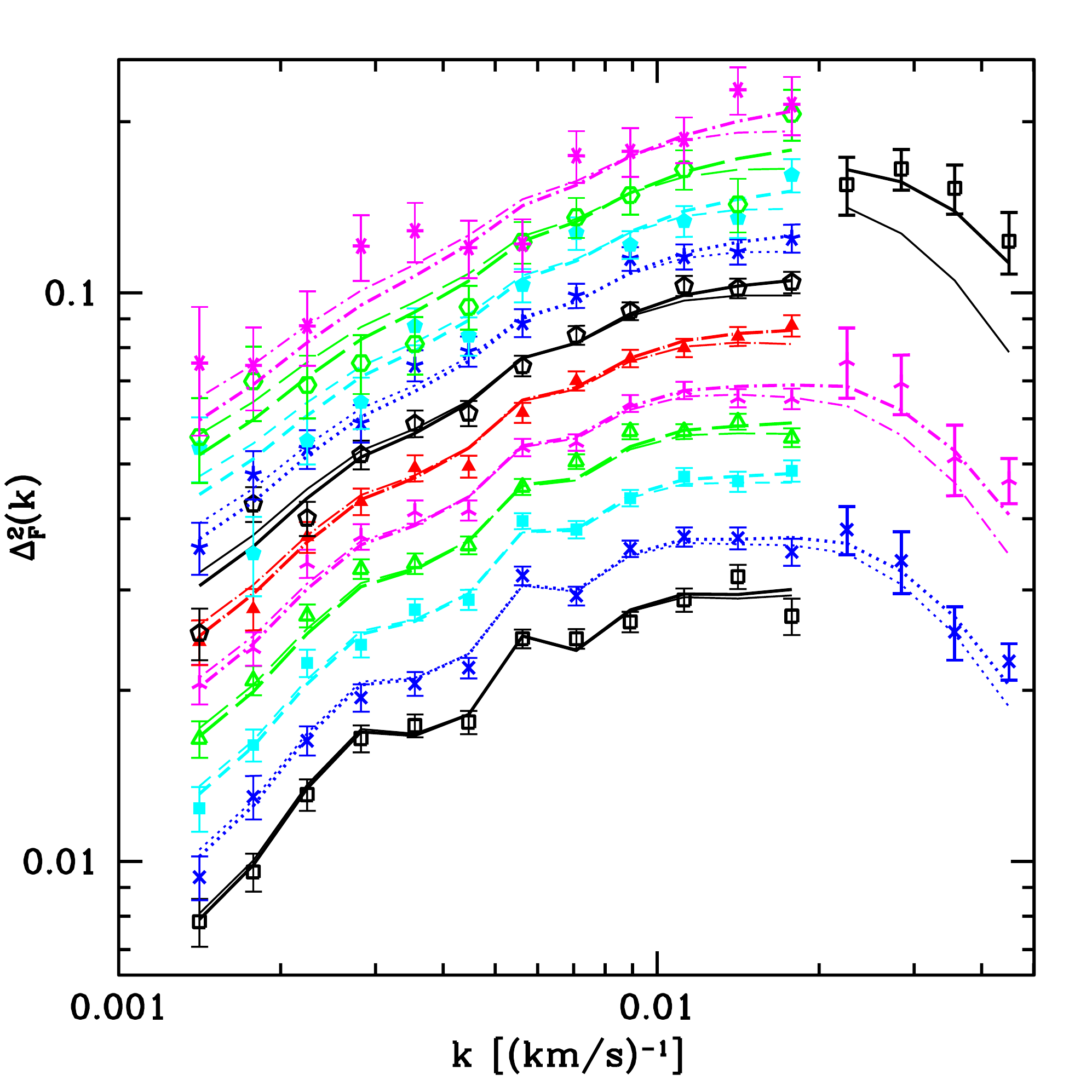} 
\end{subfigure}
    \begin{subfigure}[t]{.45\textwidth}
    \centering
\includegraphics[width=1\textwidth]{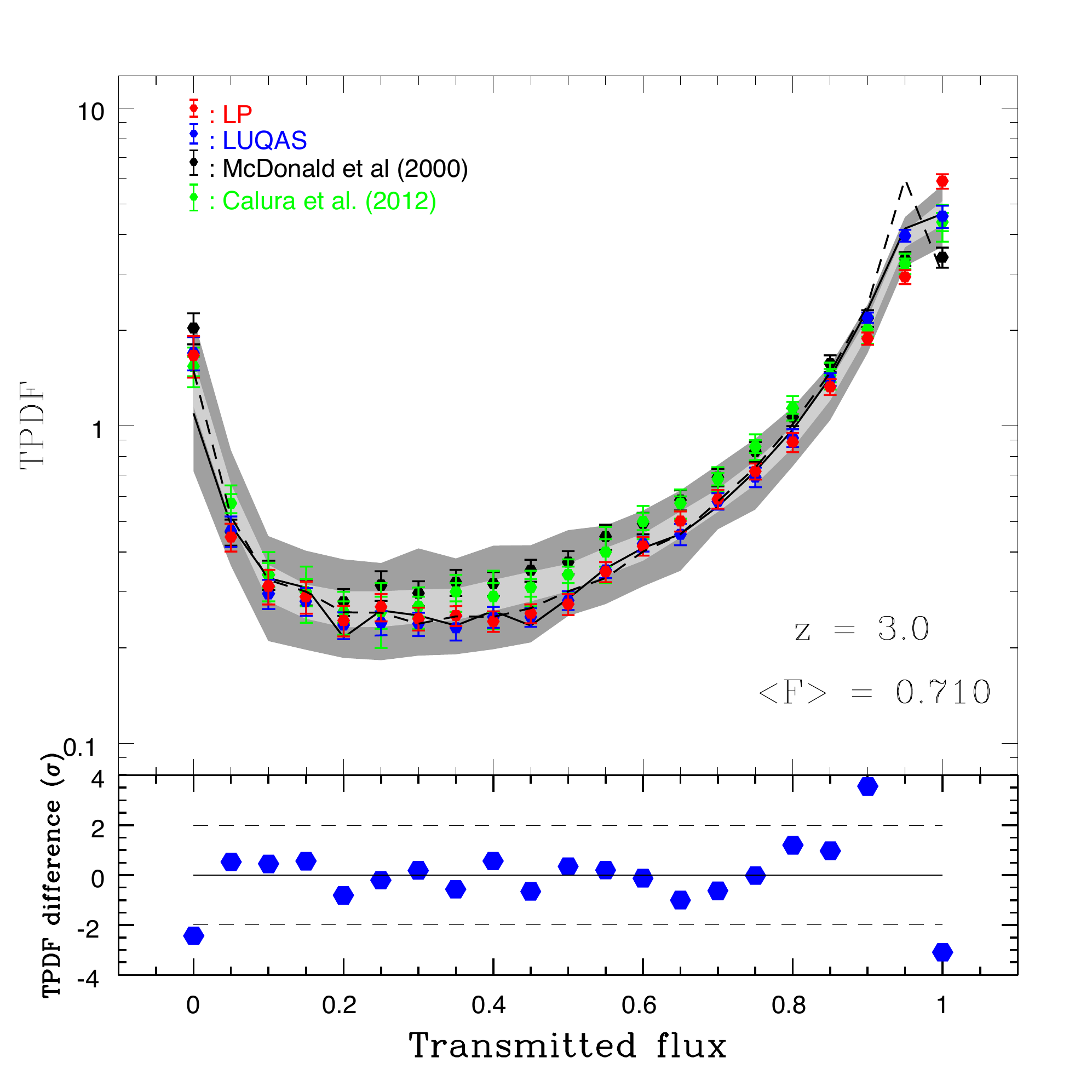}
\end{subfigure}
\caption{The two statistics that are most commonly applied to the Ly$\alpha$ forest.  {\it Left panel:}  The power spectrum of the normalized flux, showing $k P_F/\pi$, from \citet{seljak06}.  The points with error bars are measurements using $3000$ sloan quasar spectra for $k<0.02$~s~km$^{-1}$ (with the lowest set being $z=2.2$ and the highest $z=4.2$, in increments of $\Delta z =0.2$) and eight high-resolution quasar spectra at higher wavenumbers (showing $z=2.4~, 3.0$ and $3.9$).  The underlying curves are the predictions of the CDM model (thick curves) and of a model in which a $6.5~$keV sterile neutrino is the dark matter (thin curves).  {\it Righthand panels}:  The top panel shows the Ly$\alpha$ forest transmission PDF at $z=3$.  The points with error bars show three measurements of this statistic, the solid curve is this statistic estimated from a simulation, and the shaded regions represent bootstrap error estimates matching the sampling statistics of the LP measurement and calculated by sampling skewers from the simulation (with the shading indicating the $1\sigma$ and $2\sigma$ errors), from \citet{rollinde13}.  The bottom panel shows the residuals between the LP measurement and the simulations in units of the $1\sigma$ bootstrap error.
  \label{fig:Lyastat}}
\end{figure*}

 Many studies have found that, once these prescriptions are adopted, mocks extracted from simulations are able to describe the standard set of statistics applied to the Ly$\alpha$ forest data, and some have even leveraged this result to constrain cosmological parameters.  However, others have found potential discrepancies.  We now discuss the state of this comparison for the three most studied statistics.

\subsubsection{the line-of-sight power spectrum}
 The line-of-sight  Ly$\alpha$ forest power spectrum is the most studied of all Ly$\alpha$ forest statistics, defined as $P_F(k) \equiv  L^{-1} |\tilde \delta_F(k)|^2$, where $\delta_F(x) \equiv F(x)/\langle F(x) \rangle -1$ is the overdensity in the transmission at position $x$ over a sightline of length $L$ and $\tilde \delta_F(k)$ is its Fourier transform.  (Often the transmission is called the ``normalized flux''.)  The left panel in Figure~\ref{fig:Lyastat} shows for a multitude of redshifts $\Delta_F(k)^2 \equiv k P_F(k)/\pi$ -- a combination whose integral over $d \log k$ is the variance in the transmission overdensity --, from \citet{2005PhRvD..71j3515S}.   At high wavenumbers ($k \gtrsim 0.02\,$km$^{-1}$\,s), $P_F(k)$ is sensitive to the small-scale smoothing of the gas (constraining the gas temperature as well as the warmness of the dark matter), whereas smaller wavenumbers are primarily sensitive to the large-scale density distribution of matter. Many other (astrophysical) effects that could potentially affect $P_F$ have been found to be small in physically motivated models.  These include spatial variations in $T(\Delta_b)$, spatial variations in the ionizing background, and galactic feedback processes \citep{meiksin04, croft04, lai06, mcdonald05b, mcquinn11}.  This robustness to messy astrophysical processes potentially allows one to use $P_F$ to constrain cosmological parameters. 
 
Studies find that the predictions of simulations of the Ly$\alpha$ forest agree with the observed $P_F$, a confirmation of the standard model for the forest \citep{croft99, mcdonald00, 2002ApJ...581...20C, 2004ApJ...617....1T}.  This exercise has been done with an increasing level of precision, finding over $2\lesssim z \lesssim 4.5$ that, when tuning the IGM $T-\Delta_b$ relation and cosmological parameters within allowed bounds, the standard simulations yield good $\chi^2$ values even when the estimate uses $3000$ \citep[][]{mcdonald05b} and, recently, $14,000$ Sloan spectra \citep{2013A&A...559A..85P, 2015arXiv150605976P}.  The former measurement is featured in the left panel in Figure~\ref{fig:Lyastat}, showing the $P_F$ estimates and best-fit model in the range $2.2 < z < 4.2$ in increments of $\Delta z =0.2$.  These measurements are so precise that by themselves they constrain the amplitude of density fluctuations ($\sigma_8$), the matter density ($\Omega_m$), and the tilt of the primordial power spectrum ($n_s$) to $5-10\%$ \citep{viel06}.  These constraints derive from smaller comoving scales than the constraints from other cosmological probes (down to $\sim 1\,$Mpc, a decade beyond other techniques).   The agreement between cosmological parameters inferred from the forest and other methods is a fabulous consistency test of the inflation+$\Lambda$CDM paradigm, in which there is nearly zero parametric freedom in how density fluctuations on disparate scales are connected.  The bottom three rows in Figure~\ref{fig:Lyaex} illustrate how the absorption in the forest changes in the concordance cosmology relative to cosmologies with a low $\Omega_m$ and a high $\sigma_8$.   While it is debated whether to trust cosmological parameter determinations from $P_F$ (with skeptics pointing to the high value of $\sigma_8$, the seemingly unphysical temperatures that the analyses favor, or discrepancies between their mean transmission inferences and more direct measurements; \citealt{viel06, becker13}), at the $\lesssim 10\%$ level the Ly$\alpha$ forest power spectrum is consistent with the favored $\Lambda$CDM cosmology.\footnote{The 1D power spectrum is also well suited for constraining warm dark matter models that act to truncate the power at high wavenumbers.  The lower curves left panel of Fig.~\ref{fig:Lyastat} show a warm dark matter model that is clearly ruled out, from \citet{seljak06}.  See also \citet{2013PhRvD..88d3502V} for a recent analysis.}

\subsubsection{the probability distribution function of transmission}
After the Ly$\alpha$ forest power spectrum, the Ly$\alpha$ forest transmission PDF is the most studied statistic \citep[e.g.][]{mcdonald00, lidz06, kim07, lee15}.  This statistic is shown in the right panel of Figure~\ref{fig:Lyastat}.  \citet{bolton08} and \citet{2012MNRAS.422.3019C} argued that the measured transmission probability distribution function (PDF) over $2\lesssim z \lesssim 3$ disagrees with estimates from cosmological simulations.  Studies have proposed several potential resolutions, including the temperature-density relation of the IGM being ``inverted'' such that $\gamma-1<0$ \citep{bolton08,puchwein11}, that the difference owes to continuum fitting errors \citep{lee12}, or that the quoted error bars in previous analyses underestimated the sample variance \citep{rollinde13}.  The latter solution is shown in the right panel of Figure~\ref{fig:Lyastat}.  \citet{rollinde13} found that several previous independent measurements of the transmission PDF (which had used $\sim10-20$  Ly$\alpha$ forest sightlines broken into a few redshift bins) were inconsistent in each transmission bin at a few standard deviations using their quoted errors.  Furthermore, when \citet{rollinde13} measured the transmission PDF using a large simulation but emulating the sampling statistics of the measurements, their bootstrap error estimates were much larger than the errors reported on previous measurements and large enough to yield good $\chi^2$ values.  This result suggests that there is no tension between the data and simulations regarding the transmission PDF.

\subsubsection{the line-width distribution}
A final Ly$\alpha$ forest statistic that is often encountered is the line-width PDF \citep{1995AJ....110.1526H, 1997ApJ...477...21D, theuns98, 1999ApJ...517..541H}.  There have also been claimed discrepancies in this statistic between the observations and simulations, with \citet{tytler09} finding a $10\%$ difference.  We are not worried about this level of discrepancy.  The line-width PDF is very sensitive to the thermal history, which is only crudely modeled in standard simulations.  Indeed, the line widths themselves are used to measure the thermal history \citep[][see \S~\ref{ss:thermalhistory}]{schaye00, bolton13}, suggesting that with the correct thermal history the simulations' line widths may match the observed distribution. \\ 

In conclusion, our vanilla models of the forest agree with the observations to $\lesssim 10\%$ in the standard statistics, with no convincing discrepancies.  Future progress can be made by increasing the precision of the measurements and by targeting new statistics (as the standard three statistics could be insensitive to interesting effects).  One such statistic is the two-point transmission correlation function between adjacent sightlines, which studies have found can be more sensitive to astrophysical effects than the 1D statistics discussed here \citep{mcdonald07, white10, mcquinnwhite, gontcho14, pontzen14, 2015arXiv150604519A}.  

\subsection{the \HI\ column density distribution}
\label{sec:dNdz}

\begin{figure}
\includegraphics[width=6.6cm]{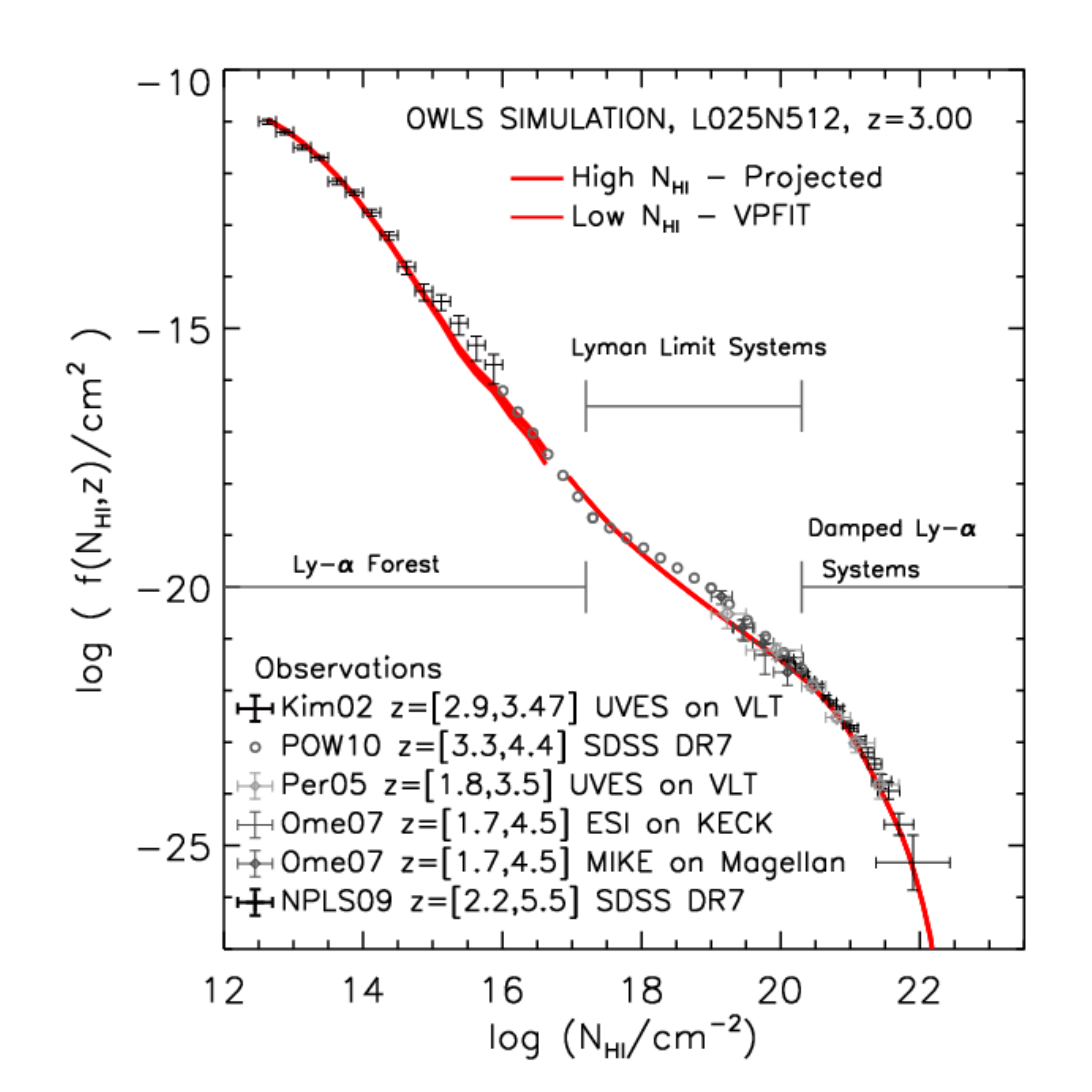}
\caption{Log$_{10}$ of the number of \HI\ systems per $dN_{\rm HI}$ per $dX \equiv (1+z)^{1/2}  \Omega_m^{-1/2} dz$, measured from cosmological simulations (red curve) and from quasar absorption spectra (points with error bars), from \citet{altay11}.
\label{fig:fNHI}}
\end{figure}

The column density distribution of \HI\ absorbers -- the number of absorbers per unit column density per unit path length -- is another statistic that is often measured from quasar absorption spectra.  The method used to measure this statistic depends strongly on column.  The lowest \HI\ columns ($N_{\rm HI} \lesssim 10^{14}$cm$^{-2}$) are measured using just Ly$\alpha$ absorption.\footnote{At low columns, the Ly$\alpha$ forest transmission PDF is a better statistic as it is crude to model the forest absorption as discrete absorbers.  With increasing column, the Ly$\alpha$ absorption lines become more distinct.}  At $10^{14} \lesssim  N_{\rm HI} \lesssim 10^{17}$cm$^{-2}$, the column density of an \HI\ absorber is inferred by also using absorption from higher Lyman-series lines.  Around  $N_{\rm HI} \sim 10^{17}$cm$^{-2}$, breaks from continuum absorption of \HI\ in the spectrum are used (or, statistically, by stacking around source-frame $912$\AA; \citealt{prochaska09}).  Finally, at  $N_{\rm HI}\gtrsim 10^{19}$cm$^{-2}$, the damping wing of the Ly$\alpha$ line is exploited.  The somewhat tedious terminology for systems in different column density ranges, which we avoid here, is given in the margin.\begin{marginnote}
\entry{Lyman-limit}{$10^{17.2} - 10^{19}$cm$^{-2}$}
\entry{super Lyman-limit}{$10^{19} -10^{20.3}$cm$^{-2}$}
\entry{damped Ly$\alpha$ (DLA)}{$N_{\rm HI} >10^{20.3}$cm$^{-2}$}
\end{marginnote}  This combination of methods has been used to measure the \HI\ column density distribution over the entire range that occurs in nature \citep[most recently by][]{kim13, rudie13, fumagalli13, worseck14b}.  The points with error bars in Figure~\ref{fig:fNHI} show select previous measurements.  

The \HI\ column density distribution is a diagnostic of the distribution of gas densities in the Universe.  A useful model for the relationship between column density, size, and physical density for overdense absorbers that cannot self-shield to the background (i.e. $\lesssim 10^{17}$~cm$^{-2}$) was proposed in \citet{schaye01}.  This study argued that the size of an absorber was on-average set by the distance a sound wave travels in the dynamical time ($t_{\rm dyn} \sim 1/\sqrt{G \rho} \sim H^{-1} \Delta_b^{-1/2}$).  This distance is also known as the Jeans' length.  Assuming a photoionization rate and a temperature allows one to then relate the density of an absorber to its column density via \citep{schaye01}
\begin{equation}
\Delta_b =  200 \left(\frac{1+z}{4} \right)^{-3}  \left(\frac{ N_{\rm HI}}{10^{17} {\rm \; cm^{-2}}}  \,\frac{\Gamma}{10^{-12} {\rm \; s^{-1}}}\right)^{2/3} \,  \left(\frac{T}{10^4~{\rm K}} \right)^{0.17}.\label{eqn:schayerho}
\end{equation}
Note that $\Gamma \sim 10^{-12} {\rm \; s^{-1}}$ is the \HI\ photoionization rate that is measured over $2<z<5$.  This model has been verified using cosmological simulations \citep{altay11, mcquinn-LL}.  This model shows that the type of system that yields a certain $N_{\rm HI}$ varies dramatically over time, with absorbers having $N_{\rm HI} \lesssim 10^{17} [(1+z)/4]^{3}$~cm$^{-2}$ corresponding to the $\Delta_b < 200$ gas that is likely to be intergalactic.  Even though they are not necessarily intergalactic, the properties of systems with $N_{\rm HI} \sim \sigma_{\rm HI}^{-1}(\nu = 1{\rm Ry}/h) = 1.6\times10^{17}~$cm$^{-2}$ are always important for the IGM because they set the mean free path of \HI-ionizing photons. 
  
Higher column systems owe to denser gas, and the denser the gas is, the closer it lies to a galaxy on average.  The closer it lies to a galaxy, the more likely astrophysical effects may enter and change the gas distribution from the numerical models that ignore such effects.  Thus, the column density distribution at higher columns than those probed by the Ly$\alpha$ forest may be a more promising diagnostic than the Ly$\alpha$ forest of galactic feedback.  Several studies have investigated how well the column density distribution is reproduced in simulations over various $N_{\rm HI}$ ranges \citep{katz96, theuns98, kohler06, mcquinn-LL, altay11, altay13b, 2013MNRAS.430.2427R}.  Figure~\ref{fig:fNHI} shows a comparison of $f(X, N_{\rm HI})$ -- the number of absorbers per $dN_{\rm HI}$ per $dX \equiv (1+z)^{1/2}  \Omega_m^{-1/2} dz$ -- between a simulation at $z=3$ (red solid curve) and a compilation of measurements, from \citet{altay11}.\footnote{As defined here, $f(X, N_{\rm HI})$ is constant with redshift if the comoving number density and physical cross section of absorbers at a given column do not evolve when $\Omega_m \approx 1$.  Curiously, the observations and simulations find that this statistic is constant at the factor of two level over the range $2<z<5$ \citep{prochaska09b, mcquinn-LL}, despite the overdensities that correspond to a given column evolving substantially (eqn.~\ref{eqn:schayerho}).}  The major features in each distribution are a transition to a power-law functional form above columns of $N_{\rm HI} \sim 10^{14}~$cm$^{-2}$ owing to the power-law density profile that develops around collapsed structures, a break in the slope at $N_{\rm HI} \sim \sigma_{\rm HI}^{-1}(\nu = 1{\rm Ry}/h) = 1.6\times10^{17}~$cm$^{-2}$ owing to systems self-shielding at higher columns \citep{2002ApJ...568L..71Z}, and a roll off at the highest columns owing to the transition to the formation of molecules and the depletion of diffuse gas onto stars \citep{2001ApJ...562L..95S}.  To better than a factor of two, the simulations and measurements agree over the columns shown (even remarkably at the higher columns that probe denser circumgalactic and galactic gas).  This result has also been shown to hold down to $z=0$ and to even be fairly robust to galactic feedback recipes \citep[see \S~\ref{ss:metals} for more description of these recipes]{altay13b, 2013MNRAS.430.2427R}.  Unfortunately, we are unaware of a quantitative comparison between simulations and measurements at columns higher than those probed by the Ly$\alpha$ forest but that are still intergalactic ($10^{14} \lesssim N_{\rm HI} \lesssim 10^{17}$cm$^{-2}$ at $z\sim3$).  Such a comparison would be interesting in light of recent, more precise measurements \citep{rudie13, kim13}.

Another useful statistic for probing the distribution of \HI\ around galaxies is to measure the amount of \HI\ absorption in quasar spectra as a function of the distance to spectroscopic galaxies.  This statistic was investigated by \citet{adelberger05} and \citet{rakic12}, who detected an enhancement in absorption out to a few physical Mpc from galaxies but with a large amount of scatter. (The scatter at $<0.2~$physical Mpc had initially been attributed to galactic feedback, generating much excitement.)  The profile of \HI\ absorption around (Lyman break-selected) galaxies is shown in the right panel of Figure~\ref{fig:metalabsorption}, from \citet{turner14}. \citet{rakic13} found that the median absorption profile in \HI\ is consistent with the predictions of standard cosmological simulations (and relatively insensitive to the galactic feedback prescriptions they employed in the simulations).

\subsection{the thermal history of the IGM}
\label{ss:thermalhistory}

\begin{textbox}
\subsubsection{Asymptotic Temperature-Density Relation}
The evolution of temperature of an unshocked ionized Lagrangian fluid element follows from the first law of thermodynamics:
\begin{equation}
\frac{dT}{dt}=-2HT+\frac{2T}{3 \Delta_b}\frac{d
\Delta_b}{dt}+\frac{2}{3k_B n_b}\frac{dQ}{dt}, \label{eqn:dTdz}
\end{equation}
where $n_b$ is the number density of all free ``baryonic'' particles in the plasma \citep[including electrons;][]{miralda94, hui97}.  In an ionized IGM, the dominant processes are photoheating and Compton cooling such that $dQ/dt \approx  \Delta E \, \Gamma n_{\rm HI} + C n_e$, where $\Delta E$ is the amount of energy per photoionization, $C(z)$ is a coefficient that describes Compton cooling off of the CMB, and in photoionization equilibrium $n_{\rm HI} \approx \alpha_A n_e/\Gamma$ (with the recombination coefficient scaling as $\alpha_A \propto T^{-0.7}$).  Using these relations, equation~(\ref{eqn:dTdz}) can be solved for arbitrary $\Delta_b(t)$ yielding
\begin{equation}
T =  \left( \left[\frac{Z^3 \Delta_b}{Z_i^3 \Delta_{b,i}}\right]^{2/3\times1.7} T_i^{1.7} e^{(Z/7.1)^{5/2} - (Z_i/7.1)^{5/2} }  + T_{\rm 0, lim}^{1.7} \Delta_b \right)^{1/1.7} ~~\underset{\rm late~times}{{\longrightarrow}} ~~~ T_{\rm 0, lim} \; \Delta_b^{1/1.7},
\label{eqn:eta_ev}
\end{equation}
where $Z \equiv 1+z$ and subscript $i$ denotes the initial state of the gas parcel at time $z_i$ \citep{mcquinn15}.  The first term in the parentheses is sensitive to the initial temperature from a passing ionization front (the exponential factor, which acts to erases memory of the initial temperature, owes to Compton cooling), and the second term encodes the limiting temperature from the balance between photoheating and cooling processes.  The function $T_{\rm 0, lim}$ has an extremely weak dependence on the prior density evolution of a gas parcel such that regardless of how the density evolves $ T_{\rm 0, lim}  \approx 10^4 [(1+z)/4]\,$K before \HeII\ reionization and twice this value after, with additionally a weak dependence on the ionizing background spectrum.  Thus, all unshocked gas with $\Delta_b \lesssim 10$ (densities below where collisional cooling is important) is driven to a single $T-\Delta_b$ relation with index $\gamma - 1 =1/1.7= 0.6$, with almost negligible dispersion \citep{hui97}.  Within a doubling of the scale factor, this evolution erases the memory of an earlier state, which just after hydrogen reionization should have had a lot of dispersion in temperature \citep{trac08,furlanetto09}, with \HeII\ reionization regenerating dispersion at $z\sim 3$ \citep{mcquinn09, compostella13}.  This behavior has motivated the power-law parameterizations of the $T-\Delta_b$ relation used in many IGM analyses.
\end{textbox}

There has been a recent resurgence in work to reconstruct the thermal history of the low density IGM over intermediate redshifts \citep{lidz10, becker10, garzili12, rudie12, bolton13, 2014MNRAS.441.1916B}, following-up the seminal investigations from over a decade ago \citep{schaye00, ricotti00, mcdonald01b, zaldarriaga01, theunsschaye02, hui03}.  Temperature measurements are interesting because they constrain different energy injection processes into the IGM and because the IGM temperature sets the minimum mass of galaxies (see sidebar The Minimum Mass of Galaxies).  Temperature measurements rely on the width of Ly$\alpha$ absorption features from the low density intergalactic medium being broader with higher temperatures, both because of thermal broadening and, to a lesser extent, the broadening owing to pressure effects.  A variety of methods to determine temperature are used, ranging from directly fitting for the width of absorption lines as a function of $N_{\rm HI}$ \citep{1999MNRAS.310...57S} to measuring the suppression at high wavenumbers in the Ly$\alpha$ forest power spectrum that owes principally to thermal broadening \citep{2000MNRAS.315..600T,  zaldarriaga01}.  All methods must be calibrated with simulations.  The spate of IGM temperature measurements near the turn of the century generally found fairly high temperatures over $2<z<4$, with temperatures at the mean density of $T_0 = (20-30)\times10^3~$K, although with a large amount of scatter and large statistical errors of $\sim(5-10)\times10^3\,$K.  Recent measurements have led to a more concordance picture with $T_0 = (10-20)\times10^3~$K, with the concordance arising because of the general agreement among several studies \citep[][albeit with significant collaborative overlap]{becker10, garzili12, bolton13, 2014MNRAS.441.1916B} and with the past study of \citet{schaye00}.  In addition, led by \citet{becker10}, some studies have chosen to estimate the temperature at the density where the variance in the estimate is minimized, rather than at $\Delta_b=1$, resulting in smaller errors.  However, there is still some tension among recent measurements \citep{lidz10, becker10}.

In the standard story, the temperature history primarily constrains the reionization history of the IGM.  When an ionization front swept through and reionized the intergalactic hydrogen, this ionization photoheated the cold IGM.  Estimates are that the IGM was heated to $(17-25) \times10^3~$K, with the exact value depending on the hardness of the incident spectrum and the velocity of the ionization front \citep{miralda94, mcquinn-Xray}.  The reionization of the second electron of helium (``\HeII\ reionization'') is another major heating event, likely driven by the harder emissions of quasars.  During \HeII\ reionization by quasars (which is discussed in more detail in \S~\ref{ss:HeIIforest}), \HeIII\ bubbles are blown around quasars and the IGM is additionally heated by $(5-10)\times10^3$K \citep{furlanetto08H, mcquinn09, compostella13}.  After each of these reionization processes the IGM cooled, mainly through the adiabatic cosmic expansion and through Compton cooling off of the CMB (which is especially important at $z\gtrsim 5$), although recombination and free-free cooling are important at the $10\%$ level.  Cooling after reionization drives most of the gas to a ridiculously-tight power-law relationship with $T = T_{\rm 0, lim} \Delta_b^{0.6}$ for $\Delta_b \lesssim 10$ within a doubling of the scale factor \citep{hui97}, as described in the sidebar titled ``Asymptotic Temperature-Density Relation''.

\begin{figure}
\includegraphics[width=1\textwidth]{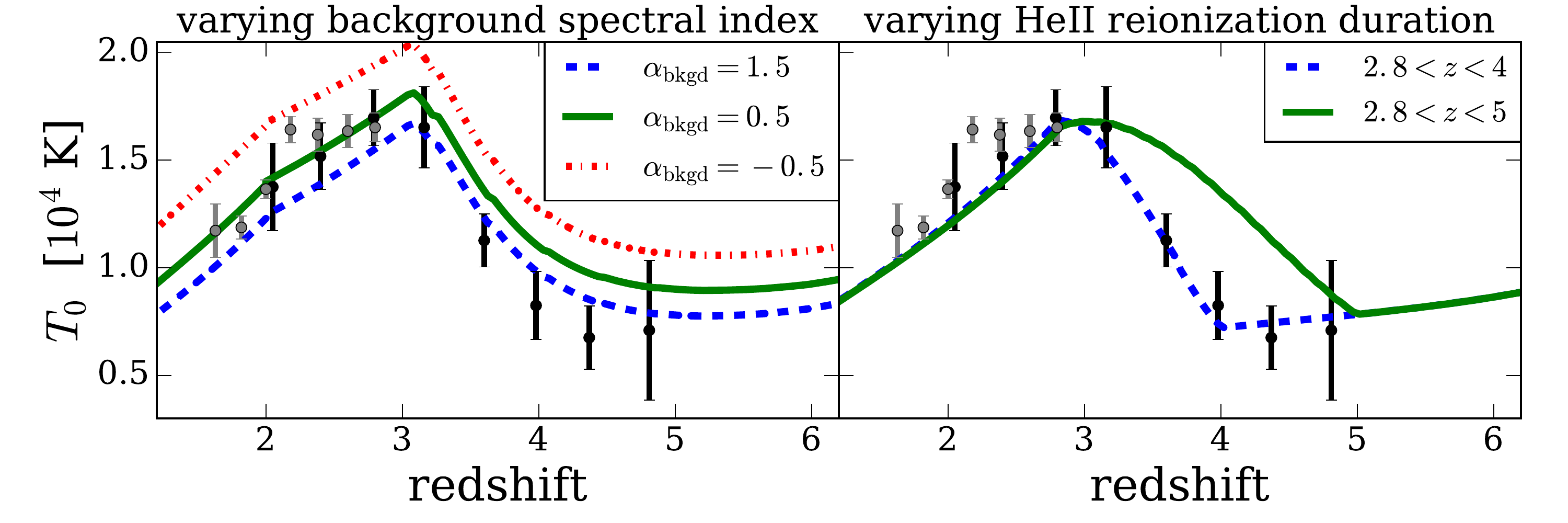}
\caption{IGM temperature measurements of \citet{becker10} and \citet{2014MNRAS.441.1916B} compared with semi-analytic model predictions for the average temperature thermal history of the IGM, varying two of the most important model parameters.  In particular, the left panel shows the dependence on the spectral index of the post-reionization ionizing background, and the right panel shows the dependence on the duration of \HeII\ reionization assuming linear-in-redshift histories.  The temperature peak in the models (and likely in the data) owes to \HeII\ reionization.  From \citet{uptonsanderbeck}.
\label{fig:temphist}}
\end{figure}

The points with error bars in Figure~\ref{fig:temphist} show recent estimates of the IGM temperature, from \citet{becker10} and \citet{2014MNRAS.441.1916B}.   Also shown is the average temperature in semi-analytic calculations that model the optically thick photoheating from reionization processes and subsequent optically thin photoheating plus cooling.\footnote{The \citet{becker10} and \citet{2014MNRAS.441.1916B} measurements do not directly constrain $T_0$, but rather the temperature at a redshift-dependent $\Delta_b$.  The $T-\Delta_b$ relation in the intermediate model in the left panel of Figure~\ref{fig:temphist} is used to extrapolate these measurements to $T_0$, although it matters little which model is chosen.}  See \citet{uptonsanderbeck} for additional details and \citet{puchwein14} for a related study.  These semi-analytic models predict a distribution of $T(\Delta_b)$ at any redshift (as  must be true in the actual IGM), but \citet{uptonsanderbeck} find that what is measured is effectively the average temperature.  The left panel in Figure~\ref{fig:temphist} shows the dependence on the spectral index of the post-reionization ionizing background, and the right the duration of \HeII\ reionization assuming linear histories spanning $2.8<z<4$ (dashed curve) and $2.8<z<5$ (solid curve) -- the two most important dependences found in \citet{uptonsanderbeck}.  The temperature peak in the models (and likely in the data) owes to \HeII\ reionization, which occurs at $z\approx 3$ without any fine-tuning of the quasar emissivity history \citep[the models in the left panel use the luminosity function of][]{hopkins07a}.  The shorter duration models (which are more consistent with most measurements of the quasar emissivity history) and softer ionizing background models are more consistent with the data.  Without \HeII\ reionization occurring at $z\sim 3$, such models would predict a monotonically decreasing temperature with time after reionization.

Thus, the standard model for the thermal history with a late reionization of \HeII\ seems consistent with the \citet{becker10} and \citet{2014MNRAS.441.1916B} measurements, which span $1.6<z<4.8$.  This consistency limits the amount of heating from mechanisms other than photoheating, such as proposals associated with TeV blazers, cosmic rays, dust, or dark matter annihilations \citep{chang12, samui05, lacki13, 2003MNRAS.341L...7I}.  This consistency does not constrain hydrogen reionization at $z>6$, as the IGM has cooled by $z=4.8$ -- the highest redshift where the temperature has been measured -- to an asymptotic temperature that does not retain much memory of earlier times.  Constraints on hydrogen reionization would be vastly improved by a temperature measurement at somewhat higher redshifts \citep{lidz14, uptonsanderbeck}.\footnote{The temperature has been measured in the proximity region of a $z= 6$ quasar \citep{bolton10}, although unfortunately this location complicates the interpretation.}   
  
  Our understanding of  the thermal history can also be improved by measuring the slope of the $T-\Delta_b$ relation, $\gamma -1$, the IGM pressure smoothing length at $\Delta_b\sim 1$, as well as fluctuations in the IGM temperature.  Forecasts are that $\gamma-1$ tends to zero or even negative values during reionization processes \citep{trac08, furlanetto09}, with $\gamma-1 \rightarrow 0.6$ afterwards (see sidebar Asymptotic Temperature-Density Relation).  Only at $z=2.4$ is $\gamma-1$ reasonably well measured with the constraint $\gamma-1 \approx 0.54\pm0.11$ \citep{rudie12, bolton13} and so there is substantial room for improvement in $\gamma-1$ determinations.  Secondly, the length scale over which fluctuations are smoothed by pressure effects is sensitive to the temperature temporally averaged over a dynamical time, which is the Hubble time at $\Delta_b\sim 1$ (rather than the instantaneous temperature that the line-of-sight forest primarily constrains).  Quasars at small angular separations potentially allow a measurement of this pressure-smoothing scale \citep{2013ApJ...775...81R, 2015arXiv150400366K}.  Finally, relic fluctuations in the IGM temperature are an inevitable byproduct of the inhomogeneous nature of reionization processes.  Unfortunately at $z<5$ the fluctuations appear not to be large enough in theoretical models to manifest in a detectable signal \citep{mcquinn11}, explaining why searches for these fluctuations have not turned up any evidence \citep{theuns00, lidz10}.  At $z>5$, a detection of temperature fluctuations in the Ly$\alpha$ forest may be more possible \citep[see \S~\ref{ss:highzLya}]{daloisio15}.

\begin{textbox}
\subsubsection{The Minimum Mass of Galaxies}
The IGM thermal history also shapes the minimum mass of galaxies.  Before reionization, the pressure of the IGM was negligible and so whether a galaxy formed in a halo was determined simply by whether the virialized halo gas had sufficient time to cool and condense.  Reionization heated the gas to $\sim 10^4$K, making the IGM Jeans' mass larger than the mass threshold for cooling \citep{1986MNRAS.218P..25R,1996ApJ...465..608T}.  Indeed, the first models for this suppression were based on comparing the halo mass with the IGM Jeans' mass or the analogous mass for an expanding universe \citep[termed the ``filtering mass'';][]{shapiro94, gnedin98, gnedin00}.  More recent studies have noted that the gas that would make it onto a halo typically does not reach a density within an order of magnitude of the mean density at collapse, and so a higher density and hence smaller Jeans' mass is more applicable \citep[although not necessarily smaller than the filtering mass;][]{hoeft06, okamoto08,  noh14}.  These studies find that the mass scale where pressure inhibits accretion is a strong function of time, suppressing $\approx 10^9 \,\Msun \; (10^{10} \, \Msun)$ halos at $z=6$ $(z=1.5)$ for a region reionized at $z\sim10$.

This suppression of accretion onto galaxies from the pressure of an ionized IGM, often attributed to the ``ultraviolet background'' or ``reionization'', is invoked to help explain the missing satellite problem:  In the cold dark matter picture, the Milky Way should have thousands of subhalos massive enough that the gas could have cooled and formed stars, but observations of satellite galaxies suggest that ultimately only a small fraction of them were able to form stars.  IGM pressure can raise the mass threshold above which subhalos can accrete gas and form stars \citep{quinn96, bullock00, 2009ApJ...693.1859B}.  Intriguingly, some of the ultra-faint dwarf satellite galaxies of the Milky Way appear to have formed their stars by the time the Universe was $\sim 1~$Gyr old \citep{brown12, 2014ApJ...789..148W}, which may indicate that accretion was shut off by the pressure of a photoionized IGM. 
\end{textbox}

\subsection{synthesis cosmic ultraviolet background models}
\label{ss:uvbmodels}

Following earlier work \citep{miralda90,giroux96}, \citet{haardt96} developed models that synthesized observations of the sources (namely quasars and galaxies) and the sinks of ionizing photons to make predictions for the properties of the extragalactic ionizing background.  Subsequent efforts have developed these models further \citep{fardal98, faucher09, haardt12}.   These models are adopted in essentially all non-adiabatic cosmological hydrodynamic simulations of the post-reionization IGM to compute the ionization state and photoheating rates of intergalactic gas,\footnote{Fortunately, the gas temperatures and, hence, the hydrodynamics in simulations (at least those that include only cooling from primordial gas) are relatively insensitive to the properties of the ionizing background.} and they also widely used in the modeling of intergalactic and circumgalactic absorption.  

\begin{marginnote}
\entry{$J_\nu$}{angularly averaged specific intensity of ionizing background}
\entry{$\Gamma$}{photoionization rate of \HI}
\entry{$\lambda_{\rm mfp}(\nu)$}{photon mean free path}
\end{marginnote}
Built into these models is the assumption that sink positions are uncorrelated and that the ionizing background is spatially uniform.  Uncorrelated positions is typically justified because the Poisson fluctuations in the number of sinks generally far exceed their clustered fluctuations.  Spatial uniformity of the background is justified at many redshifts as the mean free paths of ionizing photons, $\lambda_{\rm mfp}$, soon after the cosmological reionization of the species being ionized become much longer than the average distance between sources, a situation that suppresses variations \citep[e.g.][]{meiksin04}.\footnote{Only regions in the proximity regions of sources does this approximation break down, with only the fraction $f \approx (6\sqrt{\pi})^{-1} n^{-1/2} \lambda_{\rm mfp}^{-3/2}$ of the volume having a background that is enhanced by a factor of two over the mean, where $n$ is the number density of sources and the mean free path to be absorbed by hydrogen is $\lambda_{\rm mfp} \approx 200 [(1+z)/5]^{-4} (\nu/\nu_{\rm HI})^{3(\beta-1)}~$comoving Mpc over $2.3<z<5.5$ \citep{worseck14b}.  Even if the sources are as rare as possible -- $L*$ quasars -- $f \sim 10^{-3}  [(1+z)/5]^{6}$ at $1$~Ry.}  From these assumptions, calculating the ionizing background for a given emissivity history and \HI\ column density distribution is straightforward, although somewhat less so at wavelengths affected by \HeII\ continuum absorption.  The mathematics behind these models is described in the sidebar Uniform Ionizing Background Models.

\begin{textbox}
\subsubsection{Uniform Ionizing Background Models}
The assumption of spatial uniformity and uncorrelated absorbing clouds of fixed column allows one to calculate the background specific intensity, $I_\nu$, with two inputs, the column density distribution of absorbing clouds, $\partial^2 {\cal N}/{\partial x \partial N_{\rm HI}}$, where $x$ is the comoving distance, and the physical specific emissivity of the sources, $\epsilon(\bfx, \nu, z)$.  The spatially averaged solution to the cosmological radiative transfer equation $d [I_\nu/\nu^3]/dt = c\epsilon_\nu/[4\pi \nu^3] - c \sum_X \sigma_X n_X [I_\nu/\nu^3]$ -- just a Boltzmann equation for the (unnormalized) phase space density $I_\nu/\nu^3$ with sources and sinks on the right-hand side --, where $X \in $\{\HI, \HeI, \HeII\} and $\sigma_X(\nu)$ is the photoionization cross section, is easily derived via the method of Green's functions:
\begin{equation}
J_{\nu_0}(z_0) =  \frac{c}{4\pi} \int_{z_0}^{\infty} \frac{dz}{H(z)(1+z)} \left(\frac{1+z_0} {1+z}\right)^3 \left \langle \epsilon(\bfx, \nu, z) \right \rangle  e^{-\tau_{\rm eff}(\nu_0, z, z_0)}, 
\label{eqn:Jnu}
\end{equation} 
where $\nu = \nu_0 (1+z)/(1+z_0)$, $J_{\nu_0}(z_0) \equiv (4\pi)^{-1}\int d\Omega \, I_{\nu_0}$ is the angularly-averaged background intensity at $\nu_0$ as seen by an
observer at redshift $z_0$, and the effective optical depth, $\tau_{\rm eff}$, is defined as $e^{-\tau_{\rm eff}} \equiv \langle e^{-\tau} \rangle$, where brackets are a spatial average and $\tau = \int_{x(z)}^{x(z_0)} dx \sum_X \sigma_X n_X$.  This average yields 
\begin{equation}
\tau_{\rm eff}(\nu_0, z, z_0) = \int_{x(z)}^{x(z_0)} dx \overbrace{\int_{0}^{\infty} dN_{\rm HI}  \left( 1- e^{- \sum_{ X} \sigma_{X}(\nu) N_X(N_{\rm HI}, J_\nu)} \right) \frac{\partial^2 {\cal N}}{\partial x \partial N_{\rm HI}}}^{\equiv a \lambda_{\rm mfp}^{-1}(\nu)},
\end{equation}  
where $N_X(N_{\rm HI}, J_\nu)$ is the column density in ion $X$ \citep{1980ApJ...240..387P}.
 All previous background models calculate $N_X$ by treating the absorbers as single-density planar slabs (often using eqn.~\ref{eqn:schayerho} for $\Delta_b$).  In addition to continuum absorption, models generally treat the emission and absorption from atomic transitions of H and He.  These transitions are generally of secondary importance, with the most important being ionizing \HI\ recombination radiation, which contributes $5-20\%$ to $\Gamma$ \citep{faucher09}.
 
In the limit that photons do not redshift significantly between emission and absorption (which applies for $\nu \sim^+ \nu_{\rm HI}$ at $z>3$), equation~(\ref{eqn:Jnu}) simplifies to 
\begin{equation}
J_\nu =  \frac{1}{4\pi}  \left\langle \epsilon(\bfx, \nu, z) \right\rangle \lambda_{\rm mfp}(\nu).\label{eqn:highz}
\end{equation}
Specializing to the case of just \HI\ absorption, $J_\nu \propto \nu^{-\alpha+ 3[\beta-1]}$ for $\partial^2 {\cal N}/{\partial x \partial N_{\rm HI}} \propto N_{\rm HI}^{-\beta}$ and $\epsilon\propto \nu^{-\alpha}$.    This scaling also lets us simplify the expressions for the photoionization and photoheating rates, which, specializing to the case of \HI\ and using that $n_{\rm HI} =\alpha_A n_e/\Gamma$ and $\sigma_{\rm HI} \propto \nu^{-3}$, become
\begin{eqnarray}
\Gamma &\equiv& 4\pi \int_{\nu_{\rm HI}}^\infty \frac{d\nu}{h\nu} \sigma_{\rm HI}(\nu) J_\nu \approx  \frac{4\pi \sigma_{\rm HI}(\nu_{\rm HI}) J_{\nu_{\rm HI}}}{h(6+ \alpha-3\beta)};~~
\frac{dQ}{dt} \equiv 4\pi n_{\rm HI} \int_{\nu_{\rm HI}}^\infty \frac{d\nu}{\nu} \sigma_{\rm HI}(\nu) J_\nu (\nu - \nu_{\rm HI}) \approx  \frac{h \nu_{\rm HI} \alpha_A n_e}{5+ \alpha- 3\beta}. \label{eqn:photoheating}
\end{eqnarray} 
The photoheating rate does not depend on the amplitude of $J_\nu$, which justifies the post-processing adjustments to $\Gamma$ in Ly$\alpha$ forest analyses and limits the sensitivity of numerical simulations to the exact $J_\nu(z)$.
 \end{textbox}

\begin{figure}
\includegraphics[width=10cm]{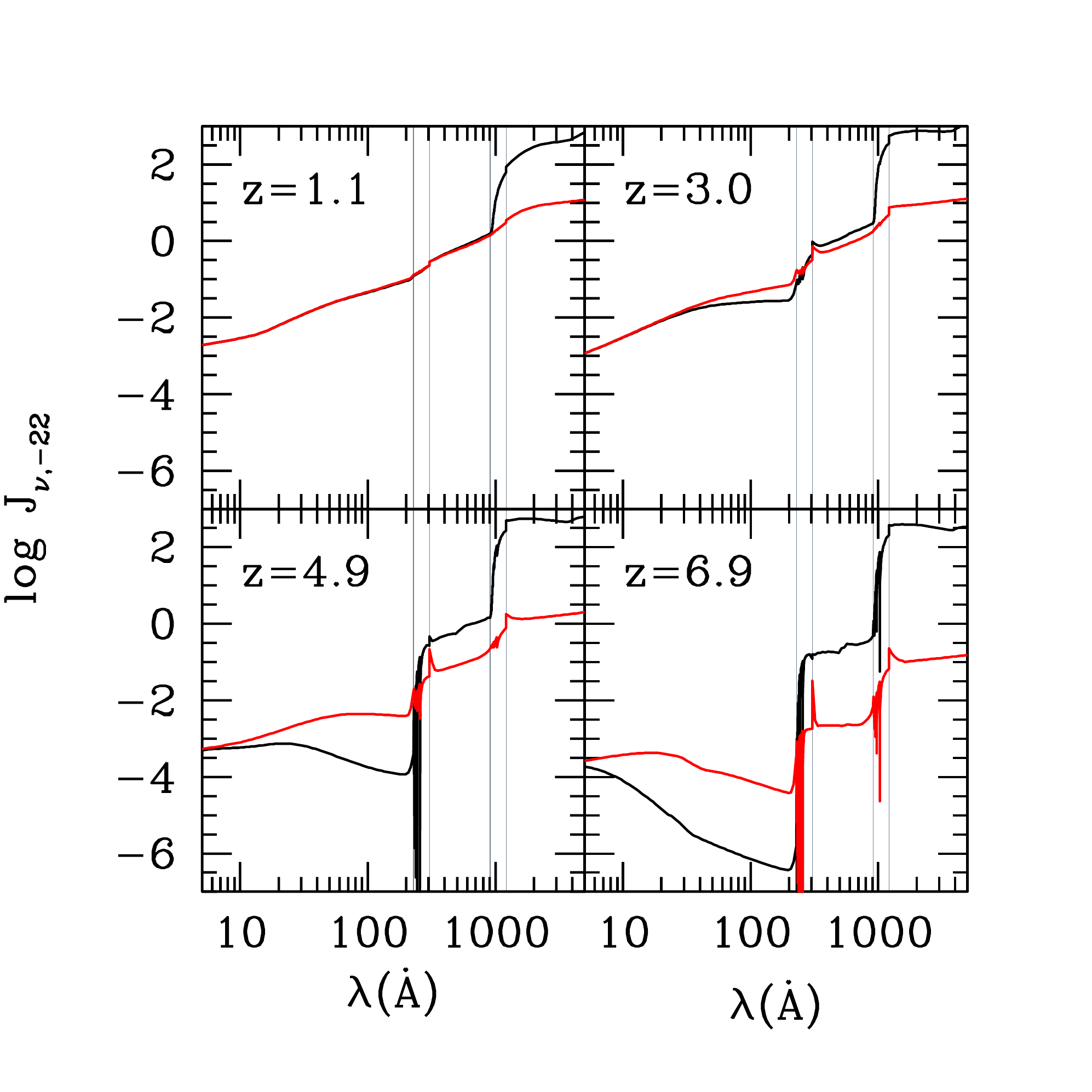}
\caption{\citet{haardt12} uniform ultraviolet background models. The black curve is their full model that includes emissions from galaxies and quasars, and the red curve is their ``quasar only'' model.  The vertical lines correspond to the Ly$\alpha$ and Lyman-continuum wavelengths of both \HI\ ($1216$\AA\ and $912$\AA) and \HeII\ ($304$\AA\ and $228$\AA).  From \citet{haardt12}. \label{fig:hm12}} 
\end{figure}

Previous background models tune the assumed emissivity for the ionizing sources (consisting of galaxies and quasars) to reproduce the transmission in the \HI\ Ly$\alpha$ forest.  The emissivity of quasars, which can be estimated to a factor of $\sim 2$ accuracy, is roughly consistent at $z\sim 2-3$ with that needed to source the entire ionizing background \citep{haardt96, faucher08,haardt12}.  However, the known population of quasars is likely unable to  maintain the backgrounds at higher redshifts, and so it is generally assumed that stars make up the deficit (an assumption discussed further in \S~\ref{sec:sources}).  These models then typically take the ionizing emissivity associated with the observed cosmological star formation history, tuning the escape fraction of ionizing photons so that the background calculation reproduces measurements of $\Gamma(z)$.  Once the emissivity history of the sources is prescribed at $\nu \sim \nu_{\rm HI}$, these models are then solved using our best guesses for the frequency dependences of the sources' spectra and using the latest measurements of the \HI\ column density distribution.  Until recently, to redshift dependent normalization factors, background models used a power-law source spectra with specific luminosity parametrized as $L_\nu \propto \nu^{-\alpha}$ and a power-law distribution of absorbers parametrized as $\partial^2 {\cal N}/{\partial x \partial N_{\rm HI}} = A \, N_{\rm HI}^{-\beta}$ where the choices of $A$ and $\beta$ are motivated by observations.  With these parameterizations, taking the limit that the mean free paths satisfy $\lambda_{\rm mfp} \ll cH^{-1}$ valid at $z\gtrsim 3$ and ignoring resonant processes (which are of secondary importance), the spectrum between $1-4\,$Ry and then also above $4\, $Ry scales as $\nu^{-\alpha + 3(\beta-1)}$ (see sidebar Homogeneous Ionizing Backgrounds).   The modeling has been improved in the most recent models by using population synthesis calculations for the stellar emission spectrum and \HI\ column density distributions that are not single power-laws \citep{haardt12}.  Figure~\ref{fig:hm12} shows the ionizing background model of \citet{haardt12} over $1 < z< 7$. The black curve is their full model that includes emissions from galaxies and quasars, and the red curve is their ``quasar only'' model.  The major breaks in the spectrum are at the ionization edges of \HI\ and \HeII\ at $912$\AA\ and $228$\AA, respectively.
   
 It is difficult to gauge the fidelity with which uniform ionizing background models describe the actual ionizing background.  Around $z\approx 2$, the shape of the ionizing background has been constrained observationally using metal line systems  in \citet{2007A&A...461..893A} and \citet{2011A&A...532A..62F}, with the latter finding agreement with some models.  At most redshifts these models have not been tested.  However, we know that uniform ionizing background models must err at certain redshifts and in certain locations.  For example, near dense systems with $N_{\rm HI} \gtrsim 10^{18}$cm$^{-2}$ local sources of radiation should be important \citep{miraldaescude05, rahmati13}.  In addition, uniform background models also break down if $n^{1/3} \lambda_{\rm mfp}(\nu) \lesssim 1$ because spatial fluctuations in $J_\nu$ will be large, where $n$ is the number density of sources.  At  $h \nu \sim 4~$Ry, energies that set the \HeII\ photoionization rate, $\lambda_{\rm mfp}$ becomes small enough ($\lesssim 100~$comoving Mpc) to result in ${\cal O}(1)$ fractional variance in $J_\nu$ at $z\gtrsim 2.5$ if quasars source this background (for which effectively $n \sim 10^{-5}$comoving Mpc$^{-3}$; \citealt{bolton06, 2009ApJ...703..702F, mcquinnHeII}).  There is also evidence that fluctuations on the order of unity or greater are occurring in the \HI\ Ly$\alpha$ forest at $z>5.5$ \citep[ \S \ref{ss:highzLya}]{becker15}.  Lastly, large spatial fluctuations should occur in $J_\nu$ for $\nu$ above the ionization potential of an ion during its reionization.

\subsection{the \HeII\ Ly$\alpha$ forest}
\label{ss:HeIIforest}

The Ly$\alpha$ resonance of \HeII\ at $304$\AA\ is the only other intergalactic absorption line that has been observed from an element created in the Big Bang, elements which of course have huge modeling advantages.\footnote{It may be possible to see intergalactic $21\;$cm in absorption during reionization \citep[e.g.][this absorption has been seen from DLAs but not the IGM]{furlanetto06}.  A 584 \AA\ forest of \HeI\ \citep{mcquinn-HeI} and an $8.7$~GHz forest of $^3$\HeII\ \citep{mcquinn-3He} are also potentially observable.}  Because the \HeII\ Ly$\alpha$ line falls blueward of $912$\AA, this spectral region is prone to foreground continuum absorption from neutral hydrogen.  Such foreground absorption means that the \HeII\ forest can only be observed at $z>2$ as the \HeII\ forest from lower redshift clouds is absorbed by the $\gtrsim 10^{19}~$cm$^{-2}$ \HI\ columns through our galaxy.  It also means that intervening systems with $N_{\rm HI} \gtrsim 10^{17}$cm$^{-2}$ absorb parts of the \HeII\ Ly$\alpha$ spectral region; quasar sightlines that by chance intersect fewer of these systems will have more useable \HeII\ Ly$\alpha$ forest spectra.  About $1\%$ of $z\sim 3$ quasars show enough transmission in the \HeII\ forest to be useful \citep{worseck11, syphers09}.  At present, the \HeII\ Ly$\alpha$ forest has been observed towards about twenty quasars spanning $2.7<z<3.8$ \citep{reimers97, hogan97, heap00, zheng04, worseck11b, syphers12}, a number that has increased significantly in the last few years with the installation of the ultraviolet-sensitive Cosmic Origins Spectrograph (COS) on the Hubble Space Telescope (HST).

\begin{figure}
\centering
\includegraphics[width=1.\textwidth]{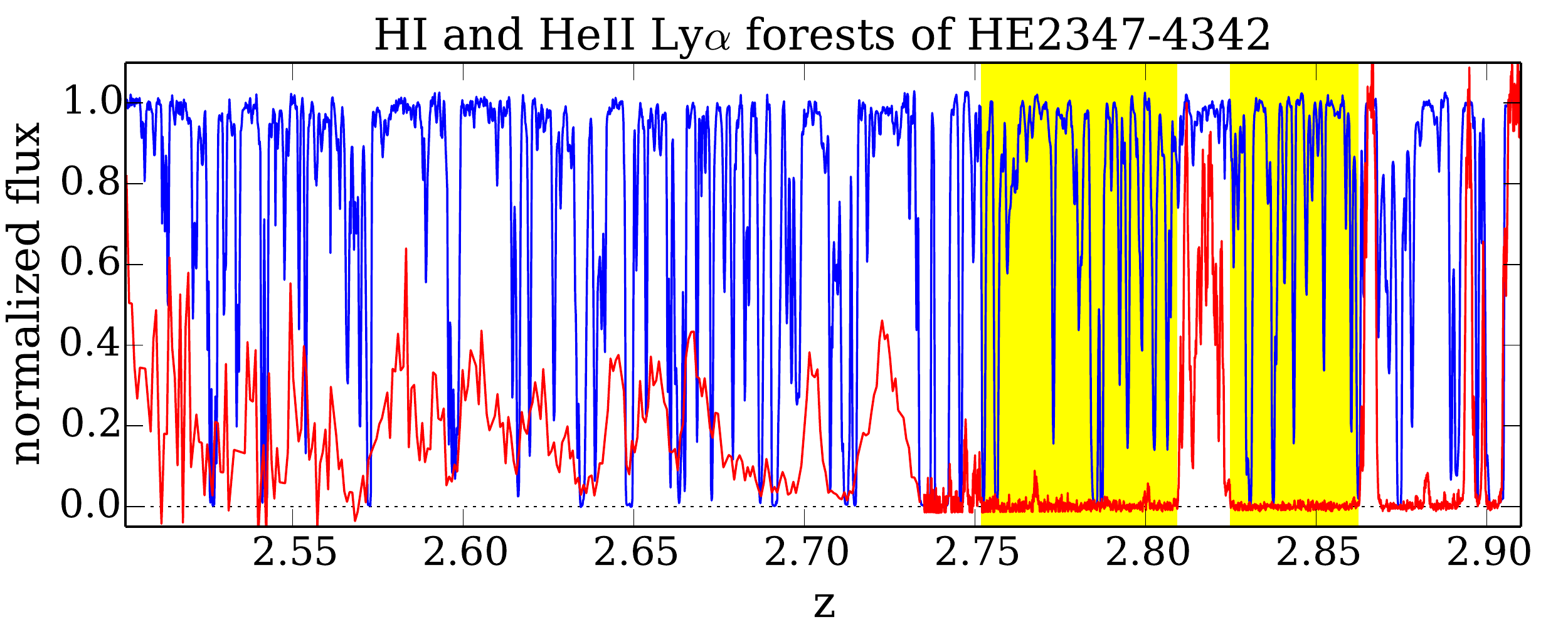}
\caption{Continuum-normalized spectra of the brightest quasar in the far ultraviolet, HE2347-4342 ($z=2.9$).  The blue curve is the \HI\ Ly$\alpha$ forest transmission of \citet{fechner07} using VLT/UVES, and the red curve is the coeval \HeII\ Ly$\alpha$ forest obtained with HST/COS by \citet{shull10} and using the reduction of \citet{worseck11b}.  This spectrum illustrates the finding of \HeII\ Ly$\alpha$ forest studies that the \HeII\ absorption is much stronger than the coeval \HI\ absorption, with large swaths of highly absorbed regions at $z\gtrsim 2.7$ (highlighted in yellow).  The HST/COS spectrum has resolution of $\lambda/\Delta \lambda\approx1500$ at $z<2.73$ and $\lambda/\Delta \lambda\approx20,000$ at higher redshifts, whereas the VLT/UVES spectrum has $\lambda/\Delta \lambda\approx 50,000$.
\label{fig:2347}}
\end{figure}

The HST/COS \HeII\ Ly$\alpha$ forest spectrum of the brightest quasar in the far ultraviolet, HE2347-4342 ($z=2.9$), is shown in Figure~\ref{fig:2347} \citep{shull10}. Also shown is  a high-resolution VLT spectrum of this sightline's coeval \HI\ Ly$\alpha$ forest from \citet{fechner07}. The \HeII\ spectrum illustrates several features generic to the existing \HeII\ Ly$\alpha$ forest sightlines.  At high redshifts, much of the \HeII\ absorption is saturated.  Indeed, Gunn-Peterson troughs, defined as regions with no detected transmission, are seen in the spectrum of HE2347-4342 at $2.7<z<2.9$, with the largest trough spanning $\Delta z \sim 0.05$ or $\sim 30$ comoving Mpc \citep{shull10}.  (The yellow highlighted regions in Figure~\ref{fig:2347} identify segments with nearly zero transmission.)  Higher redshift \HeII\ sightlines, famously Q0302-003, show even more significant troughs \citep{heap00, syphers14}.  These troughs make way for a forest of transmission at lower redshifts.  Even at these lower redshifts, there is $\approx 100$ times more \HeII\ than \HI\ at any location \citep{fechner07, worseck11b}, resulting in the \HeII\ Ly$\alpha$ transmission coming largely from the deepest voids.

\HeII\ Ly$\alpha$ forest spectra shed light on the process of \HeII$\rightarrow$\HeIII\ reionization, termed ``\HeII\ reionization'', a process that chronicles the history of intergalactic $>4\,$Ry backgrounds \citep{1993MNRAS.262..273M}.  The leading picture is that intergalactic $>4\,$Ry radiation owes to quasars.\footnote{Stars in galaxies likely cannot doubly ionize the helium. Unless the IMF is top heavy, stellar populations do not produce many $>4$Ry photons \citep{bromm-vms, 2003ApJ...584..621V}.  Even for the Wolf-Rayet and other massive stars which can produce some $>4$Ry photons, it is unlikely those photons can escape the galaxy because the escape fraction of \HeII-ionizing photons should be much less than that for \HI-ionizing photons.}  As mentioned in \S~\ref{ss:thermalhistory}, this picture is supported by most measurements of the quasar spectral energy distribution and luminosity function \citep[see \S~\ref{sec:sources} for related discussion]{hopkins07a, willot10}, which are consistent with quasars producing enough photons to reionize the \HeII\ around $z\approx 3$ \citep{wyithe03, furlanetto08H}.  The \HeII\ Ly$\alpha$ Gunn-Peterson troughs directly imply that small-scale void regions (which can be localized using \HI\ Ly$\alpha$ absorption) are $\gtrsim 1\%$ \HeII.  Because the gas is either photoionized or fully \HeII\ (and noting that the voids seen in the \HI\ Ly$\alpha$ forest are roughly a tenth of the mean density), this bound implies that the \HeII\ fraction at the mean density is $\gtrsim 10$\% \citep{mcquinnGP}.  (The $\sim 100~$kpc-scale voids in the forest cannot experience much different ionizing background than mean density regions.)  Greater than ten percent is a strong constraint; this argument shows that \HeII\ has not been reionized fully in these trough regions.  These troughs are observed to occur at $z\gtrsim2.7$ (\citealt{shull10}; although there are only two \HeII\ spectra that provide useful spectra at $z\lesssim 2.7$).  Accordingly, this redshift is often taken to mark the end of \HeII\ reionization.\footnote{The results of this exercise contrast with the inferences from the Gunn-Peterson troughs in the \HI\ Ly$\alpha$ forest at $z\sim 6$, which place the limit $x_{\rm HI} \gtrsim 10^{-4}$, because $z\sim 6$ \HI\ Ly$\alpha$ saturates much more easily \citep{fan06}.}  Studies have also attempted to understand the implications of the evolution in the mean \HeII\ Ly$\alpha$ opacity for the timing of \HeII\ reionization, which shows significant evolution at $z>2.8$ in addition to large fluctuations about the mean in $\sim10~$comoving Mpc segments \citep{worseck11b, davies14, worseck14}.  These studies have been less conclusive because the mean \HeII\ Ly$\alpha$ opacity is difficult to model.

At observable redshifts that occur after the \HeII\ reionization was complete ($2 <z\lesssim 2.7$), the \HeII\ Ly$\alpha$ forest is useful for constraining the hardness of the metagalactic ionizing background, as the ratio between \HI\ Ly$\alpha$ and \HeII\ Ly$\alpha$ optical depths directly measures the local ratio between the \HeII\ and \HI\ photoionization rates (informing ionizing background models).  Many studies have attempted this exercise using the brightest two \HeII\ quasars, HS1700+6416 and HE2347-4342.  Early work found mysterious order-of-magnitude spatial fluctuations in this ratio \citep{shull04, zheng04, shull10}, much larger fluctuations than the nearly uniform prediction of models for the post-reionization ionizing backgrounds.  This result has not been confirmed by more recent work \citep{fechner07, mcquinnHeII}, which found that the $z\sim 2.5$ data is consistent with background models as long as quasars contribute  half or more of the \HI-ionizing background.

\subsection{metal absorption lines and the enrichment of the IGM}
\label{ss:metals}

\begin{figure}
 \centering
    \begin{subfigure}[t]{0.49\textwidth}
    \centering
    \includegraphics[width=1\textwidth]{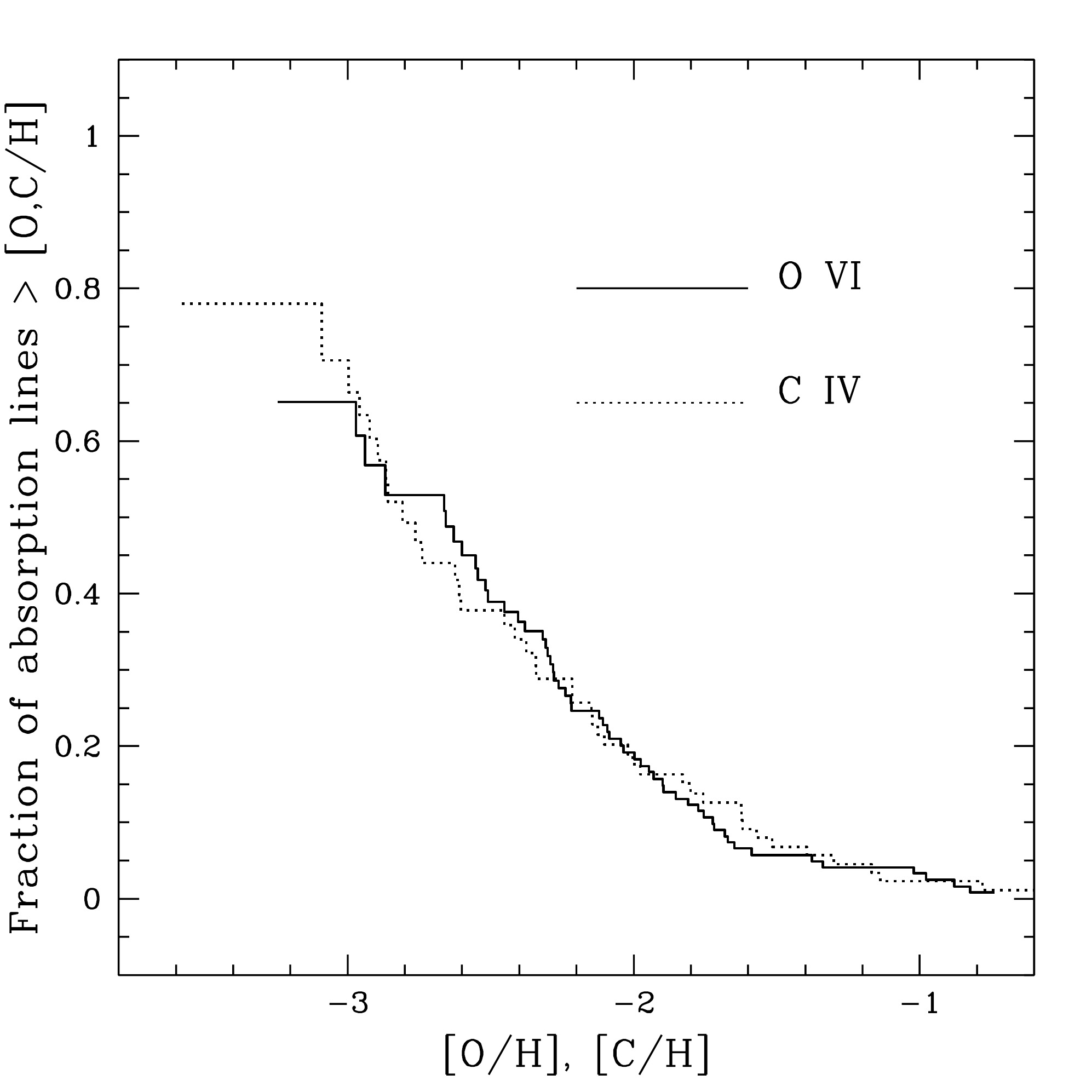}
\end{subfigure}
    \begin{subfigure}[t]{.49\textwidth}
    \centering
\includegraphics[width=1\textwidth]{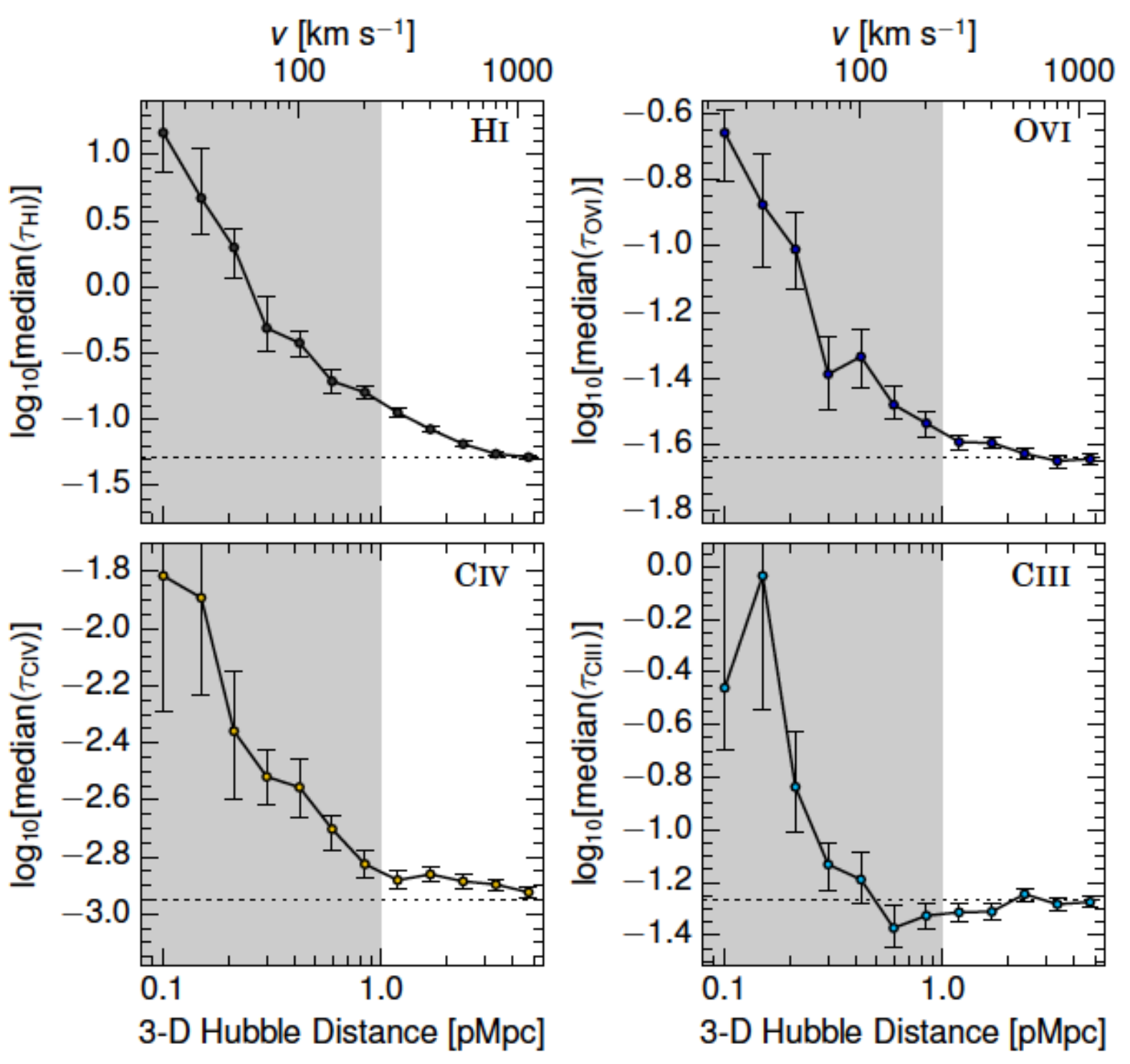} 
\end{subfigure}
\caption{{\it Left panel:}  Fraction of Ly$\alpha$ lines with $N_{\rm HI} >10^{13.6}$cm$^{-2}$ (corresponding to $\Delta_b \gtrsim 3$) that are enriched above the quoted metallicity using \OVI\ absorption (solid histogram) or with $N_{\rm HI} >10^{14}$cm$^{-2}$ ($\Delta_b \gtrsim 5$) using \CIV\ absorption (dashed histogram), from \citet{simcoe04}. {\it Right panel:} Median pixel optical depth at $z\approx 2.4$ of the named ion as a function of the physical separation between galaxy and absorber (assuming Hubble flow velocities in the line-of-sight direction, which ignores the redshift-space distortions that significantly affect the regions highlighted in grey), from \citet{turner14}.
\label{fig:metalabsorption}}
\end{figure}

Once stars formed in the Universe, not only did their ionizing emissions ionize all of the hydrogen, but their radiation pressure and their supernovae powered winds into the IGM, enriching the cosmic volume.   These winds may even have been particularly efficient at evacuating the halos of the first galaxies, which had shallow potential wells.   Observationally, galactic disks contain only a fraction of the metals synthesized by their stars \citep[e.g.][]{2007ApJ...658..941D, 2011MNRAS.416.1354D}, with the fraction of metals retained in a galaxy's ISM estimated to be $\sim20\%$ across a large range of stellar masses at $z\sim 0$ (with an uncertain percentile also residing in the circumgalactic medium; e.g. \citealt{peeples14}).  The rest must have escaped into the IGM.

Metal absorption that falls redward of the forest is annotated in the quasar absorption spectrum shown in the middle panel of Figure~\ref{fig:forestex}.  The strongest metal absorbers that appear in this figure, with optical depths $\gtrsim 1$, are likely of galactic or circumgalactic origin.\footnote{An exception is a curious population(s) of abundant, highly enriched, and diminutive overdense absorbers ($Z= 0.1-1~Z_\odot$, $L\sim100$pc, $\Delta_b\sim 100-1000$) reported in \citealt{simcoe06} and \citealt{2007MNRAS.379.1169S} that may reflect how the IGM was enriched.}  Indeed, if only familiar with the previous content in this review, one might guess that the typical metal-line optical depth in the IGM is zero:  Vanilla models for the Ly$\alpha$ forest are an astonishing success; galactic winds cannot significantly redistribute mass or this redistribution would change the statistical properties of the forest.  Contrary to this intuition, studies find that half or so by mass of the intergalactic gas has been enriched to a detectable level of [O, C/H]$>-3.5$, where e.g. [C/H] denotes $\log_{10}$ of this ratio relative to the Solar ratio.

Two methods have been used to establish this result, both utilizing the highest $S/N$ high-resolution quasar spectra.  The first is the so-called pixel optical depth method \citep{cowie98, ellison00, pod1}, which tabulates at fixed optical depth in \HI\ Ly$\alpha$ the optical depth of a given metal ion at the same location in the IGM, correcting for noise and other contaminants.  The second method involves carefully fitting individual metal absorption lines associated with an \HI\ absorber and, then, performing a survival analysis to quantify the probability of false detections  \citep{simcoe04}.  The resulting metal ion optical depths from both methods are largely in agreement, although the second method is somewhat less sensitive.   These methods have been used to infer an approximately lognormal distribution of optical depths for the prominent metal lines at fixed $\tau_{\rm Ly\alpha}$ (and hence fixed $\Delta_b$ assuming the model of \citealt{schaye01}).  In particular, \citet{pod2} found a median optical depth in \CIV\ at $z\approx 2.5$ of $\tau_{\rm CIV} = 0.1 \Delta_b^{3/2}$ for $\Delta_b \gtrsim 0.5$ and with a large standard deviation of $\approx 0.8\,$dex, and \citet{pod3, pod4} found similar optical depth distributions for \OVI\ and \SIV.  

One of the most interesting results of these analyses is the fraction of cosmic gas that has been enriched.  \citet{simcoe04} estimated from their survival analysis that $60-70\%$ of systems with $N_{\rm HI} \geq 10^{13.6}$cm$^{-2}$ (or $\Delta_b > 2.6$ using eqn.~\ref{eqn:schayerho}) show metal absorption.  The fraction of absorbers enriched to a given metallicity is shown in the left panel of Figure~\ref{fig:metalabsorption}, from \citet{simcoe04}.  If the metals for each system are well mixed, such observations imply that at least half of all gas by mass and $\approx 5\%$ by volume has been enriched to a detectable level \citep{simcoe04, 2004MNRAS.347..985P}.

In order to translate metal line optical depths to median metallicities requires assumptions about the incident ultraviolet background and the gas temperature.  All inferences use uniform background models in the vein of \citet{haardt96} and assume $\sim 10^4$K photoionized gas.  Over $2\lesssim z \lesssim 4$, \citet{pod2} and \citet{pod4} find a median metallicity of [C/H] $=-3.5 + 0.1 [z-3] +0.7 [\log(\Delta_b-1) -1]$,  an oxygen to carbon ratio of [O/C] $=0.7\pm 0.2$, and a logarithmic standard deviation of $\sigma($[C, O/H]$) \sim 0.8~$dex.  These numbers are in general agreement at $z\approx 2.4$ with the measurements of \citealt{simcoe04}, aside from the $\Delta_b$ dependence of the median metallicity, which was not detected there.  Plausible changes to the ionizing background model change the metallicity estimates by a few tenths of dex.  \citet{simcoe11} repeated the \citet{simcoe04} analysis on \CIV\ but for $z \approx4.3$ rather than $z\approx2.4$, finding a median metallicity for slightly overdense gas that is a factor of $2-3$ smaller, suggesting that much of the enrichment occurred between these redshifts.

Another approach to understand the enrichment of the IGM is to look at the statistics of absorbers in quasar spectra as a function of their distance to Lyman break-selected galaxies in the foreground of the quasar.  \citet{turner14} recently investigated this statistic at $z\approx 2.4$ using pixel optical depth methods with galaxy separation taking the place of $\tau_{\rm Ly\alpha}$.  The median absorption they measure as a function of the 3D separation  (which is calculated by adding in quadrature the transverse and line-of-sight distances assuming pure Hubble flow) from galaxies is shown in the right panel of Figure~\ref{fig:metalabsorption} for the most prominent ions.  \citet{turner14} detected correlations out to a proper Mpc for the standard metal ions.  At proper Mpc-scales these correlations likely owe to the clustering of galaxies, but at shorter distances they should reflect the extent of each galaxy's own enrichment.  However, disentangling these effects (as well as interpreting most metal line diagnostics) requires detailed modeling.

Most theoretical models of chemical enrichment use large simulations of a cosmological volume that attempt to capture galaxy formation.\footnote{As an alternative to large simulations, an interesting semi-analytic exploration of how enrichment must occur was pursued in \citet{booth12}.  This study investigated how far metals would have to be distributed around halos of a chosen mass to explain their incidence in observations.  They found that satisfying pixel optical depth constraints on how far metals propagate requires $\lesssim10^{10}\Msun$ halos to enrich a sphere of radius $\approx0.1\,$physical Mpc and that larger halos cannot be responsible for all of the enrichment.  }  These simulations adopt ``feedback'' recipes for how star formation (and sometimes quasar activity) expel gas \citep{2001ApJ...561..521A, 2001ApJ...560..599A, tescari11, 2006MNRAS.373.1265O, 2010MNRAS.409..132W, 2011MNRAS.415..353W}.  There is much freedom in how feedback is implemented in these simulations.  Some eject from galaxies one to a few solar masses in ``wind'' particles with a specified velocity for every solar mass formed in stars, tuning the exact numbers to match observations of winds, galaxy stellar masses, and other observables.  Others inject momentum/energy around star formation sites and try to follow the development of galactic winds more organically.  Generally in these simulations, the intergalactic metal yield tends to trace the star formation history, with enrichment contributed by galaxies over a broad range in stellar mass and extending hundreds of kiloparsecs from galaxies.  However, certain feedback recipes provide a better match to observations, with some able to reproduce the gross properties of intergalactic metal lines \citep[e.g.][]{2011MNRAS.415..353W}.  Describing these comparisons in detail would require its own review.

\section{the IGM at high redshifts, $z>5$, and reionization}
\label{sec:highz}

The diagnostics of the IGM at $z>5$ differ from those at lower redshifts, partly because our star diagnostic, Ly$\alpha$ forest absorption, is becoming very saturated in most pixels.  At $z>5$ it is not possible to detect the \HeII\ Ly$\alpha$ forest, and the detected metal absorbers (namely \OI\ and \CIV) are likely associated with dense gas around galaxies rather than the diffuse IGM \citep{2014MNRAS.438.1820K}.\footnote{This occurs partly because the quality of $z>5$ spectra is poorer such that the sensitivity to small optical depth absorbers decreases and partly because pixel optical depth analyses are more difficult owing to saturation in the forest.}  The focus shifts to a new set of observables that is sensitive to a neutral IGM and cosmological reionization.

\subsection{the sources of reionization}
\label{sec:sources}

\begin{figure}
  \centering
    \begin{subfigure}[t]{0.49\textwidth}
    \centering
    \includegraphics[width=.92\textwidth, height=.88\textwidth]{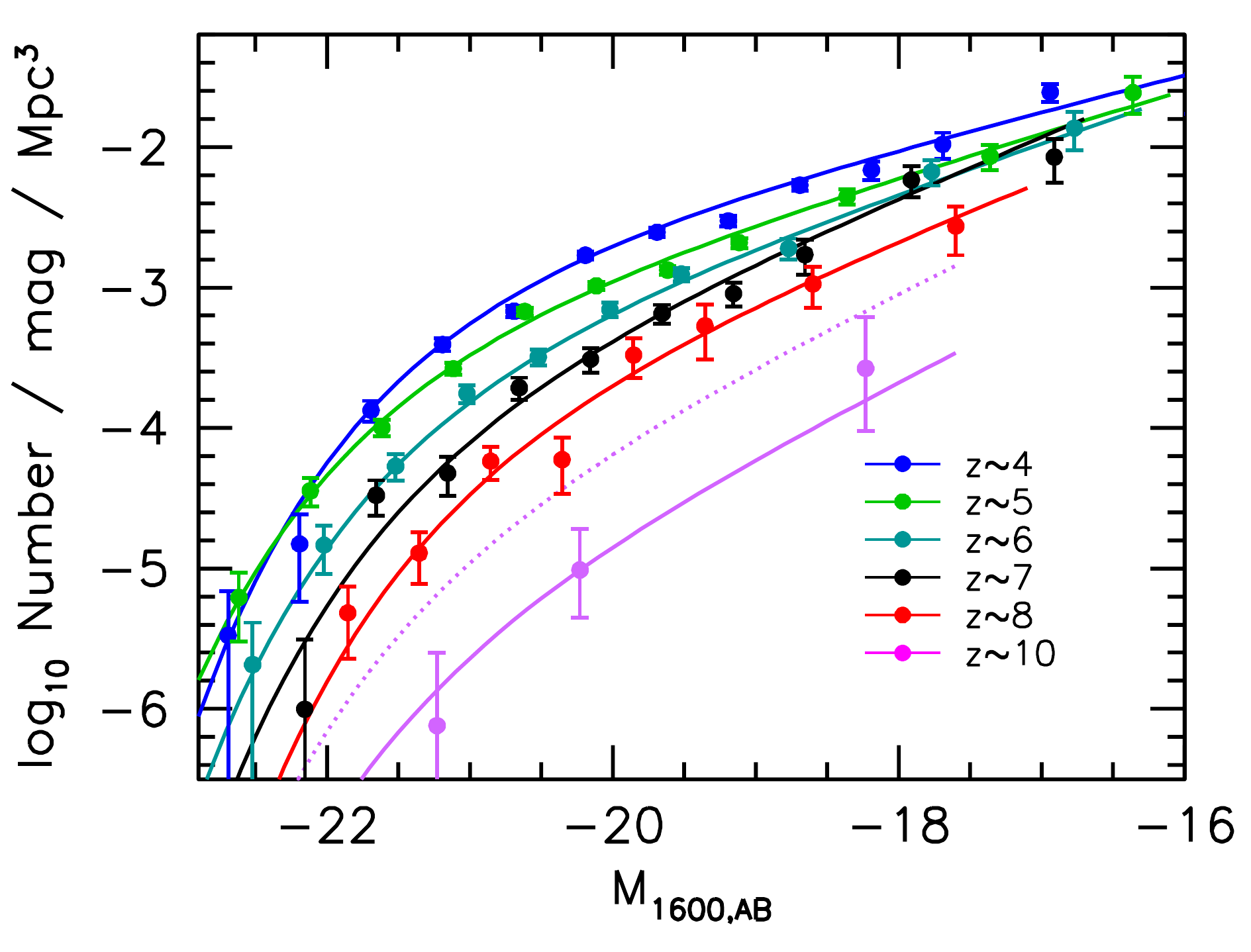}
\end{subfigure}
    \begin{subfigure}[t]{.49\textwidth}
    \centering
\includegraphics[width=1\textwidth]{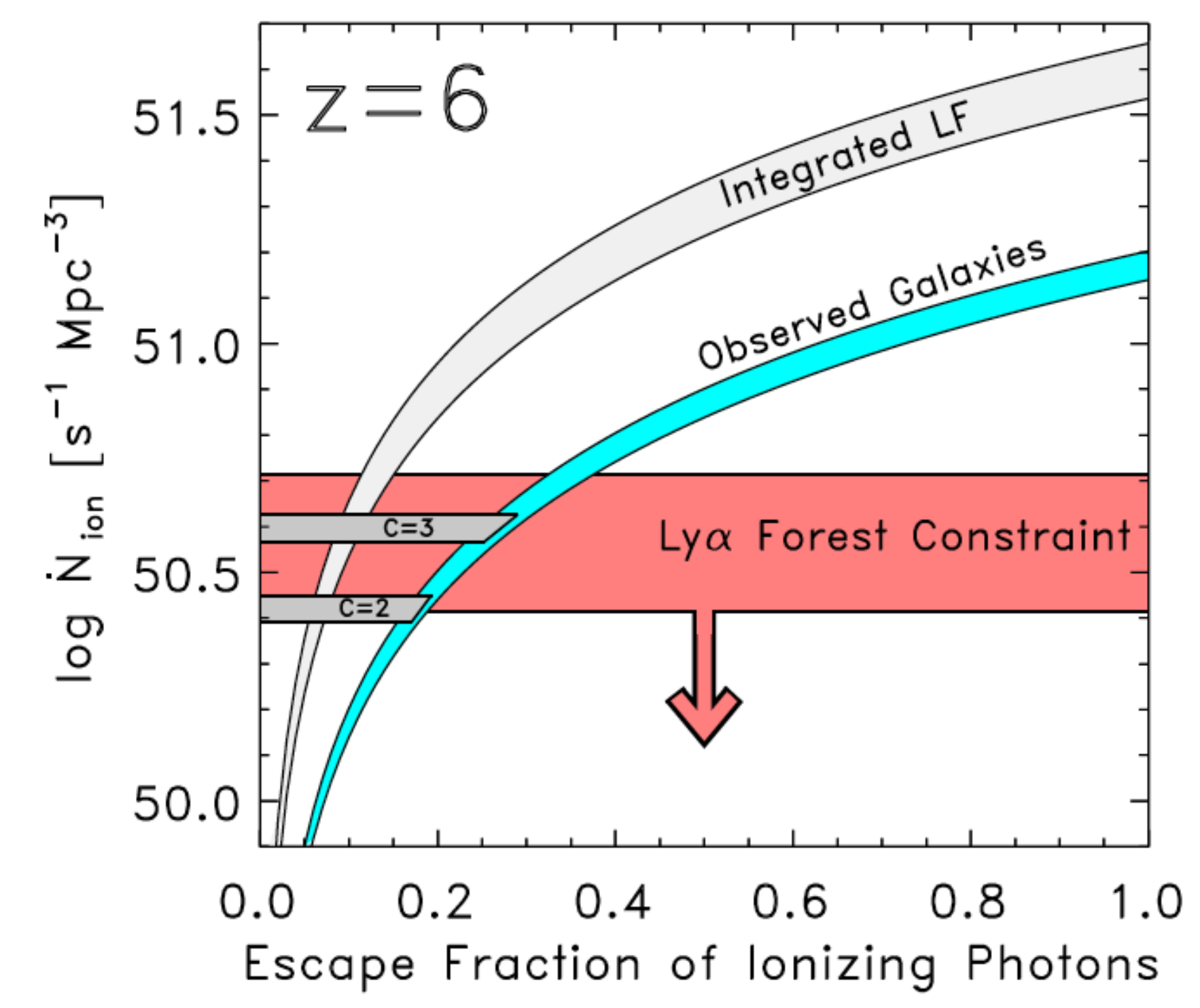} 
\end{subfigure}
\caption{{\it Left panel:}  Luminosity function of HST dropout-selected galaxies, from \citet{bouwens15}.  The points with error bars are the measurements, and the solid curves are the best-fit Schechter functions. {\it Right panel:} Comparison of $z=6$ estimates for the emissivity of ionizing photons, from the observed population of galaxies and from integrating the luminosity function below the detection threshold, as a function of the assumed $f_{\rm esc}$.  These numbers are compared against the ionizing emissivities required to maintain reionization with clumping factors of $2$ or $3$ and against the emissivities inferred in \citet{kuhlen12} using $\Gamma$ and $\lambda_{\rm mfp}$ estimates, from \citet{finkelstein12}.  Recent constraints from \citet{becker13} allow up to a factor of three higher emissivities than \citet{kuhlen12}.  \label{fig:emissivity}}
\end{figure}

The consensus model is that galaxies drove the reionization of intergalactic hydrogen \citep[e.g.][]{shapiro94, madau99, faucher08, becker13}.  This model is supported by observations of high-redshift galaxies, which find approximately enough ultraviolet emission at $z=6$ to reionize the Universe if a substantial fraction of the associated ionizing radiation escaped into the IGM \citep{robertson10, finkelstein12}.  Alternatively, the steep faint-end slope of the observed UV luminosity function also supports unseen low-mass galaxies being responsible for reionization.  (The sidebar Exotic Reionization Models discusses other models that have been previously considered.) To reionize the IGM in a time $t_{\rm SFR}$ requires a comoving galactic star formation rate density of
 \begin{eqnarray}
\dot \rho_{\rm SFR} &=  & 3.6 \times 10^{-2} N_{\gamma/b} \left(\frac{1+z}{8}\right)^{3/2} \left( \frac{t_{\rm SFR}}{0.5 t_{\rm uni}(z)} \right)^{-1} \left(\frac{f_{\rm esc}}{0.1}  \frac{\xi_{\rm ion}}{4000}\right)^{-1} {\rm \Msun \, yr^{-1}  Mpc^{-3}}\label{eqn:RI}
\end{eqnarray} 
where $t_{\rm SFR}/t_{\rm uni}(z)$ is the timescale of star formation relative to the age of the Universe, $ N_{\gamma/b}$ is the required number of ionizing photons per hydrogen atom to complete reionization (studies find $1-3$; \S \ref{ss:reionmodels}), $f_{\rm esc}$ is the highly-uncertain fraction of ionizing photons that escaped from their galactic sites of production into the IGM, $\xi_{\rm ion}$ is the number of ionizing photons that are emitted for each stellar baryon, and $\dot \rho_{\rm SFR}$ is the rate density in comoving-space.  Much numerical work has focused on the difficult problem of calculating $f_{\rm esc}$ (which requires the daunting task of resolving the multiphase ISM), but with answers essentially ranging from zero to one \citep{gnedin08, 2009ApJ...693..984W} with some recent simulations finding average values of $f_{\rm esc} \sim 10\%$ \citep{2014ApJ...788..121K, 2015MNRAS.453..960M}.   There have also been attempts to measure the escape of ionizing photons from $z\lesssim 3$ galaxies, which yield a confusing picture as well  \citep{2009ApJ...692.1287I, 2010ApJ...723..241S, siana15}.  In addition, $\xi_{\rm ion}$ equals $4000$ for an empirically motivated stellar initial mass function (IMF), although with factor of two theoretical uncertainty \citep{shull11}.  A metal-free, top-heavy IMF produces a factor of ten more ionizing photons per stellar baryon \citep{bromm-vms, 2003ApJ...584..621V}.  

Equation~(\ref{eqn:RI}) is slightly different than the more widely used metric, the critical star formation rate density to maintain reionization.  This metric compares with the comoving $\dot \rho_{\rm SFR} $ that produces enough ionizing photons to balance the number of intergalactic recombinations for a reionized IGM \citep{madau99}:
  \begin{equation}
  \dot \rho_{\rm SFR} =   3.1 \times 10^{-2} \, \left(\frac{1+z}{8}\right)^{3} \,\left( \frac{C/2}{f_{\rm esc}/0.1} \right)   \left ( \frac{\xi_{\rm ion}}{4000}\right)^{-1} ~{\rm \Msun ~yr^{-1} ~Mpc^{-3}}.\label{eqn:RIcomb}
  \end{equation}  
 Studies find a gas clumping factor of $C \equiv \langle \Delta^2 T_4^{-0.7} \rangle_V \sim 2-3$ at $z\sim 7$, where $\langle ... \rangle_V$ denotes an average over the volume and $T_4$ is the temperature in units of $10^4~$K \citep{pawlik09, mcquinn-LL, 2015ApJ...810..154K}.  Because the recombination time at $z\sim6$ is roughly the age of the Universe, equations~(\ref{eqn:RI}) and (\ref{eqn:RIcomb}) require similar $\dot \rho_{\rm SFR}$. 

There has been much discussion on whether the observed population of high-redshift galaxies was responsible for reionization.  Observationally, the luminosity function of ultraviolet dropout-selected galaxies at source-frame $\sim1500$\AA\ has been measured at $z>6$ with HST  \citep{2012ApJ...756..164F, 2012ApJ...754...83B, 2013ApJ...763L...7E, 2013MNRAS.432.2696M}.   The ultraviolet dropout technique uses the break in a high-redshift spectrum that occurs at source-frame Ly$\alpha$ owing to intergalactic Ly$\alpha$ absorption.  These measurements have been improved in recent years with the installation of the near infrared sensitive Wide Field Camera~3 on HST during the 2009 servicing mission, extending luminosity function estimates to redshifts as high as $z=10$.  The left panel in Figure~\ref{fig:emissivity} shows the luminosity function of HST dropout-selected galaxies at $z=4-10$, from \citet{bouwens15}.  There is a decline in the luminosity function over this redshift (especially at $z>6$).  Such a decline is not surprising as the abundance of the halos that host these galaxies evolves similarly at $z>6$. 

  The final step to infer the total ionizing flux of the population is to estimate the conversion from the observed flux at $\sim1500$\AA\ to the amount of ionizing flux.  This conversion involves matching the spectral slope of these galaxies' emissions to the slopes in stellar population synthesis models \citep[e.g.][]{robertson13}.  The conversion estimate combined with the best-fit $\sim1500$\AA\ luminosity function allow one to estimate the ionizing photon emissivity.  The results of this exercise are that for $f_{\rm esc}\gtrsim 0.2$ the observed population of galaxies \emph{can} maintain reionization at $z\sim 6$. The curved bands in the righthand panel of Figure~\ref{fig:emissivity} show the estimated photon emissivity from the observed galaxies as a function of their $f_{\rm esc}$, from \citet{finkelstein12}; see also \citet{2015arXiv150308228B}.  These numbers are compared against the ionizing emissivities required to maintain reionization with clumping factors of $2$ or $3$ and against the emissivity range inferred from the Ly$\alpha$ forest (a constraint discussed shortly).
  
However, it is possible that $f_{\rm esc} \lesssim 0.2$ and that less luminous galaxies than those observed reionized the Universe.  The observed population of galaxies shows a steep faint-end slope of $\alpha =-1.8\pm0.2$ \citep{kuhlen12, 2013MNRAS.432.2696M, bouwens15}.   For luminosity-independent $f_{\rm esc}$, $\alpha=-1.8$ results in about a factor of two more ionizing flux from fainter galaxies than those observed (see the ``integrated'' band in the right panel of Fig.~ \ref{fig:emissivity}), but of course $\alpha=-2$ is logarithmically divergent.  Theoretical models predict that there should be some star formation in halos down to $\sim 10^{8}\Msun$, which (if $\alpha=-1.8$ is maintained the entire way) fall $6-10$ astronomical magnitudes beyond those observed \citep{kuhlen12}.  However, it might also be surprising if galaxies in such diminutive halos reionized the Universe.  Star formation at low redshifts seems progressively less efficient as the halo mass of a system decreases because of stellar feedback.  Models in which reionization was driven by $\sim 10^8\Msun$ halos -- the smallest halos that can cool atomically -- with a standard IMF appear to be ruled out as they overproduce the observed stellar masses of the ultra-faint dwarf galaxies \citep{2014MNRAS.443L..44B}.

\begin{textbox}
\subsubsection{Exotic Reionization Models} Starting with \citet{1970ApL.....5..123A}, another source that has often been mentioned as potentially responsible for reionization is quasars.  Quasar reionization scenarios have generally been argued against because most observations show quasar numbers to be declining with increasing redshift above $z=3$ \citep[e.g.][]{hopkins07a, willot10}.  This decline seems consistent with the merger hypothesis for quasars and the decreasing abundance of massive galaxies, as the quasar luminosity scales strongly with galaxy stellar mass in models \citep[e.g.][]{2008ApJS..175..356H}.  However, \citet{2015A&A...578A..83G} recently  claimed to find a sufficient abundance of quasars at $z\sim 6$ for them to be able to reionize the Universe, partially driving a resurgence of quasar-reionization models \citep{chardin15, madau15}.  An issue with this model is that, if quasars reionized the hydrogen, their hard spectrum would also doubly ionize the helium by $z=4$ \citep{madau15}, in conflict with the \HeII\ Gunn Peterson trough detections and the IGM temperature measurements (\S~\ref{ss:HeIIforest} and \S~\ref{ss:thermalhistory}, respectively).

In addition, $X$-rays from high-mass $X$-ray binaries \citep{furlanetto06, mirabel11}, supernovae shocks \citep{johnson11}, supernova-accelerated electronic cosmic rays \citep{oh01}, and even more exotic processes such as dark matter annihilations \citep{belikov09} may contribute some fraction of the ionizations \citep{2004ApJ...613..646D, mcquinn-Xray}.  $X$-ray photons have two advantages relative to ultraviolet ones when it comes to reionization.  First, a single $X$-ray can convert as much as 30\% of its energy into ionizations (with this fraction decreasing with $x_i$; \citealt{shull85}).  Second, $X$-rays should have no problem escaping from their sites of production into the IGM, with $h \nu \lesssim 1.5 \, x_H \sqrt{(1+z)/10}\;$keV photons absorbed within a Hubble distance.  Because they penetrate much further into the IGM, early $X$-rays should have established a relatively uniform ionization floor before ultraviolet photons finished the job, with empirically motivated models predicting a floor of $0.1-1\%$ \citep{furlanetto06,pober15}, although the modeling uncertainties are immense.
\end{textbox}

In addition to directly observing high-redshift sources, the global amount of ionizing photons they emit as a function of redshift can be inferred from quasar absorption line studies.  In particular, as shown in \S~\ref{ss:uvbmodels} (c.f. eqn.~\ref{eqn:highz}), the emissivity of the sources is proportional to the often well-constrained amplitude of the hydrogen ionizing background times the mean free path of ionizing photons -- which is set by the observationally-constrained number of systems with $N_{\rm HI} \sim 10^{17}~$cm$^{-2}$.  \citet{miralda03} estimated the ionizing emissivity in this way, showing that the emissivity, even at $z=4$, was not much higher than that required to reionize hydrogen.  Later, \citet{bolton07} extended this result to $z\approx 6$, estimating $1.5-3$ ionizing photons per hydrogen atom were being emitted per Gyr (the age of the Universe at $z=6$ is $0.94\,$Gyr) and coining the phrase ``photon-starved reionization'' to describe this result.  In addition, \citet{bolton07} found an emissivity history that was remarkably constant over $z=3-6$, as \citet{miralda03} had surmised.  This result has held up in some subsequent studies \citep{faucher08, kuhlen12, becker13}.\footnote{The constancy of the emissivity with redshift is in some tension with the observed stellar population, as $\dot \rho_{\rm SFR}$ is decreasing in the observed population of galaxies, as seen in the left panel of Figure~\ref{fig:emissivity}.  Reconciling these either requires an evolving $f_{\rm esc}$ or an increasing contribution to $\dot \rho_{\rm SFR}$ from lower mass systems than have been observed.}  Recently this photon-starved result has been revisited by \citet{becker13}, using higher redshift $\lambda_{\rm mfp}$ estimates and more rigorously tracking the sources of uncertainty.  The \citet{becker13} analysis found that the data allows $3-10$ ionizing photons per hydrogen atom per Gyr at $z\approx 5$, suggesting that reionization might not be so photon starved after all.

However, there is still a compelling case for reionization to be photon starved.
Another approach to infer the emissivity is to calculate the clumping factor of ionized gas in cosmological simulations that use an empirically motivated $\Gamma$ and that account for self-shielding.  Integrating the cosmological radiative transfer equation (see text above eqn.~\ref{eqn:Jnu}) over $d^3x \,d\Omega \,d\nu \, \nu^2$ yields
\begin{equation}
\bar \epsilon_\gamma =  \overbrace{\left( \dot {\bar n}_{\rm HI} +  C \alpha_{B, 4} \; \bar n_e \bar n_{\rm HII}\right)}^{\Gamma \bar n_{\rm HI}} + 4\pi/c \dot J_\gamma,
\label{eqn:cont}
\end{equation}
where an overbar denotes a spatial average, a subscript $\gamma$ indicates the quantity in terms of the total number of ionizing photons, and $\alpha_{B, 4} \equiv \alpha_B(T=10^4{\rm K})$.  To simplify, first note that $\dot n_{\rm HI}$ is small after reionization.  Additionally, because $J_\gamma \approx \lambda_{\rm mfp} \epsilon_\gamma/4\pi$, the latter term is smaller by the factor $\lambda_{\rm mfp}/[c f t_{\rm uni}]$, where $f$ is the fraction of $t_{\rm uni}$ over which the background is evolving (the background at $2\lesssim z\lesssim6$ evolves remarkably little and so $f \sim 1$). Thus, eqn.~(\ref{eqn:cont}) simplifies to $\bar \epsilon_\gamma \approx C \alpha_4 \bar n_e \bar n_H$, such that if $C=2$ this implies $\approx 2$ ionizing photons per hydrogen atom per Gyr at $z=6$ since at this time $(\alpha_{B, 4} \; n_e)^{-1} = 1.1$~Gyr.  Simulations with radiative transfer to capture self-shielding suggest small clumping factors of only $2-3$ at $z\sim 6$ for ionizing backgrounds in the range of those allowed \citep{mcquinn-LL, 2015ApJ...810..154K}.  For regions that have been reionized within $\Delta z\approx 2$ and hence have not had time to relax, the gas can have somewhat higher clumping factors \citep{pawlik09}.

\subsection{models of reionization}
\label{ss:reionmodels}
\begin{figure}
\includegraphics[width=1.\textwidth]{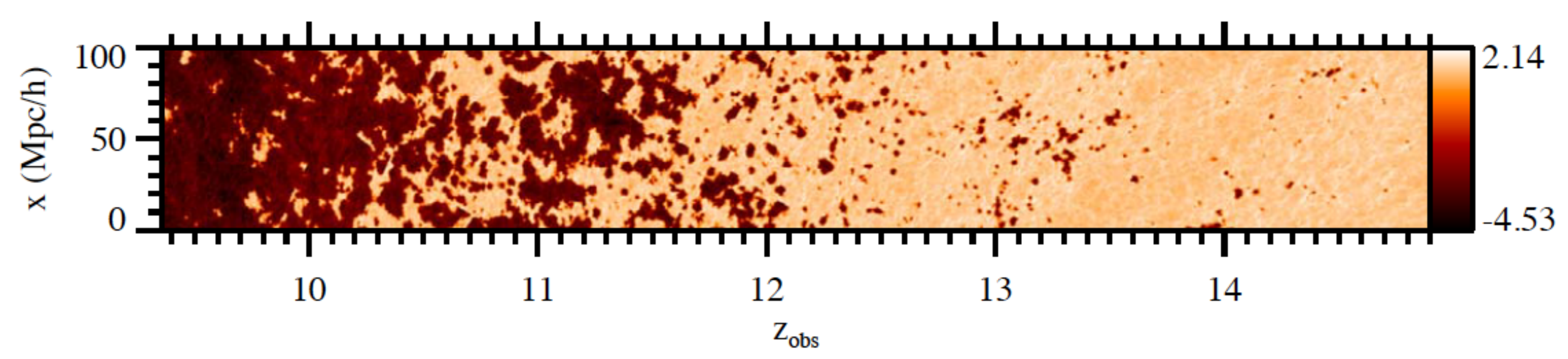}
\caption{Image of a $100/h~\Mpc$ radiative transfer simulation of reionization.  Shown is a slice through the $21$cm emission signal on the light cone (showing $\log_{10}$ of the brightness temperature).  Darker regions are ionized and brighter ones are neutral, from \citet{mellema06b} \label{fig:reionization}}
\end{figure}

Almost every observable of reionization is tied to the structure of this process, making models critical for interpreting the observations.  There has been substantial effort directed towards modeling reionization \citep{gnedin97, 2003MNRAS.343.1101C, furlanetto05, iliev06,mcquinn07, trac07, finlator09, gnedin14a, thomas09, santos10, mesinger11, 2012MNRAS.423..558C, 2015MNRAS.451.1586P}.   Existing models for the structure of reionization are almost exclusively in the prevailing galactic source paradigm.  Generically in these models, ionization fronts propagate outward from galaxies and, much like with \HII\ regions in the interstellar medium, there is a sharp boundary between where gas is ionized and where it is neutral.  This width is $\sim1 \Delta_b^{-1}~$physical~kpc during reionization, a scale that is generally much smaller than the size of the \HII\ regions themselves (and often unresolved).  Thus, a typical region in the IGM during reionization is expected to be either highly ionized or nearly neutral, leading to the term ``patchy reionization''.   The ionization fronts also heat the gas from likely hundreds of Kelvin to more than $10^4$K, causing the gas to evaporate from $< 10\;$km~s$^{-1}$ potential wells \citep[see also ``The Minimum Mass of Galaxies'' sidebar in \S~\ref{ss:thermalhistory}]{barkana99, shapiro03, okamoto08}.

Most predictions for the structure of reionization come from semi-analytic models or large-box radiative transfer simulations (typically performed with ray tracing algorithms).  ``Large box'' is used here to mean box sizes that are $\gtrsim 100~$\Mpc, roughly the scale studies have shown is required to have a sufficient sample of structures (and to not be biased by missing large-scale modes) to make accurate statistical statements about reionization \citep{barkana03, furlanetto04a, iliev06}.  The predictions for the structure of reionization in semi-analytic models and large-box simulations have been shown to be in fantastic agreement with one another \citep{zahn06, santos08, zahn11}, although these models themselves are not without controversy, as discussed shortly.  The semi-analytic models and large-box simulations predict that the structure of reionization is driven mostly by the clustering of the galactic sources.  The most striking prediction of these models is that the bubble sizes reach many megaparsecs in size, engulfing thousands and even millions of galaxies \citep{barkana03, furlanetto04a, iliev06}.  Because these overdensities of galaxies trace large-scale matter overdensities, these simulations also established that on large-scales reionization should be ``inside-out'', meaning that on megaparsec scales and greater overdense regions are ionized first, rather than outside-in as can happen on smaller scales (see sidebar The \citealt{miralda00} Model for Reionization).  Even when the global hydrogen ionized fraction is only $\sim 10$\%, these models tend to find that much of the ionized volume is subsumed by $\sim 10~$comoving Mpc bubbles.  Many semi-analytic models even find $>100~$comoving Mpc bubbles by the time the intergalactic hydrogen is $\sim90$\% ionized \citep{furlanetto04a, 2012MNRAS.425.2964Z}.  Figure~\ref{fig:reionization} shows a light cone image of a large-box reionization simulation, from \citet{mellema06b}. 

  Large-box calculations have also been used to explore how reionization models depend on the properties of the sources or on unresolved overdense structures that can act as sinks of ionizing photons.   Regarding the source properties, studies find that as more massive galaxies reionized the Universe, the sizes of \HII\ regions increase owing to the enhanced source clustering \citep{furl-models,mcquinn07}, and they also find modest changes for more complex source prescriptions such as tying some of the ionizing emissivity to major mergers \citep{2007MNRAS.374...72C}.\footnote{The detail at which the sources are followed range from prescriptions based simply on halo masses \citep{furlanetto04a, iliev06, mcquinn07} to detailed calculations based on the star formation that is happening in the simulation and following how photons escape from star-forming environs \citep{gnedin97, finlator09, 2015MNRAS.451.1586P}.  One might discount results with the former method because it is simplistic.  The argument in their support is that we do not know from which halos ionizing photons escaped (let alone resolve this process in cosmological simulations).  Because of this uncertainty, simpler, flexible models that enable exploring the allowed parameter space can be justified.}  Regarding the unresolved dense structures that act as sinks of ionizing photons, studies find that the sinks act to cap the maximum ionized bubble size at roughly the photon mean free path to intersect a sink  \citep{furlanetto05, mcquinn07, alvarez12, sobacchi14}. 
  
 A criticism of large-box models is that these dense sinks of ionizing radiation are not self-consistently captured, unlike potentially in small-box, higher spatial resolution simulations such as \citet{gnedin06}.\footnote{Conversely, the small-box simulations of \citet{gnedin06} were unable to capture the large-scale structure of reionization.  Recently, \citet{2014ApJ...793...29G} attempted to bridge the scales and capture the sinks in $\sim100~$comoving Mpc boxes.}  See the sidebar titled ``The \citealt{miralda00} Model for Reionization'' for a description of the physics that shapes the sinks.  These sinks are often imprecisely referred to as ``Lyman-limit systems'' -- which are systems defined to have $N_{\rm HI} > 1.6\times10^{17}$cm$^{-2}$, which correspond to regions with $\Delta_b \gtrsim 10 [(1+z)/10]^{-3} \Gamma_{12}^{-1}$ (eqn.~\ref{eqn:schayerho}).  (At lower redshifts, Lyman-limits contribute about half of the opacity for ionizing photons.)  Some contributions to the sinks include photoevaporating minihalos (defined as halos below the threshold to cool atomically) and other thermally relaxing gaseous structures \citep{iliev-mh, ciardi06, mcquinn06}.   An unfortunate ramification of not self consistently capturing the sinks is that large-box simulations are unable to address the final phase of reionization in which the IGM transitioned to the highly ionized state seen in the Ly$\alpha$ forest.  A related problem is that most previous simulations prescribed source emissivities that were far from being photon-starved (\S~\ref{sec:sources}).  In this very emissive source limit, the impact of dense systems/recombinations should be reduced, potentially skewing the predictions for the structure of reionization \citep{furlanetto05,2012MNRAS.423..558C, 2009MNRAS.394..960C}.

\begin{textbox}
\subsubsection{The \citealt{miralda00} Model for Reionization}
On large-scales, reionization should have been ``inside-out''  -- meaning that overdense regions that host more galaxies were ionized first.  However, as the ionized bubbles grew, the ionizing background within them increased, resulting in an ``outside-in'' process also occurring as dense, self-shielding clumps become progressively more ionized.  \citealt{miralda00} (MHR) proposed an elegant model for this outside-in process, which is able to characterize many of the essential aspects with a single parameter, the sources' ionizing emissivity $\epsilon_\gamma$.  This model makes the ansatz that all gas is ionized up to a critical density $\Delta_{b,*}$ after which it becomes neutral.  This critical overdensity is calculable using the PDF of $\Delta_b$, $P(\Delta_b)$, which can be estimated from cosmological simulations (MHR, \citealt{bolton09}), as the emissivity of the sources balances the rate density of recombinations (see eqn.~\ref{eqn:cont}): $\epsilon_\gamma = \alpha_B \bar n_e^2 \int_0^{\Delta_{b,*}} d\Delta_b P(\Delta_b) \Delta_b^2 \propto \Delta_{b, *}^{3-n},$
where the proportionality assumes $P(\Delta_b)\propto \Delta_b^{-n}$ as simulations find is applicable at high densities.  The ansatz of a critical overdensity has been subsequently tested in several studies, finding that it works very well \citep{2010ApJ...725..633F, mcquinn-LL, 2012MNRAS.427.2889Y}, and, indeed, it is often used as a simple way to incorporate self-shielding in cosmological simulations \citep[e.g.][]{2010ApJ...725L.219N}.  Next, to estimate $\lambda_{\rm mfp}$, MHR made a second  ansatz that the number density and shape of absorbers does not change with $\Gamma$, which implies that the mean free path must scale as the fraction of the volume that is neutral, $\int_{\Delta_{b,*}}^\infty d\Delta_b P(\Delta_b) \propto \Delta_{b, *}^{1-n}$, to the $-2/3$ power.  The scaling coefficient can be measured using simulations or by matching to $\lambda_{\rm mfp}$ measurements.  Finally, to close the equations requires an expression for the photoionization rate:
\begin{equation}
\Gamma = \frac{\sigma_{\rm HI}(\nu_{\rm HI}) (3\beta-3+\alpha)}{3+\alpha} \epsilon_\gamma \lambda_{\rm mfp}(\nu_{\rm HI}) \propto \Delta_{b, *}^{(7-n)/3} \propto \epsilon_\gamma^{\frac{7-n}{9-3n}} \propto \epsilon_\gamma^{1/(2-\beta)},
\label{eqn:scaling}
\end{equation}
where $\alpha$ is spectral index of the sources and $\beta$ the power-law scaling of $\partial^2{\cal N}/\partial N_{\rm HI} \partial z$ at $N_{\rm HI} = 10^{17}$cm$^{-2}$ (see \citealt{mcquinn-LL}).  MHR used this model to demonstrate that the number of recombinations per H atom during reionization is likely small.  More recently, this model has been used to (1) calculate the \HII\ bubble radii above which recombinations retard  their growth \citep{furlanetto05}, (2) to model ionizing background fluctuations \citep{2015arXiv150907131D}, and (3) to model the post-reionization IGM \citep{bolton07,2014arXiv1410.2249M}.

Interestingly, $\beta$ is found to be fairly steep with values $1.7-1.8$ in simulations \citep{mcquinn-LL, altay11} and tentatively the observations \citep{prochaska09b, rudie13}.  Using equation~(\ref{eqn:scaling}), this suggests that $\Gamma \propto \epsilon_\gamma^{3-5}$, making it even more curious why $\Gamma$ is measured to be rather constant over $2<z<5$.  However, this strong scaling potentially helps explain the quick evolution in the IGM Ly$\alpha$ opacity at $z\approx 6$ \citep{mcquinn-LL, 2014arXiv1410.2249M}.
\end{textbox}

 A generic result of modern reionization models is that $N_{\gamma/b}$ -- the ratio of ionizing photons produced to number of hydrogen atoms to complete reionization -- is within a factor of two or so of its absolute minimum value of one.  For example, \citet{2009MNRAS.394..960C} predicted values that range from $1.2-2$ for this ratio.\footnote{$N_{\gamma/b}$ is defined here to exclude the absorption of ionizing photons in the ISM/circumgalactic medium of the source galaxy.  Such absorption is handled by $f_{\rm esc}$.   This distinction explains some much higher numbers for $N_{\gamma/b}$ that have appeared previously in the literature.}  The small values of this ratio can be understood from the reionization occurring in most models primarily at $6< z<10$ and spanning a few hundred million years, a duration that is shorter than or comparable to the effective recombination timescale, $[C \alpha_B  \bar n_e(z)]^{-1}$, at these redshifts.

\subsection{observables of cosmological reionization}

The main observables of cosmological reionization are the Ly$\alpha$ forest (\S~\ref{ss:highzLya}), anisotropies in the CMB (\S~\ref{ss:CMB}), diagnostics of damping wing absorption of intergalactic hydrogen (\S~\ref{ss:DW}), and highly redshifted $21\;$cm radiation (\S~\ref{ss:21cm}).  Studies have quoted a variety of constraints on reionization, typically on the global neutral fraction, using all of the aforementioned probes.  Two of these constraints (the mean redshift of reionization from the CMB and the end redshift of reionization from the Ly$\alpha$ forest) are quite robust.  Many of the other constraints are controversial or model dependent.  Here we discuss what each of these probes may be revealing about reionization as well as each's future prospects.

\subsubsection{the $z>5$ Lyman-series forest}
\label{ss:highzLya}

\begin{figure}
\includegraphics[width=1.\textwidth]{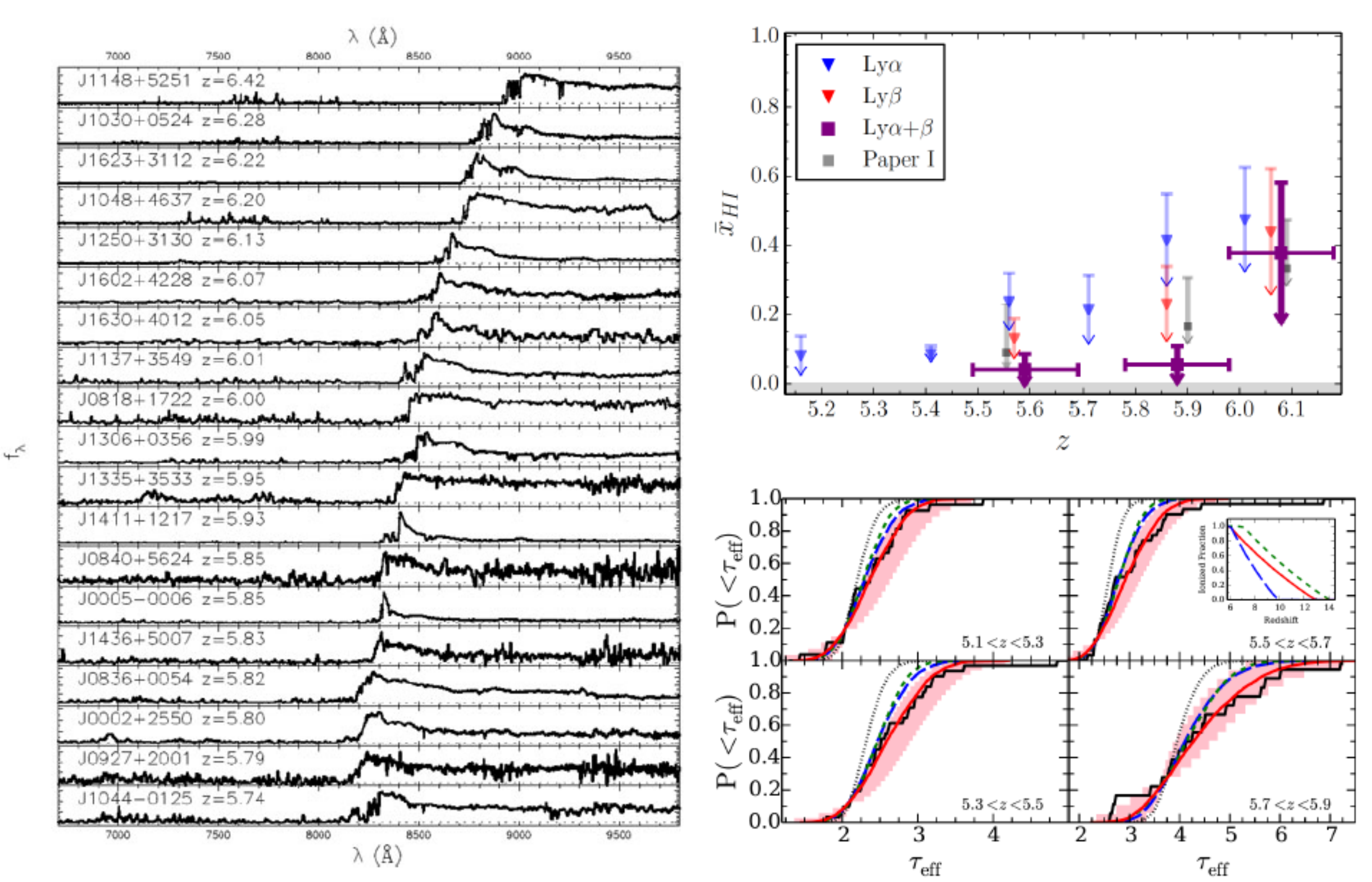}
\caption{{\it Left panel}: Ly$\alpha$ forest spectral region for many of the highest redshift quasars that have been identified, from \citet{fan06}.  {\it Top right:}  Dark gap bound on the neutral hydrogen fraction, showing the fraction of $3.3~$comoving Mpc pixels that show no transmission in Ly$\alpha$ and/or Ly$\beta$, from \citet{mcgreer15}.  {\it Bottom right:}  Panels show the cumulative PDF of $\tau_{\rm eff} = -\log({\rm transmission})$ in redshift bins that span $5.1<z<5.9$, where the transmission is calculated in $50~$comoving Mpc$/h$ pixels.  The histograms show the measurements of \citet{becker15}.  The other curves are the cumulative PDF in numerical models that include the temperature fluctuations from a patchy reionization (solid curves) and that do not include them (dotted curves), from \citet{daloisio15}.  \label{fig:highzforest}}
\end{figure}

At $z\sim 6$, the cosmic mean density saturates in Ly$\alpha$ absorption for $x_{\rm HI} \sim 10^{-5}$ (eqn.~\ref{eqn:tau}), with Ly$\beta$ and Ly$\gamma$ forests extending the range to $x_{\rm HI} \sim 10^{-4}$.  Therefore, the Lyman-series forests are insensitive to the ${\cal O}(1)$ fluctuations in the neutral fraction that define reionization.  Even though this saturation means that the Lyman-series forests cannot be used to directly detect neutral regions,\footnote{See \citet{2015ApJ...799..179M} for suggestions on how to detect neutral regions with deuterium's Ly$\alpha$ transition and with Ly$\alpha$ damping wings within the Ly$\alpha$ forest.} the $z>5$ Lyman-series forest exhibits dramatic evolution in the mean opacity as well as  extremely large spatial fluctuations in the transmission \citep{becker01, fan02, fan06, mortlock11, becker15}.  These trends must contain information about reionization and the transition to a highly ionized IGM, but there is currently little consensus in the interpretation.

The left panel in Figure~\ref{fig:highzforest} shows the Ly$\alpha$ forest towards $19$ of the highest redshift quasars that have been identified, from \citet{fan06}.  These spectra have been used to place a hard limit on the IGM neutral fraction, as any region with detected transmission cannot be fully neutral:  The top right panel in Figure~\ref{fig:highzforest} shows a measurement from such spectra of the fraction of $3.3~$comoving Mpc regions with no transmission in both the Ly$\alpha$ and Ly$\beta$ forests, from \citet{mcgreer15}.  This measurement constrains this fraction to be $<0.06 \pm 0.05$ at $z=5.9$, which translates directly to a bound on the \HI\ fraction if structures are much larger than $3.3\;$Mpc (as occurs in most reionization models).  This constraint suggests that reionization had completed or was nearly complete by $z\approx6$. 

The high-$z$ Ly$\alpha$ forest spectra in the left panel of Figure~\ref{fig:highzforest} also illustrate the steeply increasing opacity in the forest with increasing redshift (wavelength).  Compare the absorption from gas at $z=5$, which falls at $\lambda = 7300\;$\AA, to that at $z=6$, which falls at $8500\;$\AA.  The evolution is even more striking when using Ly$\beta$ and Ly$\gamma$ absorption, with \citet{fan06} finding evidence for at least a factor of two increase in the opacity over the short period from $z\approx5.8$ to $z\approx 6.2$.  It has been speculated that the fast increase in the mean opacity arises from the quick evolution that occurs when ionized bubbles overlap and reionization completes \citep{2004ApJ...610....9G}.  However, it has also been argued that the IGM opacity could evolve over this short timescale even after reionization (see sidebar The \citealt{miralda00} Model for Reionization).  However, given other indications that suggest reionization was ending at $z\sim 6$, it seems likely that this opacity increase is associated with the end of reionization.

More interesting perhaps are the large spatial fluctuations in the Ly$\alpha$ forest opacity at $z>5$.  These opacity fluctuations are larger than those expected in standard models with uniform ionizing backgrounds and power-law temperature-density relations \citep{fan06, becker15}.  Some relevant statistics that have been studied are the dark gap and peak height statistics \citep{2006MNRAS.370.1401G, fan06}.  Another is the cumulative PDF of $\tau_{\rm eff} \equiv -\log({\rm Transmission})$ calculated in in $50/h~$comoving Mpc pixels.  The latter is shown with the histograms in the bottom-right panel in Figure~\ref{fig:highzforest}.  The dotted curves are the predictions from simulations of the standard model for the Ly$\alpha$ forest as described in \S~\ref{ss:Lyaforest}.  Because these curves fall far short of explaining the measured histograms, fluctuations in either the ionizing background, $\propto \Gamma$, and/or the temperature, $T(\Delta_b)$, must source the width of this distribution because $x_{\rm HI} = \alpha(T) n_e /\Gamma$, in contrast to the $z<5$ forest where density inhomogeneities dominate the opacity fluctuations.  \citet{becker15} and \citet{2015arXiv150907131D} attempted to explain the opacity fluctuations with ionizing background fluctuations, showing that models where the mean free path, $\lambda_{\rm mfp}$, is a few smaller at $z=5.6$ than expected from lower-redshift extrapolations can explain most but not all of the width of the cumulative PDF, especially if one includes spatial fluctuations in the mean free path \citep{2015arXiv150907131D}.  Rather than a small $\lambda_{\rm mfp}$, larger fluctuations could owe to very rare sources contributing substantially to the background.  \citet{chardin15} showed that these sources would have to be bright quasars that contribute half or more of the background.  Another hypothesis comes from \citet{daloisio15}, who argued that temperature fluctuations could explain the width the of the PDF.  The solid curves in the bottom-right panel in Figure~\ref{fig:highzforest} show the effect of temperature fluctuations (incorporated using large-box calculations of patchy reionization) for the global ionization histories shown in the very top right subpanel.  They find that the signal can be explained by temperature fluctuations with relatively extended histories.  

\subsubsection{anisotropies in the CMB from reionization}
\label{ss:CMB}

A model independent constraint on reionization comes from measurements of the average Thomson scattering optical depth through reionization from cosmic microwave background (CMB) anisotropies.  If the mid-point of reionization occurred at a redshift of $z_{\rm rei}$, the free electrons generated by this process would result in a Thomson optical depth of $\tau_{\rm es} \approx 0.07 \, [z_{\rm rei}/10]^{3/2}$ for CMB photons. This electron ``screen'' suppresses the anisotropies generated at last scattering by a factor of $\exp(-\tau_{\rm es})$, and more importantly it generates polarization anisotropies at low spherical harmonic multipoles, $\ell$. [Spherical harmonics are the favored basis for decomposing the sky in CMB analyses.]  In particular, a quadrupole temperature anisotropy (which is always present after recombination owing to the Sachs-Wolfe effect) Thomson scatters off electrons generated at reionization, generating linear polarization (as the electrons shake more orthogonally to the bright axis of the quadrupole). This polarization is correlated over the horizon scale at the redshift of scattering (the correlation length of the quadrupole), which translates to polarization fraction fluctuations at multipoles of $\ell \sim \pi \sqrt{z_{\rm rei}}$.  This polarized component of the CMB has fractional amplitude $\sim \tau_{\rm es} \Delta \phi$, where $\Delta \phi \sim 10^{-5}$ is the typical size of the Sachs-Wolfe quadrupole. Comparable low-$\ell$ polarization anisotropies are not generated by other means.  Measuring this signal and its $\ell$ dependence constrains the mean redshift of reionization and sets a bound on the duration \citep{zaldarriaga97, 2003ApJ...595...13H, mortonson08, zaldarriagaCMBPol}.

This large-scale polarization ``bump'' was first detected with the WMAP satellite, estimating $\tau_{\rm es} = 0.17\pm 0.06$ using their first year data, which translates into an average redshift for reionization of $z_{\rm rei} \approx 17\pm 4$ \citep{spergel03}.  Modelers of reionization found it difficult to reproduce such an early reionization as there simply was not enough galaxy formation by this redshift in the concordance $\Lambda$CDM cosmology.  This tension led to a plethora of exotic solutions \citep{2003ApJ...591...12C, 2003ApJ...595....1H}.  This high value of $\tau_{\rm es}$ (combined with the nearly coincident discovery of $z\approx 6$ quasars with SDSS) spawned a period of significant growth in the community working on reionization.  However, it turned out that $\tau_{\rm es}$ was overestimated in the first year WMAP analysis owing to a poor foreground model, and by year three the preferred $\tau_{\rm es}$ was consistent with the ninth and final year value of $\tau_{\rm es} = 0.088\pm 0.014$ \citep{2013ApJS..208...19H}, although still $1\, \sigma$ higher than the current best fit value from the Planck satellite of $\tau_{\rm es} = 0.066\pm 0.016$ \citep{planck}.\footnote{\citet{planck} also finds consistent $\tau_{\rm es}$ when they (1) reanalyze the WMAP polarization data using the newer Planck foreground maps and, remarkably, (2) do not use any polarization data and instead use CMB lensing to break the $\tau_{\rm es} - \sigma_8$ degeneracy that is present in the temperature anisotropies (which results in just $50\%$ larger errors on $\tau_{\rm es}$).}  The Planck constraint corresponds to an instantaneous reionization with redshift $z_{\rm rei} = 8.8^{+1.7}_{-1.4}$, which may even be consistent with reionization by the observed population of galaxies \citep{robertson15}.  The Planck satellite has not yet lived up to forecasts that it would reduce the WMAP error bar on $\tau_{\rm es}$ by a factor of $2.5$ partly because systematics have prevented the full use of their low-$\ell$ polarization data.  Other than improvements in the Planck analysis (which are rumored to be forthcoming), better estimates of $\tau_{\rm es}$ will come with a future large CMB satellite \citep{zaldarriagaCMBPol}.  A cosmic variance-limited $E$-mode polarization measurement would reduce the current error bar on $\tau_{\rm es}$ and $z_{\rm rei}$ by a factor of five and would measure the duration of reionization if this process spanned a redshift interval $\gtrsim 5$ \citep{zaldarriagaCMBPol}.

The large-scale polarization anisotropies constrain the global reionization history but not the structure of this process.  However, reionization is also responsible for small-scale CMB temperature fluctuations from the kinetic Sunyaev Zeldovich effect (kSZ), which owes to Doppler scattering off the relative motions of ionized structures \citep{sunyaev80, ostriker86}.  These structures could just be post-reionization density inhomogeneities or, during reionization, ionized bubbles.  In most models, the kSZ effect is smaller than the other important source of secondary temperature anisotropies at high$-\ell$, the thermal Sunyaev Zeldovich effect (tSZ; \citealt{zeldovich69}), which owes to Compton scattering off hot gas that primarily resides in $>10^{14}\Msun$ halos.  Fortunately, the unique spectral dependence of the tSZ allows it to be separated from the kSZ.  Recently, the South Pole telescope (SPT) has reported a detection of the kSZ with $l^2 C_l/[2\pi] = 2.9\pm 1.3 \; \muK^2$ at $\ell = 3000$ \citep{2015ApJ...799..177G}, not much smaller than their constraint on the tSZ of $l^2 C_l/[2\pi] = 4.1 \pm 0.7 \,\mu$K$^2$ (a surprise given that when SPT was being planned the tSZ was predicted to be more than an order of magnitude larger).\footnote{At $\ell \approx 3000$, both the kSZ and tSZ anisotropies become comparable relative to the primordial ones.  In addition, the anisotropies in the cosmic infrared background are also comparable at the frequencies targeted by ACT and SPT, with its amplitude scaling as $\ell^2$.  Thus, ACT and SPT are most sensitive to the SZ anisotropies at $\ell \approx 3000$.  In physical scales, $\ell \approx 3000$ roughly probes angular separations of $\theta = 2\pi/\ell = 7'$ or comoving distances of $D_A(z=10) \times \theta = 20~$comoving Mpc.}

\begin{figure}
  \centering
    \begin{subfigure}[t]{0.49\textwidth}
    \centering
    \includegraphics[width=1\textwidth]{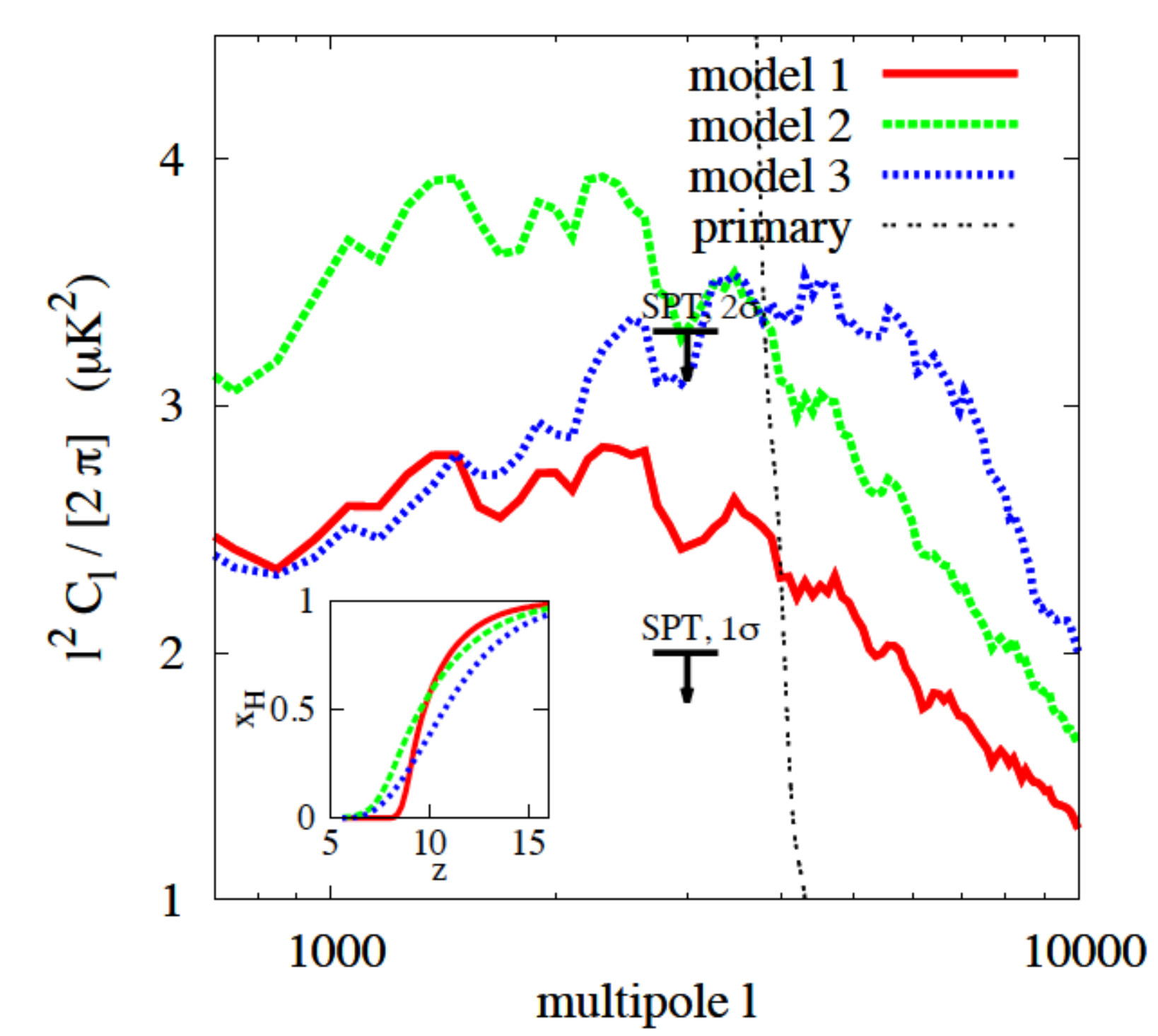}
\end{subfigure}
    \begin{subfigure}[t]{.48\textwidth}
    \centering
\includegraphics[width=.99\textwidth, height=1.\textwidth]{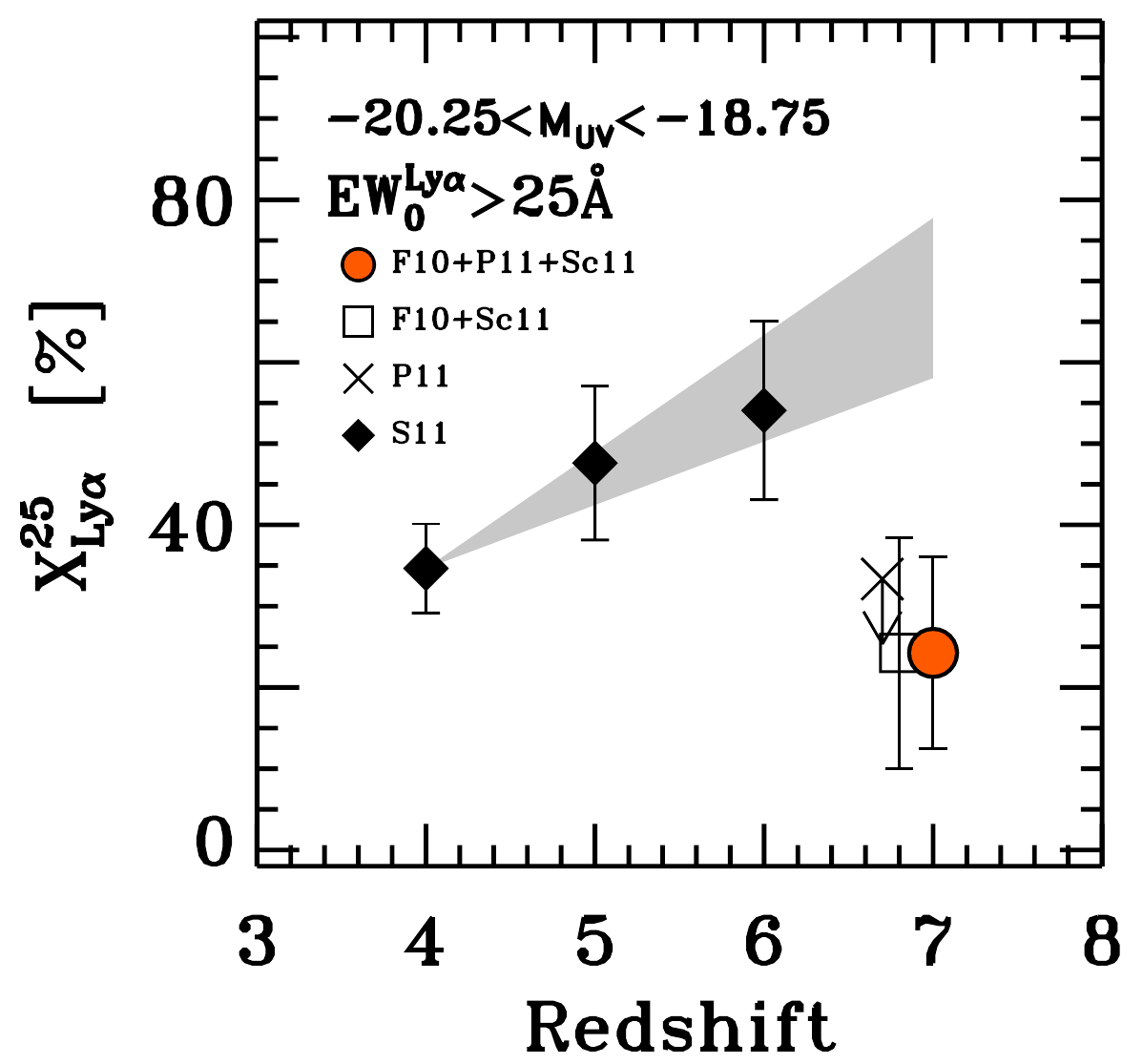} 
\end{subfigure}
\caption{{\it Left panel:} Three models for the kSZ angular power spectrum from patchy reionization presented in \citet[shown in their Fig.~4]{mesinger12}.  The upper bounds are the $1\sigma$ and $2\sigma$ limits at $\ell =3000$ on the patchy reionization contribution from SPT, assuming a minimal post-reionization kSZ of $2.2 \muK^2$.  The inset shows the global ionization histories in these models.
{\it Right panel:}  The fraction of dropout-selected galaxies within the quoted magnitude range that show Ly$\alpha$ emission with equivalent width $>25$\AA, from \citet{ono12}.  The rapid falloff at $z>6$ likely owes to reionization.
\label{fig:LAEs} \label{fig:CMB}
}
\end{figure}

Most of the kSZ signal likely derives from structures after reionization, but the prediction of reionization calculations is that there should be a significant fraction that owes to reionization \citep{1998ApJ...508..435G, 1998PhRvL..81.2004K, 2003ApJ...598..756S, zahn05a, mcquinn05, iliev07}. The kSZ contribution from after reionization, called the Ostriker Vishniac effect \citep{ostriker86, hu00}, is estimated to have amplitude $l^2 C_l/[2\pi] =2.2-3.9\muK^2$ at $\ell=3000$, with the exact value depending on how galactic feedback redistributes the baryons around galaxies \citep{2012ApJ...756...15S}.  In light of the SPT measurement, this amplitude range does not leave much room for the kSZ effect from patchy reionization.   The left panel in Figure~\ref{fig:CMB} compares this limit to three models for the kSZ from \citet{mesinger12}, assuming a minimal post-reionization kSZ of $2.2\mu$K$^2$.  All the models are ruled out at $1\sigma$ by the SPT bound, and two of the models marginally at $2\sigma$.  In most large-box models for reionization, the kSZ signal at $\ell=3000$ falls in the range $1-4~\muK^2$, with the amplitude primarily set by the duration of reionization with other parameters that affect reionization entering at a secondary level \citep{zahn12, mesinger12, 2013ApJ...769...93P}.   Assuming a  post-reionization kSZ on the low end of estimates, \citet{2015ApJ...799..177G} constrained the duration of reionization to be $\Delta z < 5.4$  at 95\% C.L. using large-box reionization models with a linear-in-redshift reionization history.  This bound on the duration is perhaps slightly longer than the duration of reionization in typical models.  Constraints on the kSZ should improve significantly in the next couple years with ACTPol and SPTPol efforts \citep{2010SPIE.7741E..1SN, 2012SPIE.8452E..1EA}.

Other CMB anisotropies from reionization have also been investigated, namely the non-Gaussianities induced by angular fluctuations in $\tau_{\rm es}$ \citep{2009PhRvD..79d3003D, su11}, small-scale polarization anisotropies  \citep{hu00, dore07}, and linear-order Doppler anisotropies that are buried under the primary \citep[][the kSZ is second order but dominates over linear order at $\ell \gtrsim 1000$]{2006ApJ...647..840A, 2008MNRAS.384..291A}.  In standard reionization models, these other anisotropies require at the very least mammoth efforts for a significant detection.

\subsubsection{the \HI\ Ly$\alpha$ damping wing of a neutral IGM}
\label{ss:DW}

If the neutral hydrogen fraction in the $z\sim 8$ IGM is $x_{\rm HI}$, the damping wing of the IGM's Ly$\alpha$ line scatters radiation with an optical depth of $\tau_{\rm DW} \sim x_{\rm HI}$ at $10^3$km~s$^{-1}$ from the resonance \citep{miralda98, mcquinn-GRB, mesinger-dw}.  Thus, scattering from an intergalactic damping wing can extend redward of source-frame Ly$\alpha$, outside of the highly absorbed Ly$\alpha$ forest.  Furthermore, the damping wing optical depth from gas in the Hubble flow scales with frequency as $\nu^{-1}$ rather than the $\nu^{-2}$ for a static cloud \citep{miralda98}, potentially allowing intergalactic damping wing absorption to be distinguished.  The amount of scattering also depends on the size of the ionized bubble around the source, with significant scattering redward of Ly$\alpha$ if the bubble is $\lesssim 1$~physical Mpc (i.e. $10^3$km~s$^{-1}$ of Hubble flow at $z\sim 8$).
 \begin{marginnote}
 Ly$\alpha$ damping wing optical depths:
\entry{static cloud}{$\tau_{\rm DW} = 3 \, \frac{ N_{\rm HI}}{10^{20}/{\rm cm}^2 } (\frac{ 10^3{\rm km/s}}{ \Delta v})^{2} $}
\entry{neutral IGM}{$\tau_{\rm DW} = x_{\rm HI} \left( \frac{1+z}{8.5} \right)^{\frac{3}{2}} \frac{ 10^3{\rm km/s}}{ \Delta v} $}
\end{marginnote}
  
  This effect has been used to place constraints on the neutral fraction at $z=6.3$ using the gamma ray burst observed on 05/09/04 \citep{totani06}.  There have also been claimed detections of the damping wing from a neutral IGM in the spectra of high-redshift quasars \citep{mesinger04}, most notably ULASJ1120+0641 at $z=7.1$, the current redshift record holder \citep{mortlock11}.  Our assessment of the literature is that these constraints and detections are not sufficiently robust to place much weight.\footnote{For gamma ray bursts (GRBs), it is difficult to separate IGM absorption from absorption from \HI\ within the host galaxy without a high quality near infrared spectrum and a relatively small amount of \HI\ in the host galaxy. The case in favor of a successful separation for the GRB on 05/09/04 is not compelling \citep{mcquinn-GRB}.  For quasars the intrinsic shape of their broad Ly$\alpha$ line needs to be known to isolate the damping wing absorption.  Claims that the Ly$\alpha$ line shape is sufficiently understood for the quasar at $z=7.1$ have come under question \citep{2015MNRAS.452.1105B}.}  Eventually we are likely to get lucky and have a spectrum that yields a more convincing detection of a damping wing from neutral patches in the IGM.  

The \HI\ Ly$\alpha$ damping wing from a neutral IGM would also scatter Ly$\alpha$ emission lines from galaxies \citep[e.g.][]{loeb99}, potentially suppressing the observed abundance of galaxies selected by this often-very-luminous emission line \citep[such galaxies are called ``Ly$\alpha$ emitters'';][]{1999ApJ...518..138H, 2001ApJ...563L...5R}.  This effect may explain the rapid decrease in the number density of Ly$\alpha$ emitters detected above $z=6$ \citep{2006Natur.443..186I}.  Damping wing scattering should also suppress the Ly$\alpha$ line of dropout-selected galaxies.  Several groups have detected a decline with redshift between $z=6$ and $7$, and again between $z=7$ and $8$, in the fraction of dropout galaxies that show significant Ly$\alpha$ emission \citep{schenker12, ono12, 2014ApJ...793..113P, 2014ApJ...794....5T},  after an increase in this quantity between $z=4$ and $z=6$ (see the right panel in Fig.~\ref{fig:LAEs}).  These studies have attributed this decrease to neutral intergalactic gas scattering the Ly$\alpha$ emission line.  While the observed decline likely owes to reionization, the total amount of decline is not yet well constrained, especially because sample variance owing to the modulation of emitters by patchy reionization can be quite large in the small fields in which this decline has been measured \citep{taylor14}.  If the full trend holds, standard reionization models require an uncomfortably rapid evolution in the neutral fraction to explain it \citep[to $x_{\rm HI} > 0.4$ by $z=7$; ][]{mesinger15}, although there is debate over whether the dense absorption systems that many of these studies had ignored allows for less rapid evolution \citep{boltonLya,choudhury14, mesinger15}.  The interpretation of trends in Ly$\alpha$ emission is complicated by the fact that Ly$\alpha$ photons can be blue-shifted as the photons scatter out of the host galaxy or ``absorbed'' by flows onto the host, effects that influence the amount of scattering from a neutral IGM \citep{santos04, 2007MNRAS.377.1175D}.\footnote{A more robust probe may be the enhanced clustering of Ly$\alpha$ emitters that models predicts should occur during reionization \citep{furlanetto04c, mcquinn-Lya, mesinger08, sobacchi15}.  In particular, LAEs should only be seen in bubbles with size $\gtrsim 1~$physical Mpc, potentially leading to significant spatial fluctuations during reionization because the LAEs are modulated by the sizes of the ionized bubbles that they sit in.  At $z=6.6$ in the Subaru Deep Field (the highest redshift where there is a sufficient number density for this exercise), no significant excess clustering has been detected \citep{mcquinn-Lya, ouchi10}.}

\subsubsection{redshifted $21\,$cm radiation}
\label{ss:21cm}
\begin{figure}
  \centering
    \begin{subfigure}[t]{0.49\textwidth}
    \centering
    \includegraphics[width=1\textwidth]{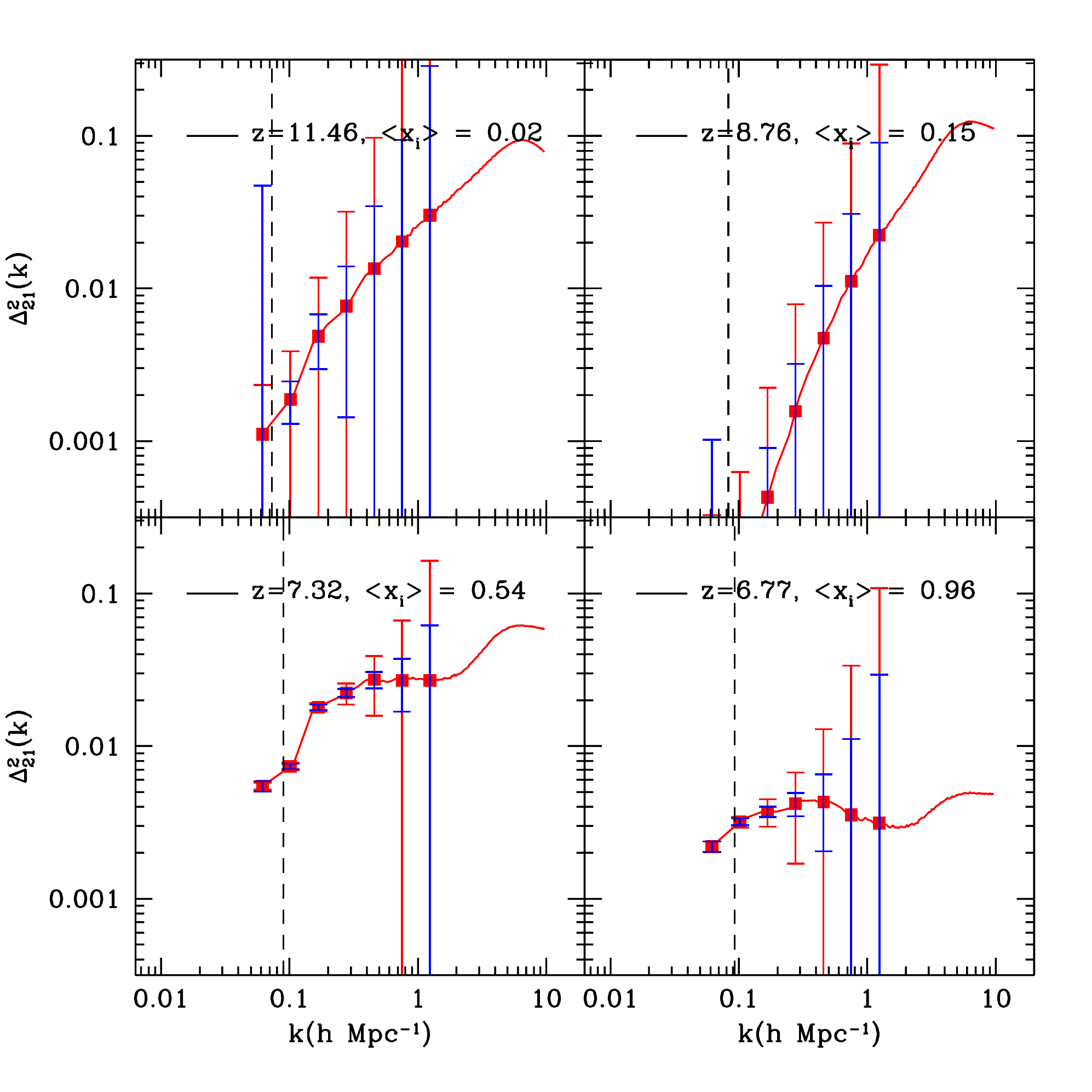}
\end{subfigure}
    \begin{subfigure}[t]{.47\textwidth}
    \centering
\includegraphics[width=.99\textwidth, height=1.\textwidth]{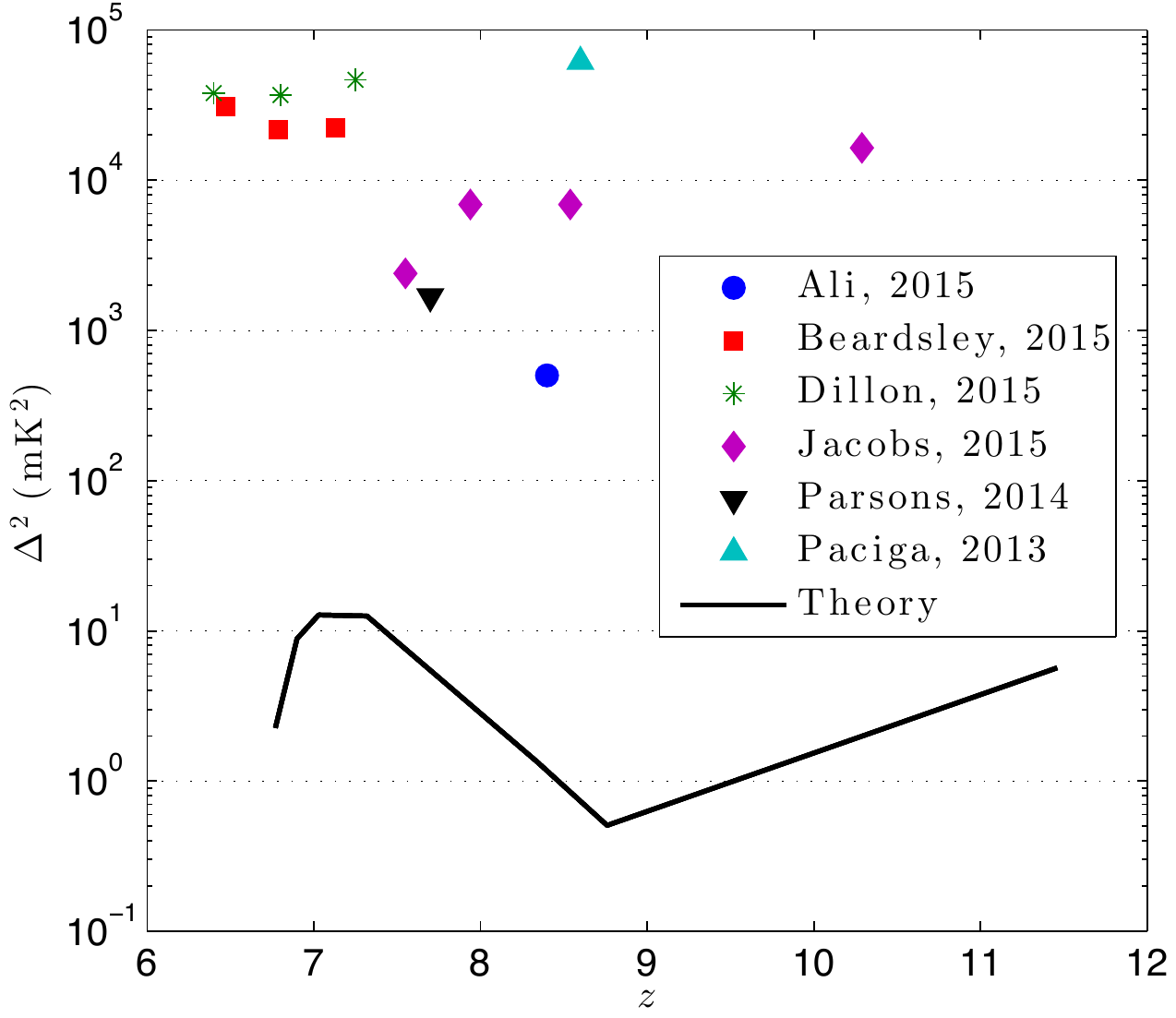} 
\end{subfigure}
\caption{
{\it Left panel}: Curves show the evolution of the $21\,$cm dimensionless power spectrum, $\Delta^2_{21} \equiv k^3 P(k)/2\pi^2$, in a large-box reionization simulation.  The red error bars are forecasted errors for a $1000$\,hr integration using an instrument that is five times more sensitive than the current MWA and assuming perfect foreground removal, from \citet{2008ApJ...680..962L}.  {\it Right panel}:   Published upper bounds on the $21\,$cm dimensionless power spectrum at $k\approx 0.2~$Mpc$^{-1}$ -- the wavenumber at which previous analyses are most sensitive --, from \citet{beardsleythesis}. The bounds fall at least two orders of magnitude above the solid theory curve, which is from the same model as in the left panel.  \label{fig:21cm}}
\end{figure}

The grand hope of reionization aficionados is that the structures during reionization will be mapped using the $21\,$cm hyperfine transition of atomic hydrogen \citep{madau97, 2003ApJ...596....1C, zaldarriaga04}. This observable has the potential to yield vastly more information about reionization than any other probe (and, with additional effort, the era that preceded reionization; \citealt{furl-rev, 2010ARA&A..48..127M, 2012RPPh...75h6901P}). While in principle the $21\,$cm signal provides a tomographic image of reionization like that shown in Figure~\ref{fig:reionization}, the first generations of instruments do not have the sensitivity to make images, and we must hope instead to achieve statistical detections of the signal \citep{mcquinn06}.\footnote{Imaging may be possible if there exist $>100$Mpc \HII\ bubbles (as in many semi-numeric models) especially with LOFAR \citep[which currently has the largest collecting area; ][]{2012MNRAS.425.2964Z}.}  There is a worldwide effort to detect the $z\sim 6-10$ $21\,$cm signal using both specialized and multipurpose interferometers, such as the GMRT in India, MWA in Australia, PAPER in South Africa, LOFAR primarily in the Netherlands, and 21cmArray in China (with planning underway for the next generation instruments HERA and the SKA-low in South Africa and Australia).\footnote{\url{http://gmrt.ncra.tifr.res.in}, \url{http://www.mwatelescope.org}, \url{http://eor.berkeley.edu}, \url{http://www.lofar.org/astronomy/eor-ksp/epoch-reionization}, \url{http://reionization.org}, \url{https://www.skatelescope.org}}  With the exception of the Netherlands, these locations have been chosen to be as isolated as possible from anthropogenic radio transmissions that can easily contaminate this signal, a signal that from $z=6-10$ falls in the well trafficked $130-210~$MHz band.  There are also efforts afoot attempting to detect the sky-averaged (or ``global'') $21\,$cm signal from $z=6-30$.\footnote{\url{http://www.haystack.mit.edu/ast/arrays/Edges/}, \url{http://www.tauceti.caltech.edu/leda/}, \url{http://lunar.colorado.edu/dare/}}  Most of the global instruments consist of a single well-calibrated dipole antenna, in stark contrast to the thousands of dipoles that comprise the MWA, for example.

The challenges associated with detecting the redshifted $21\,$cm signal are daunting.  The foregrounds scale roughly in brightness temperature as $T_b \sim 500 [(1+z)/10]^{2.7}$K, dwarfing the 
\begin{equation}
T_b^{21}({\hat n}, z) = 9 \, x_{\rm HI}\Delta_b \left(1+z \right)^{1/2} \left(1 - \frac{T_{\rm CMB}(z)}{T_{\rm S}} \right)\left(\frac{dv/dx}{H(z)/(1+z)} \right)^{-1}~~ {\rm mK}
\label{eqn:T21}
\end{equation}
signal.  All variables on the right-hand side of equation~(\ref{eqn:T21}) for which an argument is not given are evaluated in direction $\hat n$ and at redshift $z$, and $T_{\rm S}$ is the ``spin temperature'' -- the temperature that characterizes the ratio in the excited to ground hyperfine states \citep[e.g.][]{furl-rev}.  Fortunately, all appreciable foregrounds (synchrotron and bremsstrahlung) are smooth in frequency, in contrast to the $21$cm signal, allowing the foregrounds to be cleanly separated \citep{2011MNRAS.413.2103P}.  (Such a separation is even thought to be possible for the more slowly varying global $21\,$cm signal; \citealt{2015ApJ...799...90B}.)  Realizing this separation with a real instrument is the largest challenge for $21\,$cm efforts.   Even once the $21\,$cm signal has been isolated, current instruments geared towards the fluctuating signal (whose sensitivity owes to $\sim 10^3-10^4$m$^2$ of effective collecting area in a few hundred meter core) require integration times of hundreds of hours to reach sensitivity to the signal in standard reionization models \citep{morales05, mcquinn06, parsons12, beardsley13}.  

The number of publications quoting upper bounds on the $21\,$cm power spectrum has been rapidly increasing in recent years (most recently for the GMRT, PAPER, and MWA efforts in \citealt{GMRT, ali15, dillon15}, respectively).  A compilation of constraints around a wavenumber of $k= 0.2~$comoving Mpc$^{-1}$ are shown in the right panel of Figure~\ref{fig:21cm}, adapted from \citet{beardsleythesis}. (The instrumental sensitivity tends to be maximized at $k= 0.2~$Mpc$^{-1}$ as the foreground removal procedure reduces sensitivity at lower wavenumbers, whereas the errors from thermal noise blow up at higher wavenumbers.)  Currently PAPER has quoted the strongest upper bound on the $21\,$cm power spectrum \citep{ali15}, but the PAPER bound is still $\sim 2$ orders of magnitude below models in the most likely limit of $T_{\rm S} \gg T_{\rm CMB}$.  Nevertheless, the PAPER bound rules out some reionization models in which $T_{\rm S} \ll T_{\rm CMB}$, which could occur if the $X$-rays associated with high-redshift star formation (which are responsible for heating the IGMd) fall on the lower range of the values found at low redshifts \citep{pober15}.  Also of note, the EDGES global $21\,$cm instrument has placed a bound on the reionization epoch duration of $\Delta z > 0.06$ \citep{2010Natur.468..796B}. Improvements on $21\,$cm constraints will come with a better understanding of the instrument and foregrounds, with integrating for longer durations (with errors on the power spectrum decreasing {linearly} with time\footnote{The current PAPER bound at $k=0.2$Mpc$^{-1}$ is not far from the theoretical thermal noise floor.}), and with building larger arrays (such as HERA and the SKA).  

 Once a detection of the $21\,$cm signal from reionization is claimed, it will likely require some convincing for the astrophysical community to believe that the signal is indeed cosmological $21\,$cm radiation.  For the fluctuating signal efforts, such a validation could potentially derive from multiple independent measurements, from a detection in cross-correlation with a co-spatial galaxy survey \citep{2007ApJ...660.1030F, 2009ApJ...690..252L}, or from a detection of temporal evolution in the signal that is consistent with the predictions of reionization models \citep{2008ApJ...680..962L}. 
 
 A definitive detection of $21$cm radiation from reionization will surely motivate searches for this signal from earlier times as, unlike other probes of the $z>5$ IGM, $21$cm radiation is not just a reionization-era probe and has the potential to probe all the way to $z\sim 200$  \citep{madau97, chen04, furl-rev}. 
 Unfortunately, the higher the redshift that is targeted, the brighter the Galaxy, requiring collecting areas that scale as $\sim \lambda^{3}$ to maintain the same thermal noise sensitivity to $T_b^{21}$.  The physics that sets $T_{\rm S}$ and, hence, $T_b^{21}$ is quite rich \citep{field58, madau97, chen04, furl-rev}.  Once the galaxies formed and the Universe achieved comoving star formation rate densities of $\sim 10^{-3}{\rm \Msun ~yr^{-1} ~Mpc^{-3}}$ (with only a weak IMF dependence), enough $10.2-13.6$ eV photons were produced to couple $T_{\rm S}$ to the gas kinetic temperature, $T_{\rm k}$, through the Wouthuysen-Field effect \citep{wout, field58, furl-rev, mcquinnoleary}.  Once $T_{\rm S}$ coupled to $T_{\rm k}$, the $21$cm signal becomes sensitive to how the IGM was pre-heated before reionization, likely by $X$-rays associated with the first stellar deaths \citep[with the signal changing significantly for heating at the level of $\Delta T = T_{\rm CMB}$]{madau97, chen04, furlanetto06}.\footnote{Shocking of the cold IGM, another hypothesized heating source \citep{gnedin04}, has been shown to be small in the low density regions that dominate the signal \citep{mcquinnoleary}.}  The amount of $X$-ray heating from high-redshift galaxies is very uncertain \citep{2014Natur.506..197F, 2014MNRAS.443..678P}, but scalings based on the $X$-ray to star formation rate in low-redshift stellar populations suggests that significant heating likely occurred after $T_{\rm S}$ was coupled to $T_{\rm k}$ but before reionization \citep{furlanetto06, mcquinnoleary}.  Thus, 21cm potentially allows the study of the cosmic gas once the first galaxies turn on, lighting up the gas that is henceforth known as the IGM.

\section{the low-redshift IGM ($z<2$, focusing on $z\sim0$)}
\label{sec:lowzIGM}

\begin{figure}
 \includegraphics[width=1\textwidth]{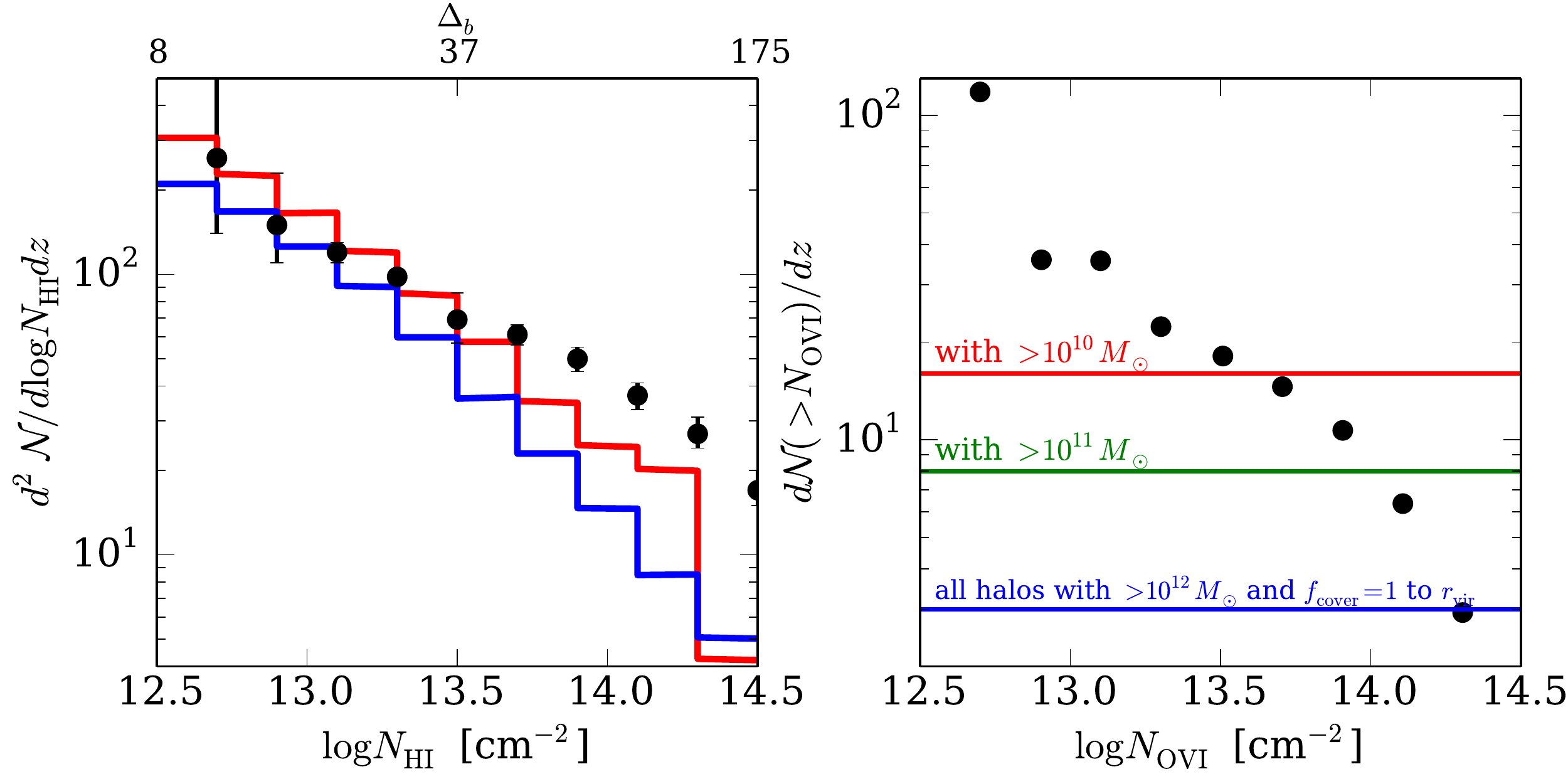}
\caption{Column density distributions of \HI\ (differential) and \OVI\ (cumulative) at $0 <z<0.4$.  The black dots are the measured distributions from \citet{danforth14} using HST/COS data.  The histograms in the left panel show the the \HI\ distribution from the numerical simulations in \citet[red]{kollmeier14} and \citet[blue]{shull15}.   A second $x$-axis is included in the left panel to show the estimated density at a given column using the \citet{schaye01} model.   The horizontal lines in the right panel indicate the expected number of absorbers per $dz$ if halos above the specified mass threshold have a covering fraction of unity out to one virial radius.   
\label{fig:lowz}}
\end{figure}

At lower and lower redshifts, the Ly$\alpha$ forest becomes progressively more transmissive as the Universe is further diluted by cosmological expansion, with an average transmission of $\exp[-\tau_{\rm eff}]$ with $\tau_{\rm eff} = 0.016(1+z)^{1.1}$ over $0<z<1.2$, compared to $\tau_{\rm eff} = 0.36$ at $z=3$ \citep{meiksin06}.  Since this evolution is gradual, this review's boundary of $z=2$ between ``low'' and ``intermediate'' redshifts is of no physical significance.  (It is the case that at $z\lesssim2$ the most prominent lines, the Lyman-series and  \OVI\ $\lambda\lambda 1032, \; 1038$\AA, fall in the ultraviolet, requiring a space telescope.)    Estimates find that only $30\pm 10\%$ of the $z\sim 0$ gas is seen in Ly$\alpha$ absorption \citep{shull12} and that $\sim 10\%$ of the baryons lie within galaxies or reside as hot gas inside galaxy clusters.  This accounting leaves a large fraction that are ``missing''.  Our inability to observe most of the baryons at $z\sim 0$ is referred to as the ``missing baryon problem''.  See \citet{2007ARA&A..45..221B} for a recent review of this problem.
\begin{textbox}[h]
\subsubsection{The $z\sim 0$ \HI-ionizing background} The increasing transmissivity of the \HI\ Ly$\alpha$ forest with decreasing redshift occurs despite \HI\ photoionization rate, $\Gamma$, steadily decreasing at $z<2$ as star formation and quasar activity shut off (in contrast to redshifts of $2<z<5$ during which $\Gamma$ is more or less constant).  The effect on $\Gamma$ of this shutoff is partly compensated by the mean distance ionizing photons travel transitioning to the photon horizon below $z=2$, rather than the mean distance to intersect a Lyman-limit system as at higher redshifts \citep{madau99, faucher09}.  The net result is a decrease in $\Gamma$ by more than an order of magnitude from that inferred at $2<z<5$.  From tuning $\Gamma$ in simulations so that they match the observed \HI\ column density distribution, \citet{shull15}  measured $\Gamma(z) = 5 \times 10^{-14} (1+z)^{4.4}$s$^{-1}$ over $0 < z < 0.5$.   Studies find that this value is consistent with plausible models in which either the ionizing background owes both to quasars and galaxies with $f_{\rm esc} \approx 0.05$ \citep{shull15, 2015MNRAS.451L..30K} or just to quasars if $\Gamma$ falls on the lower end of recent estimates \citep{2015MNRAS.451L..30K}.
\end{textbox}

It is not a surprise that a large fraction of the $z\sim0$ baryons is difficult to observe.  Even for photoionized gas, Ly$\alpha$ absorption results in detectable optical depths only for gas with $\Delta_b \gtrsim 10$.   In addition, numerical simulations show that $30-50\%$ of the baryons by mass (but only $10\%$ by volume) have been shock-heated into a ``warm/hot''  $10^5-10^7$K phase \citep[with the fraction of gas roughly constant per interval in $\log T$ over $10^5-10^7$K at $z=0$;][]{cen99, dave01, croft01, cen06}, temperatures where the hydrogen becomes further ionized by collisions.  Most of the warm/hot intergalactic medium (WHIM) is overdense with $\Delta_b\gtrsim10$, with lower density gas still predominantly at  temperatures characteristic of photoionization.  The WHIM is the result of nonlinear structure formation happening at $z\sim 0$ on scales where the potential wells have depths of $\sim [ 100 \, {\rm km~s}^{-1}]^2$.  (There is a relatively tight correlation between potential well depth and comoving scale in our cosmology, and on-average larger scales become nonlinear later.)  As a result, the baryons are likely to shock at $\sim 100 \, {\rm km~s}^{-1}$, producing $\sim 10^6$K gas \citep{2004ApJ...611..642F}.  The fraction by mass of warm/hot gas is a strong function of redshift with $\{48, ~36, ~21, ~8\}\%$ by mass at $z=\{0, ~1, ~2, ~3\}$ in the ``galactic super-wind'' simulation of \citet{cen06}.   These fractions are even more strongly decreasing with increasing redshift for a higher temperature threshold than $10^5~$K.

Studies find that outflows from galactic feedback can increase the warm/hot fraction by tens of percent \citep{dave01, cen06}. Indeed, the most important problem pertaining to the low-$z$ IGM is not whether the baryons have disappeared between $z=1100$ and $z=0$.  Instead, it is in understanding how galactic feedback redistributes gas around galaxies (and how this redistribution in turn affects how the IGM feeds galaxies).  Only a small fraction, $\sim 5\%$ of the gas that should have funneled onto galaxies by $z=0$ has done so and formed stars \citep{2004ApJ...616..643F}.  Much of this gas has been ejected and redistributed by galactic feedback in a manner that we have yet to fully understand.  Currently, the primary observables to understand the impact of feedback on the $z\sim 0$ IGM are \HI\ Lyman-series and \OVI\ $\lambda\lambda 1032, \; 1038$\AA\ absorption (other detectable absorption lines probe denser circumgalactic and galactic gas).

The column density distribution of both \HI\ (left panel) and \OVI\ (right panel) are shown in Figure~\ref{fig:lowz} for $z\sim0.2$ using the HST/COS measurements of \citet{danforth14}.  Two simulation predictions for the \HI\ are plotted in the left panel of Figure~\ref{fig:lowz}, from \citet{kollmeier14} and \citet{shull15}, using backgrounds with $5\times$ and $\approx 3\times$ higher $\Gamma$ than that of \citet[ \S~\ref{ss:uvbmodels}]{haardt12}, respectively (see sidebar Uniform Ionizing Background Models).  Much like at higher redshifts, the \HI\ column density distribution at $N_{\rm HI} \lesssim 10^{14}$cm$^{-2}$ is found to be only weakly affected by present feedback implementations in simulations \citep{dave10, shull15}.  However at $N_{\rm HI} \gtrsim 10^{14}$cm$^{-2}$, neither simulation shown in Figure~\ref{fig:lowz} agrees well with the data, which in the adaptive mesh refinement (AMR) simulations in \citet{shull15} likely owes to resolution but it may reflect differences between the feedback prescription and reality in the smooth particle hydrodynamics (SPH) simulations of \citet{kollmeier14}, also see \citet{dave01}. This comparison at higher columns should be further investigated, especially in light of these recent COS observations. The upper $x$-axis in left panel of Figure~\ref{fig:lowz} shows an estimate for the overdensity that corresponds to a given column in the model of \citet{schaye01}.\footnote{This model's prediction that $N_{\rm HI}\lesssim 10^{14.5}$cm$^{-2}$ have $\Delta_b < 100$ and hence are intergalactic is consistent with observations that stack on impact parameter to  around galaxies, which show that essentially all absorbers with $N_{\rm HI}\gtrsim 10^{15}$cm$^{-2}$ (and $N_{\rm OVI} \gtrsim 10^{14}$cm$^{-2}$) come from within the virial radius of halos \citep{johnson15}.}  

Turn to the right panel in Figure~\ref{fig:lowz}, which considers \OVI. The horizontal lines show the cross section of the virial radius of halos of different masses.  For the most likely scenario in which \OVI\ owes to $\gtrsim10^{11}\Msun$ halos that are able to efficiently form stars, the galactic winds must reach a couple virial radii. \citet{2005ApJ...623L..97T} performed a similar analysis, finding that to explain the \OVI\ absorption galaxies down to $0.01-0.1~L_*$ are required to enrich the IGM out to $\sim200~$kpc.  Observations show that $N_{\rm OVI}$ has little or no correlation with $N_{\rm HI}$.  This lack of correlation owes to either \OVI\ probing a different gaseous phase of the IGM or to metals not being well mixed.  The simulations of \citet{shull12} and of \citet{2011MNRAS.413..190T} suggest that \OVI\ probes a mixture of phases (both warm photoionized gas and the WHIM).  Aside from at the highest metal columns (which are likely probing circumgalactic material), modern simulations are able to reproduce the \OVI\ column density distribution at the factor of a few level \citep{shull12, 2015MNRAS.446..521S}, suggesting that their galactic wind prescriptions result in enriched ejecta that travel roughly the correct distance from galaxies.  Such simulations predict that much of the enrichment occurred at relatively high redshifts, with \citet{2010MNRAS.409..132W} finding that at least half of the $z=0$ metals were ejected at $z>2$ (with metals residing in lower density gas having been ejected earlier).

Going forward, significant improvements in the quality of the \HI\ and \OVI\ measurements over HST must wait for a future ultraviolet-sensitive space telescope such as HDST \citep{2015arXiv150704779D}.  An unexplored frontier in understanding the WHIM is mapping the unseen $\gtrsim 10^6$K phase.  (Collisional ionizations destroy the \OVI\ by these temperatures.)  Such mapping may be possible with (1) observations of the soft $X$-ray background \citep[predicted to constitute $1-10\%$ of the total observed background;][]{cen99, croft01,2001ApJ...548L.119K}, (2) with isolating the tSZ CMB anisotropy that owes to the WHIM \citep{2006ApJ...643....1A, 2013MNRAS.432.2480G, 2013A&A...550A.134P}, and (3) with studies of \OVII\ and \OVIII\ absorption \citep{1998ApJ...509...56H, 2007ARA&A..45..221B}.  Limits on WHIM emission from the soft $X$-ray background are currently on the upper end of model forecasts, and there have been purported detections of field \OVII\ absorption with Chandra and XMM-Newton \citep{2007ARA&A..45..221B}.  However, a statistical sample of robust \OVII\ and \OVIII\ absorption detections requires a future spectroscopic $X$-ray satellite.   The aforementioned probes of $\gtrsim 10^6$K gas will be most sensitive to dense gas either within or at the outskirts of galactic halos, further informing models of galactic feedback and flows between galaxies and the IGM.  Such information may even solve the great mystery of why an increasing number of galaxies at low redshifts are red and dead, with no recent star formation.

\section{concluding remarks}

 The success of our models at reproducing the \HI\ Ly$\alpha$ forest at $z=2-5$ shows that much is known about the low-density IGM at  intermediate redshifts.  The same models also appear to be consistent with the \HI\ column density distribution (all the way to galactic columns) and with some recent IGM temperature measurements (assuming a late \HeII\ reionization), although these comparisons are less exact.  Despite these successes, there is plenty of room for additional precision in the measurements, in the models, and in their comparison over $z=2-5$.  There are several established and hypothesized processes that could source a detectable departure from our vanilla IGM models.
 
 At much lower and higher redshifts (as well as near galaxies at all redshifts) our understanding of the IGM is far from complete.  At low redshifts and around galaxies, likely the most important unresolved questions are 1) how galactic feedback shapes the IGM and 2) how this in turn affects galaxy formation.  We currently have many measurements that constrain the answers, including of the statistics of \HI\ and \OVI\ absorbers at $z\sim 0$, of the distribution of metal absorbers around massive galaxies at $z\sim 2.5$, and (while beyond the scope of this review) of circumgalactic medium absorption \citep[e.g.][]{tumlinson11, stocke13, werk14}.  However, there is still much to do on both the observational and theoretical fronts as highlighted throughout this review.  At high redshifts, the IGM is shaped by the process of reionization.  We have tenuous indications from several distinct observables that hydrogen reionization ended around $z=6$ and that it extended to higher redshifts.  Yet, we know little about the structure and duration of this important process.  Future observations as well as more in-depth study of the high-redshift Ly$\alpha$ forest, of secondary anisotropies in the CMB, of Ly$\alpha$ emitters, and (ultimately) of redshifted $21\,$cm radiation promise to provide the details that complete this missing chapter in our cosmological narrative.

\section{acknowledgments}
We thank G. Altay, A. Beardsley, A. D'Aloisio, G. Becker, R. Bouwens, X. Fan, C. Fechner, S. Finkelstein, F. Haardt, J. Hennawi, A. Lidz, P. Madau, I. McGreer, G. Mellema, A. Mesinger, Y. Ono, J. Prochaska,  E. Rollinde, P. Upton Sanderbeck, J. Schaye, U. Seljak, R.~Simcoe, M. Turner, M. Walther, D. Weinberg, G. Worseck, and J. Yeh for sharing figures, spectra, and/or providing comments on the manuscript. 
This work is supported from NSF grants AST-1514734 and AST-1312724 as well from NASA through the Space Telescope Science Institute grant HST-AR-13903.00. Some of this work was performed at the Aspen Center for Physics, which is supported by National Science Foundation grant PHY-1066293.

\bibliographystyle{Astronomy}
\bibliography{References}
\end{document}